\documentclass[pra,floatfix,footinbib,notitlepage]{revtex4-1}
\usepackage{amsmath,amssymb,mathtools}
\usepackage[makeroom]{cancel}
\usepackage[euler]{textgreek}
\usepackage[bbgreekl]{mathbbol}
\usepackage{graphicx}
\usepackage{bm,color}
\usepackage{array}
\usepackage{natbib,hyperref}
\usepackage[capitalise]{cleveref}
\usepackage{qcircuit}
\usepackage{enumitem}
\usepackage{amssymb}
\usepackage[explicit]{titlesec}
\hypersetup{breaklinks=true,colorlinks=true,linkcolor=blue,urlcolor=blue,citecolor=blue}

\newcommand{\be}{\begin{equation}}
\newcommand{\ee}{\end{equation}}

\newcommand{\vphi}{\varphi}

\newcommand{\bra}[1]{\ensuremath{\langle#1 |}}
\newcommand{\ket}[1]{\ensuremath{ |#1\rangle}}

\newcommand{\braket}[2]{\ensuremath{\langle#1 \vphantom{#2} |  #2 \vphantom{#1} \rangle}}

\newcommand{\opmatrix}[3]{\ensuremath{\langle#1 \vphantom{#3} | #2 | #3 \vphantom{#1} \rangle}}
\newcommand{\norm}[1]{\left|\left| #1 \right|\right|}
\newcommand{\absv}[1]{\left| #1 \right|}

\def\O{{\cal{O}}}
\def\t{\theta}

\def\P2{\tilde{P}^2}
\def\X2{\tilde{X}^2}
\def\tphi{\tilde{\phi}}
\def\tPhi{\tilde{\Phi}}
\def\tPi{\tilde{\Pi}}
\def\sinc{\text{sinc}}

\def\dH{\tilde{\cal{H}}}

\def\hphi{\hat{\phi}}
\def\hf{\hat{f}}

\def\tF{\tilde{\cal{F}}}
\def\bphi{\bm{\vphi}}
\def\tvphi{\tilde{\vphi}}
\def\tkappa{\tilde{\kappa}}
\def\bk{\bm{\kappa}}
\def\bi{\bm{i}}
\def\bK{\bm{K}}
\def\bF{\bm{F}}

\begin{document}
\title{Bosonic field digitization for quantum computers}
\author{Alexandru Macridin, Andy C.~Y.~Li, Stephen Mrenna, Panagiotis Spentzouris}
\affiliation{Fermilab, P.O. Box 500, Batavia, Illinois 60510, USA}
\date{\today}

\begin{abstract}

 Quantum simulation of quantum field theory is a flagship application of quantum computers that promises to deliver capabilities beyond classical computing.
The realization of quantum advantage will require methods that can accurately predict error scaling as a function of the resolution and parameters of the model and that can be  implemented efficiently on quantum hardware.
In this paper, we address the representation of lattice bosonic fields in a discretized field amplitude basis, develop  methods to predict error scaling, and present efficient qubit implementation strategies.
A low-energy subspace of the bosonic Hilbert space, defined by a boson occupation number cutoff, can be represented with exponentially good accuracy by a low-energy subspace of a finite size Hilbert space. The finite representation construction and the associated errors are directly related to the accuracy of the Nyquist-Shannon sampling and the Finite Fourier transforms
of the boson number states in the field and the conjugate-field bases.
We analyze the relation between the boson mass, the discretization parameters used for wavefunction sampling and the finite representation size. Numerical simulations of small size $\Phi^4$ problems demonstrate that the
boson mass optimizing the sampling of the ground state wavefunction is a good approximation to the optimal boson mass  yielding the minimum  low-energy subspace size. However, we find that accurate sampling of general wavefunctions does not necessarily result in  accurate representation.
We develop methods for validating and adjusting the discretization parameters to achieve more accurate simulations.

\end{abstract}

\maketitle

\section{Introduction}
Numerical simulations of systems with continuous variables, whether
classical or quantum, require digitization and truncation
approximations.
For a simulation to be useful, it is essential to know the limit and
effect of these approximations.
The impact of discretization is especially important when the
computational resources required for simulation are scarce.
This is a concern for present and near-future quantum computations and
classical simulation of complex systems. For example, in the case of strongly correlated systems and lattice field theories, complex schemes are developed~\cite{privman_finite_size,cardy_finite_size}
%,KENNA2004292,lv_nwaa212,degrand_prd_2009}
to extrapolate  the finite size results to the thermodynamic and continuous limits.
Unlike the parameters defining the physical problem under
investigation, the parameters defining the algorithm (discretization
parameters, cutoffs, number of iterations, \emph{etc.}) should be
chosen by the user to optimize the efficiency of the simulations.  To
do this, criteria are needed to assess whether the choice of these
parameters is valid and procedures are needed to adjust them for higher accuracy when necessary. In this paper, we present digitization procedures for bosonic fields, investigate the errors introduced by these procedures and the errors' dependence on the discretization's parameters, and introduce a guide for validating and adjusting the discretization's parameters using feedback from quantum simulations.

Quantum computing offers a change of paradigm for numerical
simulations. Many-body and field theory simulations, severely limited
on classical computers by the exponentially large memory requirement
or the insurmountable Monte Carlo sign problem, might be feasible on future quantum computers.  Nevertheless, due to the characteristics of the hardware used for quantum computations, quantum algorithms require
a radically different way of storing, manipulating  and measuring the
information compared to classical computations. As a consequence,
specific methods are needed for error analysis, benchmarking and validation.

In a commonly used approach for the numerical simulation of continuous field theories, especially for High Energy Physics problems, the space (or the time-space) coordinates are discretized and the
continuous theory is mapped to a lattice field theory.
The lattice field problem is solved numerically
with the best methods available. The continuous field results are obtained by extrapolating
the lattice spacing to zero.
This procedure is well studied in the literature
and is \textit{not} the subject of this work.  In condensed matter problems, the lattice is given by the physical crystalline structure, and this procedure might not even be necessary.
A different approach, which \textit{is} the focus of this paper, involves the discretization and the truncation of the field amplitude and the representation of the lattice field with qubits.

Systems with bosonic degrees of freedom arise
in the Standard Model (Higgs field, gauge fields) and in the low-energy effective models describing collective excitations in condensed
matter physics (phonons, magnons, plasmons, \emph{etc.}). One challenge
in developing quantum algorithms for bosonic systems is
related to the truncation of the Hilbert space, since, unlike fermion
or spin systems,  boson systems can have an unbounded occupation
number.  While it is easy to map a truncated Hilbert space onto
the qubit space in a boson number basis, it is difficult to efficiently implement
the evolution operator in this basis
for many models of interest (such as relativistic scalar field models
and electron-phonon systems).
For this reason, truncation and discretization in the field amplitude
basis has been considered.   The first  quantum algorithm for scalar field theories
using field amplitude discretization was proposed by Jordan
\textit{et al.}~\cite{jordan_science_2012,jordan_qic_2014}. Their error
analysis, based  on the Chebyshev's inequality for estimating the
probability to have large amplitude fields, implies a number of
discretization points per site that scales as $\O(\epsilon^{-1})$,
where $\epsilon$ is the field truncation error.
In fact
~\cite{somma_qic_2016,macridin_prl_2018,macridin_pra_2018,klco_pra_2019},
the number of the discretization points scales exponentially better
than this, \textit{i.e.} $\O\left(\log\left(\epsilon^{-1}\right)\right)$, when
the wavefunction is restricted to a low-energy subspace
defined by a boson number cutoff. This is a consequence of the
properties of the Hermite-Gauss
functions~\cite{macridin_prl_2018,macridin_pra_2018} when using
Nyquist-Shannon sampling.

The main focus of this paper is the representation of the lattice bosonic fields on
the finite space of the quantum hardware.
By {\em representation} of a bosonic field on qubits, we mean two things: \textit{i)} a mapping of the bosonic  wavefunctions to
qubit wavefunctions and, \textit{ii)} an isomorphic mapping of the bosonic field operators to discrete field operators acting on the qubit space.

The paper starts with a general overview of the main results and concepts, in \cref{sec:overview}.

\Cref{sec:low-energy-subspace} builds upon  the work presented in Refs~\cite{macridin_prl_2018,macridin_pra_2018} and addresses
the construction of the finite representation in the field amplitude
basis.  It extends the previous work by providing
a thorough analysis of the errors associated with this construction and investigating the relation between the sampling errors of the field-variable wavefunction and the boson truncation.
By \emph{errors} in this paper, we mean only the theoretical errors
  related to the boson field representation on qubits.  We do not
  consider other errors specific to quantum simulations
that arise from Trotterization, qubit decoherence, gate fidelity, control noise, \emph{etc.}
The construction of the finite Hilbert space is possible because:
\textit{i)}  the boson number wavefunctions both in the field and the
conjugate-field bases can be accurately sampled in a finite number of
points, which is a consequence of the Nyquist-Shannon sampling theorem applied to \textit{almost} band-limited and field-limited functions~\cite{jaming_2016,slepian_ieee_1976, Landau_Pollak_3_1962} and, \textit{ii)} the field and the conjugate field sampling sets
can be accurately connected via a finite Fourier transform.
The accuracy of the finite representation depends upon the errors
arising from sampling, the Finite Fourier transform and the  truncation introduced by the boson number cutoff.
The dimension of the finite Hilbert space is the same as the number of the sampling points. The low-energy subspace is spanned by the boson
number states below a cutoff. For a fixed cutoff, the errors decrease exponentially with increasing number of the sampling points.
Empirically, we find that an accuracy $\epsilon\approx 10^{-4}$
requires a finite Hilbert space dimension that is $2$ times larger than the
dimension of the low-energy subspace. Many
interesting problems, including the broken symmetry phase of the
$\Phi^4$ field model and the intermediate and the strongly coupled
regimes of electron-phonon systems,
can be addressed with no more than $6$ qubits per lattice site.
However, a word of caution is appropriate. While accurate representation implies  accurate sampling, the converse statement is not true. We present examples of functions that can be sampled  with great accuracy but have a significant component outside the low-energy subspace. The action of the discrete field operators on
states outside the low-energy subspace yields uncontrollable
errors.  Therefore, a measurement of the boson distribution is necessary
to ensure that the wavefunction in
a quantum simulation belongs to the low-energy subspace.

The second part of the paper (\cref{sec:bmass}) addresses the
choice of the discretization parameters in quantum simulations.
Different choices of the discretization and sampling intervals correspond to different choices of the boson mass and boson vacuum. The optimal choice of the boson mass corresponds to the minimal boson
number cutoff since  this choice also implies the minimal size of the
finite Hilbert space and implicitly the smallest number of required
qubits for implementation. The optimal boson mass is
interaction-dependent and it is not known \emph{a priori}.
While finding the optimal boson mass by
minimizing the boson number cutoff is impractical, finding the boson
mass that maximizes the accuracy of the wavefunction's sampling is
feasible, requiring only local field measurements.
By employing exact diagonalization methods for small size $\Phi^4$ problems in different parameter regimes,
we find that the boson mass providing optimal sampling corresponds to the optimal boson mass.

In the third part of this paper (\cref{sec:val}),
we describe measurement methods for the local field and the
conjugate-field distributions and additionally  for the local boson
distribution. We also introduce a practical guide for adjusting  and
validating the discretization parameters using the feedback from
quantum simulation measurements. The guideline follows a simple
procedure. First, based on the field distribution measurements, the
sampling intervals are adjusted to minimize the sampling errors.   The
optimal sampling intervals determine the number of discretization
points and the boson mass  to be used in further simulations, provided
that these parameters yield a measured boson distribution below the cutoff. Otherwise, the number of the  discretization points is increased.
Note that the boson distribution
measurement is not needed during the optimization process,
but only as a final check after
the discretization parameters are adjusted.

In \cref{sec:disc} we discuss the applicability
of the discretization method presented here to quantum problems
written in the first quantization formalism and the  challenges for implementing bosonic algorithms on present and future quantum computers.

\Cref{sec:conclusions} contains our conclusions.

\section{Overview}
\label{sec:overview}

The objective of our work is to present a comprehensive study of  bosonic field digitization on quantum computers.  We present our methodology in great detail to allow the readers to build their own models and perform calculations for specific problems.  However, in this section we present a general overview of the main results and concepts.

A general assumption for our method is that
the problem of interest can be addressed
accurately by restricting the  Hilbert space to a finite low-energy subspace defined by a cutoff of maximum $N_b$ bosons per lattice site.

While qubit encoding of the boson number states is straightforward (employing, for example, a binary representation of the boson number), the implementation in the boson number basis of the Trotter step operators corresponding to the field dependent interaction terms  requires a lengthy decomposition in single and two qubit gates, as discussed in \cref{sec:fbasis}.
The implementation of these Trotter steps is much simpler in the field amplitude basis, since the Hamiltonian's field dependent terms are diagonal in this  basis. However, representing
the truncated low-energy subspace in the field amplitude basis has its challenges, caused mainly by the fact that  the field amplitude basis is a continuous and unbounded set. Controlled discretization and truncation procedures are required.  We address the construction of the bosonic field representation in the field amplitude basis in \cref{sec:Fieldrep}.

We start constructing the representation of a local Hilbert space in \cref{ssec:local}
and then, in \cref{ssec:latH}, the representation for the lattice field is constructed as a direct product of local (one at each lattice site) representations. The construction of the local representation
is based on the discretization properties of
the Hilbert space's vectors in the field amplitude basis. In this basis the vectors are equivalent to square integrable functions. Their weight at large argument decreases fast with increasing the argument. The same statement is true for the Fourier transform of these functions. The Nyquist-Shannon sampling theorem  can be employed to approximate these functions and, as well, their Fourier transforms. A field variable wavefunction can be reconstructed with $\O(\epsilon)$ accuracy from its value in a finite set of sampled points. Analogous the   Fourier transform of the wavefunction can be reconstructed with $\O(\epsilon)$ accuracy from its values in a finite set of conjugated-field sampled points. The set of field sampling points and the set of conjugate-field sampling points are related with $\O(\epsilon)$ accuracy via a Finite Fourier Transform. The error $\O(\epsilon)$ can be decreased by increasing
the width of the field and conjugate-field sampling windows. In \cref{app:tails,app:FFTerror} we calculate upper bounds for the sampling errors, relating these bounds to the wavefunction's weight outside the field and conjugate-field sampling windows.

To construct the local representation we focus on the sampling properties of boson number states written in the field amplitude basis.
Both the  boson number states in  the field amplitude basis and their Fourier transforms are proportional to Hermite-Gauss functions.
For a cutoff $N_b$ and
an accuracy $\epsilon$ a finite number of discretization points $N_\vphi > N_b$ can be chosen such that all boson states with $n<N_b$ can be sampled with $\O(\epsilon)$ accuracy
in $N_\vphi$ field-variable points or $N_\vphi$ conjugate-field-variable points. The sampling and the recurrence properties of the Hermite-Gauss functions allows us to define a $N_\vphi$ finite size Hilbert space $\dH$ and discrete version of the field and conjugate field operators,
$\tPhi$ and $\tPi$, acting on $\dH$. On the subspace of $\dH$ spanned by the first $N_b$ eigenvectors of the discrete harmonic oscillator Hamiltonian ({\em i.e.} constructed with the discrete field operators, $\tPhi$ and $\tPi$, see \cref{eq:dhosc}) the discrete field operators obey the canonical commutation relation $\left[\tPhi,\tPi\right]=iI+\O(\epsilon)$. For a problem of interest, as long as $N_b$ is taken large enough such that the contribution of the boson states with $n>N_b$
can be neglected,  the infinite Hilbert space can be replaced by $\dH$ and the field operators $\Phi$ and $\Pi$ can be replaced by $\tPhi$ and $\tPi$ with $\O(\epsilon)$ accuracy. The number of the qubits required for a local representation is $n_q=\log\left(N_\vphi\right)$.
The representation for a $N$ site lattice field,  requires $N \log \left(N_{\vphi}\right)$ qubits.

In practice it is essential to quantify and control the errors.  In the last part of \cref{ssec:local}  a numerical analysis of the errors involved in the construction of the finite representation is presented. For $N_\vphi=64$, $N_\vphi=128$ and
$N_\vphi=256$ we calculate the sampling errors and the error associated with the commutations relation of the discrete field operators. These errors are proportional to the tail weights of the boson number states outside sampling interval windows.
For a fixed $N_b$ the representation error can be reduced exponentially by increasing the number $N_\vphi$ of the discretization points. The ratio
$N_b/N_\vphi$ belongs to $[0.3, 0.7]$ when the error is in the range $\left[10^{-5}, 10^{-3}\right]$. For example, a finite representation with an accuracy of order $10^{-4}$ can be obtained by taking $N_\vphi =2 N_b$. Encoding this representation requires only one extra qubit (per site) when compared to the encoding in the boson  number basis.

The relation between the sampling accuracy of a general wavefunction and its projection onto the low-energy subspace defined by the boson number cutoff is further addressed in \cref{ssec:leval}.
While belonging to the low-energy subspace implies accurate sampling (consequence of the representation's construction described in \cref{sec:Fieldrep}), we find that the converse is not true. We present two examples of functions with small tail weights  outside sampling intervals which can be discretized with very good accuracy but have significant weight onto the subspace spanned by boson states with $n>N_b$. As a consequence, the discrete field operators acting on these functions produce uncontrollable errors. Accurate discretization of bosonic field wavefunctions is not enough to ensure the accuracy of the numerical simulations. Boson number distribution measurements are required to ensure the wavefunction belongs to the low-energy subspace.

The construction of the field amplitude representation depends on the definition of bosons, which is not unique. The boson creation and annihilation operators depends on the mass parameter.
Different mass bosons are related by a  squeezing operator (Bogoliubov transformation).
Different choices of the boson mass correspond
to different representations. A representation
which requires the smallest truncation cutoff $N_b$ for a given accuracy is optimal, since it requires the smallest amount of resources for algorithm implementation.

In principle the optimal boson mass can be determined by optimizing the boson distribution as a function of the mass parameter. However, this approach is impractical, since boson distribution measurement is expensive in quantum simulations. On the other hand the measurements of the local field and conjugate-field distribution is straightforward (as discussed in \cref{ssec:meas}). Calculating the sampling windows which minimize the sampling errors of the wavefunction is much easier than optimizing the boson mass  for the smallest cutoff $N_b$. In \cref{sec:bmass} we investigate the relation between the optimal sampling intervals and the optimal boson mass.

For a given number of the discretization points, the sampling and Finite Fourier Transform errors are the smallest when the weight of the wavefunction outside the field sampling interval $F$ equals the weight of the wavefunction's Fourier transform outside the conjugate-field sampling interval $K$. For this choice of the sampling intervals, is the ratio $K/F$, which equals the representation's boson mass,  the same as the optimal boson mass? While we don't know the answer in general, numerical simulation for small size lattices find the answer to be {\em yes} in many cases. Several examples are presented.

The harmonic oscillator case is illustrated first in \cref{ssec:squeezedb}. The optimal boson mass is equal to the harmonic oscillator mass parameter $m_0$, since in this case the ground state is the vacuum state. When  the boson mass $m_1$ is larger (smaller) than $m_0$, for a fixed truncation error,  the cutoff number $N_b$ increases linearly with increasing the  ratio $m_1/m_0$ ($m_0/m_1$).  The optimal boson mass can be obtained by optimizing the sampling errors. The ratio $K/F=m_0$ when $F$ and $K$ are chosen such that the the weight of the wavefunction outside the interval $F$ equals the  weight of the wavefunction's Fourier transform outside the interval $K$.

Two examples of interacting systems, a local $\phi^4$ scalar field  (\cref{ssec:1phi4}) and a two-site $\phi^4$ scalar field with imaginary mass (\cref{ssec:2phi4}) are also presented.
In both cases the ground state local field distribution is narrower than the local conjugate-field distribution. Optimal sampling requires the ratio $K/F$ to be larger than the Hamiltonian mass parameter. The ratio $K/F$ determined this way agrees with the optimal boson mass obtained by optimizing the boson number distribution.

 In order to enhance the fidelity of applications using our methodology, procedures for validating and adjusting the discretization parameters $N_\vphi$ and $m$ for optimal performance, using feedback from quantum simulations,
are presented in \cref{sec:val}.
The procedures require measurements of the local field distribution, the local conjugate-field distribution and the local boson distribution. These measurements, described in \cref{ssec:meas}, are local, involving  the register of $\log(N_\vphi)$ qubits assigned to encode the bosonic field at one lattice site. The field and conjugate-field distributions require a direct measurement of the qubits. The boson distribution measurement is more laborious. We present two methods for the boson distribution measurement. The first one employs quantum state tomography~\cite{Altepeter2004,nielsen_chuang_2010} of the local qubit register of size $\log(N_\vphi)$. The second method is done by employing Quantum Phase Estimation method~\cite{nielsen_chuang_2010,cleve_1998_procRsocA} for a local harmonic oscillator and requires an ancillary register of $\log(N_\vphi)+1$ qubits. The boson distribution can be measured with great accuracy  since the energy levels of a harmonic  oscillator are equidistant.
The probability of having bosons above the cutoff $N_b$ is given by the probability to measure
integers larger than $N_b$ in the ancillary register.

Finally, to support efficient utilization of compute resources, a practical guide for adjusting the discretization parameters in order to improve quantum simulation's performance is proposed in \cref{ssec:guide}. The initial discretization intervals are determined by assuming a mean-field value for the boson mass. Simulations are run and the local field and conjugate-field distributions are measured. The sampling intervals are adjusted to optimally cover the regions where the field and the conjugate-field distribution have significant support. New simulations which measure the boson distribution are run. If the number of  bosons above the cutoff $N_b$ is negligible ({\em i.e.} it is of the order of the desired accuracy) the discretization parameters are good and the simulation's results can be trusted. Otherwise the number of the discretization points $N_\vphi$ should be increased to accommodate for a larger cutoff $N_b$.

\section{Low-energy subspace representation}
\label{sec:low-energy-subspace}
The Hilbert space of a lattice bosonic field is a direct product of local Hilbert spaces
at each lattice site.  Every local Hilbert space  is infinite  dimensional,
but for most problems can be represented by a finite subspace that contains
the relevant degrees of freedom.
In general, the relevant degrees of freedom
depend on the problem under investigation.
In this work, we study the low-energy physics of a field theory
where a cut off $N_b$ on the boson occupation number can be imposed at each site, such that the
states with more than $N_b$ bosons per site can be safely neglected.
First we briefly discuss the problems associated with the  representation of the bosonic field in the boson occupation number basis. Then we address the bosonic field representation in the field amplitude basis.

\subsection{Representation in the occupation number basis}
\label{sec:fbasis}

The lattice boson number states are a direct product of single site boson number states.  At each site the boson number states $|n\rangle$ are eigenstates of the harmonic
oscillator Hamiltonian:
\begin{align}
\label{eq:hosc}
H_{h}=\frac{1}{2}\Pi^2+\frac{1}{2} m_0^2 \Phi^2= m_0 \left(a^{\dagger} a +\frac{1}{2} \right).
\end{align}
The creation and the annihilation operators, $a^{\dagger}$ and $a$, are related to the field operators by
\begin{align}
\label{eq:phi_aa}
\Phi = \frac{1}{\sqrt{2 m_0}} \left(a+a^{\dagger}\right)~~\text{and}~~~
\Pi = -i\sqrt{\frac{m_0}{2}} \left(a-a^{\dagger}\right),
\end{align}
\noindent and $\ket{n}= \frac{1}{\sqrt{n!}} a^{\dagger n} \ket{0}$, where $\ket{0}$ is the boson vacuum state.

The boson number  basis has been used extensively for numerical simulations of bosonic
fields on classical computers. For field theories, it is intuitive
to define a low-energy subspace by introducing a cutoff $N_b$ in the boson number states.
The cutoff is chosen such that the states with
more than $N_b$ bosons have a negligible contribution to the low-energy physics. In
general, the cutoff $N_b$  depends on the interaction type and strength, but also on
the boson mass parameter $m_0$, as can be seen in  \cref{eq:phi_aa}.
A particular choice of the boson mass $m_0$ makes the most efficient use of the
computational resources, as we will discuss in Section~\ref{sec:bmass}.

At each site, boson number states truncated to a cutoff $N_b$  can be easily encoded on $n_q=\log(N_b)$ qubits of a quantum computer. For example, a binary representations of the integer number $n$ can be used. Different encodings are also possible~\cite{sawaya_npj_2020}. However, quantum
computation using the boson number representation is
difficult to implement in models with field amplitude dependent coupling when the cutoff $N_b$ is of the order of $10$ or larger ({\em i.e.} when $n_q>3$). For example,
let's consider coupling terms such as  $\sum_{\langle j,l \rangle} \Phi_{j} \Phi_{l}$  present in $\Phi^4$ theory or in the phonon models, where $j$ and $l$ are nearest-neighbor lattice site indices. The correspondent Trotter step unitary operator,
\begin{align}
e^ {-i \t \Phi_j \Phi_{l}} =e^{-i  \frac{\t}{2m_0}\left(a_j^{\dagger}a_l^{\dagger}+ a_j^{\dagger} a_l + a_j a_l^{\dagger} +a_j a_l
\right) },
\end{align}
\noindent have a dense matrix representation. Since a general unitary of size $k$ requires $\O(4^{k})$ CNOT gates~\cite{Barenco_pra_1995,shende_pra69_2004,krol_2021_arxiv} this Trotter step requires a lengthy decomposition with $\O(4^{2n_q})$ two-qubit gates (in this case $k=2n_q$ because bosons at two different sites are involved). Similarly, the Trotter step operators for  $\frac{\lambda}{4!}\sum_j \Phi_j^4$ interaction in $\Phi^4$ theory or for electron-phonon coupling in phonon models requires a decomposition with $\O(4^{n_q})$ two-qubit gates (in this case  bosons at only one site are involved, hence $k=n_q$).

For weakly interacting problems with
small number of bosonic excitations, quantum algorithms implemented using a boson number representation for the  bosonic field might be feasible. The study of different encoding schemes presented in Ref~\cite{sawaya_npj_2020} finds that the efficiency of a particular encoding is heavily dependent on the model and on the truncation cutoff.  In this study we propose a finite representation suitable for quantum algorithms addressing both weakly and strongly interacting field theories.

\subsection{Representation in the field amplitude basis}
\label{sec:Fieldrep}

We consider first the local field construction and then we extend it to lattice field.

\subsubsection{Representation of the local Hilbert space}
\label{ssec:local}

In this section, we address the finite representation of local Hilbert space at a particular lattice site.
The local Hilbert space is specified by the field and the conjugate-field operators,
$\Phi$ and $\Pi$, satisfying the canonical commutation relation
\begin{align}
\label{eq:ho_comm}
\left[\Phi, \Pi \right]=iI.
\end{align}

The local Hilbert space admits continuous bases, such as
the field and the conjugate-field variable ones, and  denumerable bases.
In the field variable basis, the local Hilbert space
is the space of the square integrable functions,
$L^2( \mathbb{R} )$.
The boson number states, discussed in \cref{sec:fbasis}, are an example of a denumerable basis.

Considering the difficulties associated with the implementation of Trotter step operators for
field amplitude dependent interaction terms in the boson number basis,
a more convenient basis for quantum computation  is
the field amplitude basis $\{\ket{\varphi}\}$. Here $\{\ket{\varphi}\}$ are the
eigenvectors of the field operator, {\em i.e.}
$\Phi\ket{\varphi}= \varphi \ket{\varphi}$.
The field dependent interaction terms and the corresponding Trotter step operators are diagonal in this basis and easy to implement in a quantum algorithm~\cite{macridin_pra_2018,macridin_prl_2018,Li_2021}. However, the eigenvectors $\{\ket{\varphi}\}$ are Schwartz distributions
and  not proper vectors of the Hilbert space. The eigenspectrum of the
field operators is continuous and unbounded,
but a representation suitable for quantum computation
requires discretization and truncation procedures.
An apparent difficulty to introducing a finite representation
for field operators is caused by their commutation relations.
It is known (see for example Ref~\cite{Gieres_2000}) that the canonical commutation relations cannot be satisfied on a finite dimensional space,
 since on a finite dimensional space the trace of the left hand side of   \cref{eq:ho_comm} is zero and the trace of the right hand side is not. However,
we  construct (see Section~\ref{para:constr}) a finite Hilbert space $\tilde{ {\cal{H}} }$ with
a dimension $N_\vphi$ larger than the boson number cutoff $N_b$
to represent the low-energy subspace of dimension $N_b$.
We define the field operators
$\tilde{\Phi}$ and $\tilde{\Pi}$  on the finite Hilbert space such that $\left[\tPhi, \tPi\right]I_{N_b}=iI_{N_b}$,
where $I_{N_b}$ is the projector operator onto the low-energy subspace spanned by the first $N_b$ eigenvectors of the harmonic oscillator Hamiltonian. The algebra generated by the operators $\tilde{\Phi}$ and $\tilde{\Pi}$ is isomorphic with the algebra generated by $\Phi$ and $\Pi$, when both are restricted to the low-energy subspace.

The construction of the finite representation in the field amplitude basis is based on the discrete sampling of the square integrable functions, which  is discussed in the next section.
\\

\paragraph{Nyquist-Shannon sampling of wavefunctions}
\label{para:NSth}
The field amplitude representation of the low-energy subspace is directly related to
the discretization and the truncation of wavefunctions  belonging to $L^2(\mathbb{R})$ space.
The discretization procedure takes advantage of the fact that the weight of the square integrable functions
 at large argument is small and decreases with increasing argument.

To simplify our analysis we consider  arbitrary wavefunctions $f(\vphi)\in S(\mathbb{R})$, where $S(\mathbb{R})$ is the Schwartz space
 containing the smooth and rapidly decaying functions. The Schwartz space
is dense in $L^2(\mathbb{R})$~\cite{becnel_2015,schwartz_space,schwartz_space_math}.
The Fourier transform
\begin{align}
\label{eq:ftdef}
\hf(\kappa)&=\frac{1}{\sqrt{2 \pi}}\int_{-\infty}^{\infty}  f(\vphi) e^{-i \kappa \vphi} d \vphi,
\end{align}
\noindent also belongs to $S(\mathbb{R})$.

We introduce the field limiting projector on the interval $\left[-F,F\right]$
\begin{align}
\label{eq:PF}
P_F=\int_{-F}^{F}  \ket{\vphi}\bra{\vphi} d\vphi
\end{align}
\noindent and the tail vector
\begin{align}
\label{eq:tailF}
\ket{w_F^f}&=\left(1-P_F \right) \ket{f} \equiv Q_F \ket{f},
\end{align}
with $\ket{f}=\displaystyle\int f(\vphi)\ket{\vphi} d \vphi $.
The norm of $\ket{w_F}$ is equal to the tail weight of $f(\vphi)$
outside the interval $\left[-F,F\right]$,
\begin{align}
\label{eq:tail_f}
||w_F^f|| =\left(\int_{-\infty}^{-F} |f(\vphi)|^2 d\vphi + \int_{F}^{\infty}  |f(\vphi)|^2 d\vphi\right)^{\frac{1}{2}}.
\end{align}
Similarly, we introduce the conjugate-field limiting (we will also call it  band-limiting borrowing a signal
processing common nomenclature) projector on the interval $\left[-K,K\right]$,
\begin{align}
\label{eq:QK}
P_K=\int_{-K}^{K} \ket{\kappa}\bra{\kappa} d\kappa
\end{align}
\noindent and the tail vector
\begin{align}
\label{eq:cutk_function}
\ket{w_K^f}&=\left(1-P_K\right) \ket{f} \equiv Q_K \ket{f}.
\end{align}
\noindent The norm of $\ket{w_K^f}$ is equal to the tail weight of $\hf(\kappa)$
outside the interval $\left[-K,K\right]$,
\begin{align}
\label{eq:tail_hf}
||w_K^f|| = \left(\int_{-\infty}^{-K} |\hf(\kappa)|^2 d\kappa + \int_{K}^{\infty}  |\hf(\kappa)|^2 d\kappa\right)^{\frac{1}{2}}.
\end{align}
The tail weight of $f(\vphi)$ outside the interval $\left[-F,F\right]$ can be made as small as desired by increasing $F$. In the
literature~\cite{jaming_2016,slepian_ieee_1976,Landau_Pollak_3_1962},
functions with $\epsilon$ small tail weigh
are called \textit{almost} field-limited functions.
Analogously, the tail weight of $\hf(\kappa)$ outside the interval
$\left[-K,K\right]$ can be made as small as desired by increasing $K$.
The function $f(\vphi)$ is \textit{almost} band-limited.

When $||w_K^f||$ is small, the
vector $\ket{f}$ can be considered band-limited to a good approximation, {\em i.e.} $\ket{f} \approx P_K\ket{f}$.
The Nyquist-Shannon sampling theorem~\cite{Shannon_1949} for band-limited functions can be employed.
The following approximation for $f(\vphi)$ (see
Appendix~\ref{app:NShlf}) follows:
\begin{align}
\label{eq:NS_K}
f(\vphi) \approx \opmatrix{\vphi}{P_K}{f}=\sum_{i=-\infty}^{\infty} \opmatrix{\vphi_i}{P_K}{f} u_K(\vphi-\vphi_i)
\approx \sum_{i=-\infty}^{\infty} f(\vphi_i) u_K(\vphi-\vphi_i),
\end{align}
\noindent where
\begin{align}
\label{eq:phiiuk}
\vphi_i=i \Delta_{\vphi},~~~
%\label{eq:deltax}
\Delta_{\vphi}=\frac{\pi}{K},~~~\text{and}~~~
%\label{eq:uk}
u_K(\vphi)=\sinc\left( \frac{\vphi}{\Delta_{\vphi}} \right) \equiv
\frac{\sin \left(\pi \frac{\vphi}{\Delta_{\vphi}}\right)}{ \pi \frac{\vphi}{\Delta_{\vphi}} }.
\end{align}
\noindent
Moreover, $f(\vphi)$ is small for $|\vphi|>F$ when $F$ is large.
The summation in  \cref{eq:NS_K} can be restricted to a finite number $N_\vphi$ of points
\begin{align}
\label{eq:NS_app1}
f(\vphi) \approx \tilde{f}_\vphi(\vphi)=\sum_{i=-\frac{N_\vphi-1}{2}}^{\frac{N_\vphi-1}{2}} f(\vphi_i) u_K(\vphi-\vphi_i),
\end{align}
\noindent when the condition  $N_\vphi \Delta_\vphi \ge 2F$ is fulfilled, {\em i.e.} when
the sampling points cover the window interval $\left[-F,F\right]$ where $f$ has significant support.
Note that the Nyquist-Shannon theorem commonly described  in the literature
considers the summation index $i$ in  \cref{eq:NS_K} to take
integer values, but this is easily generalized to half-integer values
(see Appendix~\ref{app:NShlf}),
which are more convenient for an even number of
discretization points (as required by a qubit representation).

According to  \cref{eq:NS_app1}, the wavefunction $f(\vphi)$ can be approximated by a finite  expansion
of {\em sinc} functions with the coefficients equal to the value of the function in
\begin{align}
\label{eq:Ncphi}
 N_{\vphi}=\lceil \frac{2}{\pi}F K \rceil
\end{align}
\noindent equidistant points.
In  \cref{eq:Ncphi} the notation $\lceil x\rceil$ means the {\it ceiling function} applied to the real number $x$, and is equal to the least integer greater than or equal to $x$.  Finding analytical bounds for the accuracy of this approximation
is not straightforward, see for example Ref~\cite{Landau_Pollak_3_1962}.
We claim that (see Appendix~\ref{app:tails}) a bound for  \cref{eq:NS_app1} is:
\begin{align}
\label{eq:1error}
||f-\tilde{f_\vphi}|| \lesssim  ||w_K^f||+||w_F^f|| + \frac{\pi r_K^f }{2K} +
\sqrt{\frac{\pi}{2K}\left(|f(-F)|^2+|f(F)|^2\right)}
\end{align}
\noindent where $r_K^f$ is the  weight of $\kappa\hf(\kappa)$ outside the interval
$\left[-K,K\right]$,
\begin{align}
\label{eq:rfk2}
r_K^f=\bigg(\int_{-\infty}^{-K} \kappa^2 |\hf(\kappa)|^2 d \kappa+\int_{K}^{\infty} \kappa^2 |\hf(\kappa)|^2 d \kappa\bigg)^{\frac{1}{2}}.
\end{align}
\noindent All terms in  \cref{eq:1error}
vanish rapidly in the limit of large $F$ and $K$ for the rapidly decaying functions belonging to the Schwartz space.

Using the same reasoning, the conjugate-field variable functions can approximated
by  a finite  expansion of $N_\vphi$ {\em sinc} functions
\begin{align}
\label{eq:NS_fk}
\tilde{f}_\kappa(\kappa) = \sum_{p=-\frac{N_\vphi-1}{2}}^{\frac{N_\vphi-1}{2}} \hf(\kappa_p) u_F(\kappa-\kappa_p),
\end{align}
\noindent with
\begin{align}
\label{eq:kappapp}
\kappa_p=p \Delta_{\kappa},~~~
%\label{eq:deltak}
\Delta_{\kappa}=\frac{\pi}{F},~~~\text{and}~~~
%\label{eq:uF}
u_F(\kappa)=\sinc\left( \frac{\kappa}{\Delta_{\kappa}} \right).
\end{align}
\noindent The vector $\ket{\tilde{f}_\kappa}$ differs from $\ket{f}$ by
\begin{align}
\label{eq:2error}
||f-\tilde{f}_\kappa|| \lesssim  ||w_K^f||+||w_F^f|| + \frac{\pi r_F^f }{2F} +
\sqrt{\frac{\pi}{2F}\left(|\hf(-K)|^2+|\hf(K)|^2\right)}
\end{align}
\noindent where $r_F^f$ is the weight of $\vphi f(\vphi)$ outside the interval $\left[-F,F\right]$,
\begin{align}
\label{eq:rfF2}
r_F^f=\bigg(\int_{-\infty}^{-F} \vphi^2 |f(\vphi)|^2 d \vphi+\int_{F}^{\infty} \vphi^2 |f(\vphi)|^2 d \vphi\bigg)^{\frac{1}{2}}.
\end{align}

The accuracy of both approximations of $\ket{f}$, $\ket{\tilde{f}_\vphi}$ and
$\ket{\tilde{f}_\kappa}$ are determined by the values of $f(\vphi)$ and $\hf(\kappa)$
outside the intervals $\left[-F,F\right]$ and $\left[-K,K\right]$, respectively. Note that $\ket{\tilde{f}_\vphi}$
is a band-limited function and
$\ket{\tilde{f}_\kappa}$ is a field-limited function, while $\ket{f}$ isn't necessary
band-limited or field-limited. An approximation of $\ket{f}$ that is both band-limited and field-limited
does not exist, since no analytical function, except the zero function,
can be simultaneously band-limited and field-limited~\cite{Landau_Pollak_3_1962,slepian_ieee_1976,Engelberg_2008}.

%%\paragraph{Finite Fourier transform of sampled points.}
%%\label{para:FFT}

The vector $\ket{f}$ can be reconstructed from a
set containing the field sampled values $\{f(\vphi_i)\}_i$
or from a set containing the conjugate-field  sampled values $\{\hf(\kappa_p)\}_p$.
The accuracy of the reconstruction is determined by the values of $\ket{f}$
outside the field and conjugate-field sampling intervals.
 However, accurate sampling is only a
necessary condition for the representation of the bosonic field on quantum hardware.
A quantum algorithm also requires implementation of unitary operators that
can describe accurately the evolution of the system. While the field and conjugate-field functions $f(\vphi)$
and $\hf(\kappa)$ are related by a continuous Fourier transform, the representation for bosonic fields on qubits is based
on the assumption that a Finite Fourier Transform (FFT) connects the sampling sets
$\{f(\vphi_i)\}_i$ and $\{\hf(\kappa_p)\}_p$ with high precision, as will be discussed in Section~\ref{para:constr}.

The difference between the FFT $\tF$ of the field sampling set $\{f(\vphi_i)\}_i$ denoted by $\{(\tF f)(\kappa_p)\}_p$ and the function's  Fourier transform in the conjugate-field sampling points $\{\hf(\kappa_p)\}_p$
is determined by the  weight of the function outside
the sampling windows and decreases with increasing $F$ and $K$. In \cref{app:FFTaliasing} we find that
\begin{align}
\label{eq:fftdif1}
 \Delta_\kappa \sum_{p=-\frac{N_\vphi-1}{2}}^{\frac{N_\vphi-1}{2}} |(\tF f)(\kappa_p)-\hf(\kappa_p)|^2 &\lesssim 2 \left(||w_F^f||^2+ ||w_K^f||^2\right)+\frac{\pi}{K} \left(|f(-F)|^2+|f(F)|^2\right)+ \frac{\pi}{F} \left(|\hf(-K)|^2+|\hf(K)|^2\right).
 \end{align}
 \noindent Similarly, the difference between
 the inverse finite Fourier transform of the set $\{\hf(\kappa_p)\}_p$,  denoted by $\{(\tF^{-1} \hf)(\vphi_i)\}_i$,
 and the function at the field sampling points, $\{f(\vphi_i)\}_i$, is given by
 \begin{align}
\label{eq:fftdif2}
\Delta_\vphi \sum_{i=-\frac{N_\vphi-1}{2}}^{\frac{N_\vphi-1}{2}} |(\tF^{-1} f)(\vphi_i)-f(\vphi_i)|^2 &\lesssim 2 \left(||w_F^f||^2+ ||w_K^f||^2\right)+\frac{\pi}{K} \left(|f(-F)|^2+|f(F)|^2\right)+ \frac{\pi}{F} \left(|\hf(-K)|^2+|\hf(K)|^2\right).
\end{align}
\noindent  The definition of  $\tF$
and $\tF^{-1}$ is given by \cref{eq:fft1,eq:fft2}
in \cref{app:FFTaliasing}.
\\
~
\paragraph{Finite representation construction}
\label{para:constr}

In this section, we define the discrete field operators and construct
the finite Hilbert space of the representation based on the
discretization properties of the boson number states.
This section ends with a detailed analysis of the errors generated by the approximations used in this construction.

\subparagraph{Sampling of Hermite-Gauss functions.}
The wavefunctions' sampling procedures discussed in the previous section are applied here to the boson number states in the field amplitude basis.
The boson number states form a denumerable basis for the local Hilbert space
and provide an intuitive way to introduce the relevant low-energy subspace
 for the problem under investigation.

  In the field amplitude basis the boson number state $\ket{n}$ is the Hermite-Gauss
(HG) function of order $n$,
\begin{align}
\label{eq:hgfunction}
\braket{\vphi}{n}\equiv\phi_n(\vphi)=\left(\frac{m_0}{\pi}\right)^{1/4}\frac{1}{\sqrt{2^n n!} } e^{-\frac{m_0 \vphi^2}{2}}H_n(\sqrt{m_0}\vphi),
\end{align}
\noindent where $H_n$ is the Hermite polynomial of order $n$. The Fourier transform of $\phi_n(\vphi)$
 to the conjugate-field variable $\kappa$ is also proportional to a Hermite-Gauss function of
 order $n$~\cite{Gradshteyn_1980},
\begin{align}
\label{eq:ft_hgfunction}
\braket{\kappa}{n}\equiv\hat{\phi}_n(\kappa)=\frac{\left(-i\right)^n}{\pi^{1/4} m_0^{1/4}\sqrt{2^n n!} }
e^{-\frac{ \kappa^2}{2 m_0}}H_n(\frac{\kappa}{\sqrt{m_0}}).
\end{align}
The recurrence properties of the HG functions (see also  \cref{eq:phi_aa}) imply
\begin{align}
\label{eq:hg_rec_phi}
\vphi \phi_n(\vphi)&=\opmatrix{\vphi}{\Phi}{n}=\frac{1}{\sqrt{2 m_0}}\left(\sqrt{n}\phi_{n-1}(\vphi)+\sqrt{n+1}\phi_{n+1}(\vphi)\right)\\
\label{eq:hg_rec_kappa}
\kappa \hat{\phi}_n(\kappa)&=\opmatrix{\kappa}{\Pi}{n}=-i\sqrt{\frac{m_0}{2}}\left(\sqrt{n}\hat{\phi}_{n-1}(\kappa)
-\sqrt{n+1}\hat{\phi}_{n+1}(\kappa)\right).
\end{align}

\begin{figure}[tb]
    \begin{center}
        \includegraphics*[width=5in]{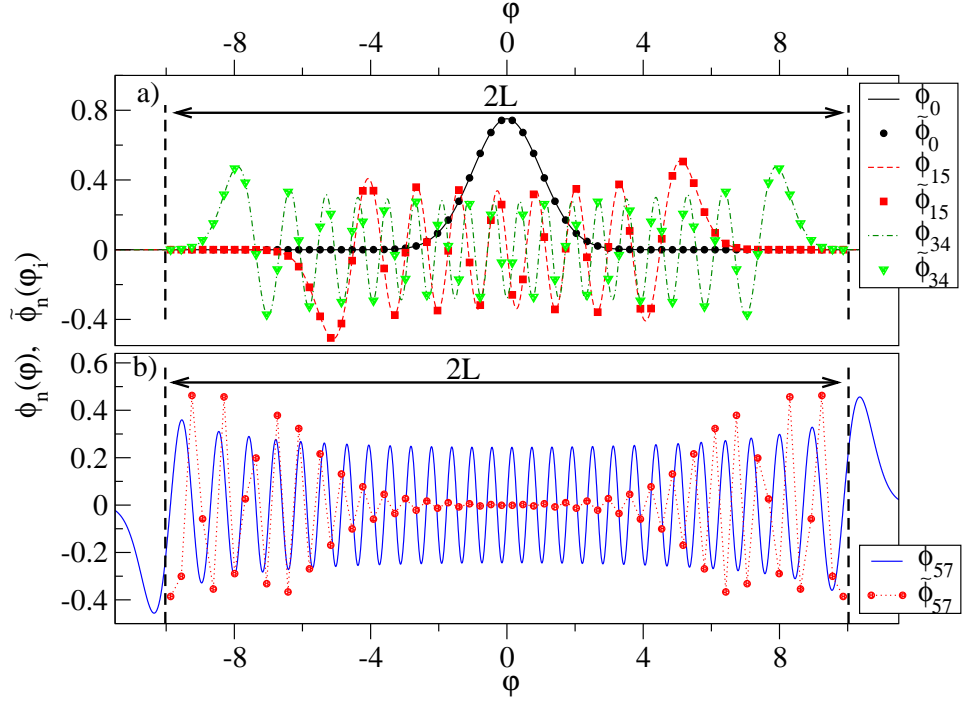}
        \caption{a) Hermite-Gauss functions $\phi_0(\vphi)$, $\phi_{15}(\vphi)$ , and $\phi_{34}(\vphi)$
        (solid, dashed and, respectively, dot-dashed lines) and the discrete harmonic oscillator (with $m_0=1$) eigenstates
        $\tphi_0(\vphi_i)$, $\tphi_{15}(\vphi_i)$, and $\tphi_{34}(\vphi_i)$ (circle, square and, respectively, triangle symbols)
         for a finite Hilbert space with  $N_\vphi=64$ discretization points.
         Within $\O(10^{-4})$ accuracy the support of the HG functions with $n \le N_b=34 $ is
         inside the interval $\left[ -L, L\right]$, where $L=\sqrt{\pi N_\vphi/2}$ (see  \cref{eq:Nvphi_L}). These HG functions are
        sampled accurately by the discrete harmonic oscillator eigenvectors.
        b) The support of $\phi_{57}(\vphi)$ (solid line) has significant weight outside $\left[ -L, L\right]$ and the
        function cannot be sampled by the eigenvector $\tphi_{57}(\vphi_i)$ (circle symbols) of the discrete harmonic oscillator.}
        \label{fig:hgdisc}
    \end{center}
\end{figure}

The HG functions have significant weight on an interval centered on zero and are
exponentially small at large argument, as can be inferred from~\cref{eq:hgfunction,eq:ft_hgfunction}.
The width of the window needed to contain a significant weight increases with increasing order $n$.
Several HG functions are shown in \cref{fig:hgdisc} for illustration.

For a boson state $\phi_n(\vphi)$, the sampling errors appearing in \cref{eq:1error,eq:2error,eq:fftdif1,eq:fftdif2} can be written in terms of the tail weights
$||w^{\phi_n}_{F}||$ and $||w^{\phi_n}_{K}||$. This can be understood by noting that $||w^{\phi_n}_{F}||$ and  $||w^{\phi_n}_{K}||$ are monotonically decreasing with increasing $F$, respectively $K$, when $F$ and $K$ are large enough.
Therefore the dependence
$F=F(||w^{\phi_n}_{F}||)$,
and $K=K(||w^{\phi_n}_{K}||)$ can be found, \textit{i.e.}
the sampling interval widths can be expressed as function of the tail
weights. As a consequence, all of the terms $r^{\phi_n}_F$, $r^{\phi_n}_K$, $|\phi_n(F)|^2$ and $|\hphi_n(K)|^2$
can be written in terms of the tail weights.

For HG functions, a parameter  $L$  can be defined  that relates the
field and conjugate-field sampling windows when $||w^{\phi_n}_{F}||=||w^{\phi_n}_{K}||$:
\begin{align}
\label{eq:Ldef}
F=\frac{L}{\sqrt{m_0}} \text{, }~~~K= L \sqrt{m_0}.
\end{align}
\noindent The HG function $\phi_n(\vphi)$ and its Fourier transform
$\hphi_n(\kappa)$, can be sampled with a finite set of points
\begin{align}
\label{eq:Nvphi_L}
N_\vphi=\lceil \frac{2}{\pi}FK\rceil=\lceil\frac{2}{\pi}L^2 \rceil,
\end{align}
\noindent
and an error  determined by the function tail weights,
\begin{align}
\label{eq:tail_hgn}
\epsilon_w(n,L) \equiv ||w^{\phi_n}_{F}||=||w^{\phi_n}_{K}||.
\end{align}

 By considering only the leading term  $2^n \vphi^n$ of the Hermite
polynomial $H_n(\vphi)$, employing partial integration, and applying
Stirling's formula, it can be shown that
\begin{align}
\label{eq:wncst}
\epsilon_w^2(n,L) \lesssim \frac{1}{L\sqrt{\pi}} \frac{2^n L^{2n}}{n!} e^{-L^2 }
\approx
e^ {-L^2 -\ln \sqrt{2}L\pi }  e^{n \ln\frac{2eL^2}{n}}e^{-\frac{1}{2}\ln n}.
\end{align}

For a fixed $n$, the tail weight $\epsilon_w(n,L)$ decreases exponentially
 with increasing $L$.
For a fixed $L$ and $n \ll 2eL^2$, the tail weight $\epsilon_w(n,L)$ increases with increasing $n$.
Thus, for a cutoff $N_b$ and an error $\epsilon$, a parameter $L(N_b, \epsilon)$ can be chosen  such that
\begin{align}
\label{eq:Lphi}
\epsilon_w(n,L)< \epsilon \text{ ~~~~~for all  } n<N_{b}.
\end{align}
\noindent By increasing $L$, the error $\epsilon$ can be decreased exponentially, {\em i.e.}
$\epsilon \propto e^ {-\frac{  L^2 } {2} +\left(N_{b}-\frac{1}{2}\right) \ln L}$,
as can be inferred from  \cref{eq:wncst}.

Equations~\eqref{eq:NS_app1},~\eqref{eq:1error} and~\eqref{eq:Lphi}  imply that,  for $n<N_b$,
\begin{align}
\label{eq:NS_phir}
\phi_n(\vphi) = \sum_{i=-\frac{N_{\vphi}-1}{2}}^{\frac{N_{\vphi}-1}{2}} \phi_n(\vphi_i) u_K(\vphi-\vphi_i) + \O(\epsilon),
\end{align}
\noindent where
\begin{align}
\label{eq:delta_x}
\vphi_i = i \Delta_{\vphi}
~~~\text{and}~~~
\Delta_\vphi=\sqrt{\frac{2 \pi} {N_\vphi m_0}}.
\end{align}

Similarly, \cref{eq:NS_fk,eq:2error,eq:Lphi}  imply that, for $n<N_b$,
\begin{align}
\label{eq:NS_phi_k}
\hat{\phi}_n(\kappa) = \sum_{p=-\frac{N_{\vphi}-1}{2}}^{\frac{N_{\vphi}-1}{2}} \hat{\phi}_n(\kappa_p) u_F(\kappa-\kappa_p) + \O(\epsilon),
\end{align}
\noindent where
 \begin{align}
 \label{eq:deltakappa}
\kappa_p = p \Delta_{\kappa}
~~~\text{and}~~~
 \Delta_\kappa=\sqrt{\frac{2 \pi m_0} {N_\vphi} }.
 \end{align}

 The orthogonality properties of the {\em sinc} functions,
 \begin{align}
 \label{eq:sincortho}
 \int  u_K(\vphi-\vphi_i)u_K(\vphi-\vphi_j) d\vphi =\Delta_\vphi \delta_{ij},
 \end{align}
 \noindent and HG functions
 yield the following
 orthogonality relation
 \begin{align}
 \label{eq:dnm}
 \Delta_{\vphi}\sum_{i=-\frac{N_{\vphi}-1}{2}}^{\frac{N_{\vphi}-1}{2}} \phi_n(\vphi_i) \phi_m(\vphi_i) =\delta_{nm}+\O(\epsilon)
 ~~\text{  for  } n,m<N_b.
 \end{align}
Finally, \cref{eq:fftdif1,eq:fftdif2,eq:Lphi} imply that,
 for $n<N_b$, the field sampling set  $\{\phi_n(\vphi_i)\}_{i}$ and the conjugate-field
 one $\{\hat{\phi}_n(\kappa_p)\}_{p}$ are related via a finite Fourier transform~
 \begin{align}
 \label{eq:fftkappam}
 \sqrt{\Delta_{\kappa}}\hat{\phi}_n(\kappa_p)= \frac{1}{\sqrt{N_\vphi}}
 \sum_{j=-\frac{N_{\vphi}-1}{2}}^{\frac{N_{\vphi}-1}{2}} \sqrt{\Delta_{\vphi}} \phi_n(\vphi_j)e^{-i \frac{2 \pi j p} {N_{\vphi}}}
 +\O(\epsilon).
 \end{align}
\\

 \subparagraph{Finite Hilbert space construction.}

The low-energy subspace of dimension $N_b$ can be represented by a Hilbert space $\tilde{ {\cal{H}} }$ of
dimension $N_{\vphi} > N_b$, spanned by a set of orthogonal
 vectors $\{\ket{\tvphi_i}\}_i$. On $\dH$, we define the discrete field operator
 \begin{align}
 \label{eq:tphi}
 \tilde{\Phi}\ket{\tvphi_i}&=\vphi_i \ket{\tvphi_i}, ~~~\text{  with } \vphi_i = i \Delta_{\vphi}= i \sqrt{\frac{2 \pi}{N_{\vphi} m_0}}
 ~~~\text{  and  } i = -\frac{N_{\vphi}-1}{2},-\frac{N_{\vphi}-1}{2}+1,...,\frac{N_{\vphi}-1}{2},
 \end{align}
 \noindent and the discrete conjugate-field operator
 \begin{align}
 \label{eq:pi_field}
 \tPi=m_0 \tF \tPhi \tF^{-1},
 \end{align}
 \noindent where $\tF$ is the finite Fourier transform,
 \begin{align}
 \label{eq:fft}
  \tF =\frac{1}{\sqrt{N_\vphi}}\sum_{j,p=-\frac{N_{\vphi}-1}{2}}^{\frac{N_{\vphi}-1}{2}} e^{i \frac{2 \pi}{N_\vphi} jp}\ket{\tvphi_j} \bra{\tvphi_p}.
 \end{align}
 \noindent Note that the vectors $\{\ket{\tkappa_p}\}_p$, obtained by applying
 a finite Fourier transform on $\{\ket{\tvphi_i}\}_i$
 \begin{align}
 \label{eq:fft_im}
 \ket{\tkappa_p} \equiv \tF\ket{\tvphi_p}=\frac{1}{\sqrt{ N_{\vphi} } }
 \sum_{j=-\frac{N_{\vphi}-1}{2}}^{\frac{N_{\vphi}-1}{2}} \ket{\tvphi_j}e^{i \frac{2 \pi j p}{N_{\vphi}} },
 \end{align}
  \noindent are eigenvectors of $\tPi$,
 \begin{align}
 \label{eq:tpi}
 \tilde{\Pi}\ket{\tkappa_p}&=\kappa_p\ket{\tkappa_p} ~~~\text{  with } \kappa_p = p \Delta_{\kappa}= p \sqrt{\frac{2 \pi m_0}{N_{\vphi}}}
 ~~~\text{  and  } p =-\frac{N_{\vphi}-1}{2}, -\frac{N_{\vphi}-1}{2}+1,...,\frac{N_{\vphi}-1}{2}.
 \end{align}

\textbf{\emph{
 The subspace of $\dH$ spanned by the first $N_b$ eigenvectors, $\{\ket{\tphi_n}\}$, of the
 discrete harmonic oscillator Hamiltonian
\begin{align}
\label{eq:dhosc}
\tilde{H}_{h}&=\frac{1}{2}\tilde{\Pi}^2+\frac{1}{2} m_0^2 \tilde{\Phi}^2,
 \end{align}
\noindent is a representation of the low-energy subspace of the full Hilbert space with $\O(\epsilon)$  accuracy,
provided  that $N_\vphi \Delta_\vphi \ge 2F$, where $F$ is large enough
that the  weight of the $n = N_b+2$ Hermite-Gauss function outside
the interval $\left[-F, F\right]$ is $\O(\epsilon)$ small.}}

To validate our construction, consider the  subspace of $\dH$ spanned by
 the vectors $\{\ket{\tilde{n}}\}_{n<N_b+2}$ defined as
 \begin{align}
 \label{eq:tn}
 \ket{\tilde{n}} \equiv \sqrt{\Delta_{\vphi}} \sum_i \phi_n(\vphi_i) \ket{\tvphi_i} =\sqrt{\Delta_{\kappa}} \sum_p \hat{\phi}_n(\kappa_p) \ket{\tkappa_p}+ \O(\epsilon),
 \end{align}
 \noindent  (see \cref{eq:fftkappam,eq:fft_im}). Note that the
 ability to relate accurately the field and conjugate-field sampling points of HG functions of order ${n<N_b+2}$
 by the finite Fourier transform is essential for \cref{eq:tn}.
 The set $\{\ket{\tilde{n}}\}_{n<N_b+2}$ is orthogonal and normalized (within  $\O(\epsilon)$ accuracy),
  as implied by \cref{eq:dnm}.
 Moreover \cref{eq:hg_rec_phi,eq:hg_rec_kappa} imply
 \begin{align}
 \label{eq:rec_phi}
 \opmatrix{\tvphi_i}{\tilde{\Phi}}{\tilde{n}}&=\vphi_i \braket{\tvphi_i}{\tilde{n}} =\frac{1}{\sqrt{2 m_0}}\left(\sqrt{n}\braket{\tvphi_i}{\widetilde{n-1}}+\sqrt{n+1}\braket{\tvphi_i}{\widetilde{n+1}}\right)
 + \O(\epsilon)\\
 \label{eq:rec_kappa}
 \opmatrix{\tkappa_p}{\tilde{\Pi}}{\tilde{n}}&=\kappa_p\braket{\tkappa_p}{\tilde{n}}=-i\sqrt{\frac{m_0}{2}}
 \left(\sqrt{n}\braket{\tkappa_p}{\widetilde{n-1}}
 -\sqrt{n+1}\braket{\tkappa_p}{\widetilde{n+1}}\right)+ \O(\epsilon),~~~\text{  when } n+1<N_b+2,
 \end{align}
 \noindent since, as can be deduced from  \cref{eq:tn}, $\braket{\tvphi_i}{\tilde{n}} \propto \phi_n(\vphi_i)$ and $\braket{\tkappa_p}{\tilde{n}}\propto \hat{\phi}_n(\kappa_p)$.
 \Cref{eq:rec_phi,eq:rec_kappa} can be written as
 \begin{align}
 \label{eq:rec_phi1}
 \tilde{\Phi}\ket{\tilde{n}}& =\frac{1}{\sqrt{2 m_0}}\left(\sqrt{n}\ket{\widetilde{n-1}}+\sqrt{n+1}\ket{\widetilde{n+1}}\right)+ \O(\epsilon)\\
 \label{eq:rec_kappa1}
 \tilde{\Pi}\ket{\tilde{n}}&=-i\sqrt{\frac{m_0}{2}}
 \left(\sqrt{n}\ket{\widetilde{n-1}}
 -\sqrt{n+1}\ket{\widetilde{n+1}}\right)+ \O(\epsilon),~~~\text{  when } n+1<N_b+2.
 \end{align}
 Using \cref{eq:rec_phi1,eq:rec_kappa1},
 it  can be shown that
 \begin{align}
 \label{eq:dhosc_ev}
 \tilde{H}_{h}\ket{\tilde{n}}&=m_0 \left(n+\frac{1}{2}\right) \ket{\tilde{n}} +\O(\epsilon) ~~~\text{  when } n+2<N_b+2.
 \end{align}
 \noindent The vectors $\{\ket{\tilde{n}}\}_{n<N_b}$ are approximations
 of order $\O(\epsilon)$ of the eigenstates of the discrete harmonic oscillator.  For illustration, in \cref{fig:hgdisc}-(a), we show several
 eigenvectors $\{\ket{\tphi_n}\}_{n<N_b}$ of $\tilde{H}_{h}$ (circle, square and triangle symbols),
 obtained by exact diagonalization. As can be seen, they sample very well the HG functions plotted with lines.

%
% \begin{align}
% \Phi^2 \ket{n}&=&\frac{1}{2}\left(\sqrt{\left(n-1\right)n}\ket{n-2} + \sqrt{\left(n+1\right)\left(n+2\right)}\ket{n+2} \right)&+\frac{1}{2} \left(2n+1\right)\ket{n}\\
% \Pi^2 \ket{n}&=&\frac{-1}{2}\left(\sqrt{\left(n-1\right)n}\ket{n-2} + \sqrt{\left(n+1\right)\left(n+2\right)}\ket{n+2} \right)&+\frac{1}{2} \left(2n+1\right)\ket{n}\\
% \Phi \Pi \ket{n}&=&\frac{-i}{2}\left(\sqrt{\left(n-1\right)n}\ket{n-2} - \sqrt{\left(n+1\right)\left(n+2\right)}\ket{n+2} \right)&+\frac{i}{2} \ket{n}\\
% \Pi \Phi \ket{n}&=&\frac{-i}{2}\left(\sqrt{\left(n-1\right)n}\ket{n-2} - \sqrt{\left(n+1\right)\left(n+2\right)}\ket{n+2} \right)&-\frac{i}{2} \ket{n}
% \end{align}
% \begin{align}
% \Phi^4 \ket{n}&=\frac{1}{4}\sqrt{(n-3)(n-2)(n-1)n}\ket{n-4}+\frac{1}{4}(4n-2)\sqrt{(n-1)n}\ket{n-2} +\frac{1}{4}(6n^2+6n+3)\ket{n} \\ \nonumber
%   &+\frac{1}{4}(4n+6)\sqrt{(n+1)(n+2)}\ket{n+2} + \frac{1}{4}\sqrt{(n+1)(n+2)(n+3)(n+4)}\ket{n+4}
% \end{align}

\begin{figure}[tb]
    \begin{center}
        \includegraphics*[width=5in]{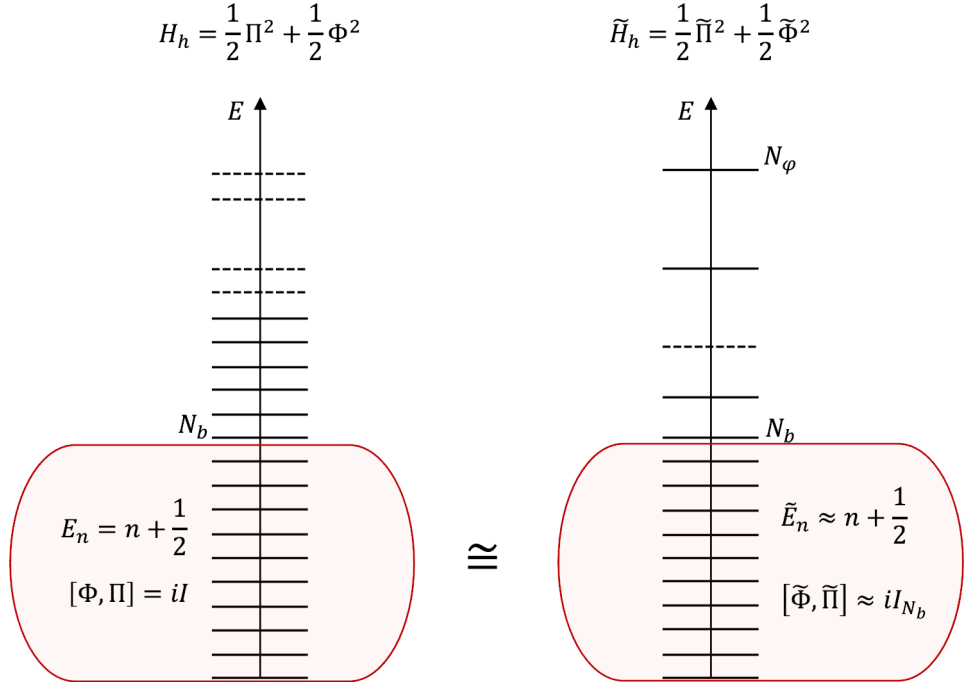}
        \caption{Within $\O(\epsilon)$ accuracy,  the algebra generated
        by the field operator $\Phi$ and $\Pi$ restricted to
        the $N_b$ size low-energy subspace  of the
        harmonic oscillator Hamiltonian~\eqref{eq:hosc}  (shaded region, left side)
        is isomorphic with the algebra generated by the discrete field operators $\tPhi$ and $\tPi$
        restricted to the the $N_b$ size low-energy subspace  of the discrete harmonic oscillator Hamiltonian~\eqref{eq:dhosc} (shaded region, right side).
        The accuracy increases exponentially with increasing the
        size $N_\vphi$ of the finite Hilbert space, see  \cref{eq:epsNphiNb}.
        }
        \label{fig:lediag}
    \end{center}
\end{figure}

 Using  \cref{eq:rec_phi1,eq:rec_kappa1}  to calculate the commutator of the discrete field operators, one gets
 \begin{align}
 \label{eq:comm}
 \left[\tilde{\Phi}, \tilde{\Pi}\right] \ket{\tilde{n}} =i \ket{\tilde{n}}+ \O(\epsilon),  ~~\text{   for  } n<N_b.
 \end{align}
 \noindent Thus the operators
 $\tilde{\Phi}$ and $\tilde{\Pi}$ obey (within the error $\O(\epsilon)$) the same commutation relation
 as $\Phi$ and $\Pi$ (see  \cref{eq:ho_comm}) on the subspace spanned by the vectors $\{\ket{\tilde{n}}\}_{n<N_b}$.

 As long as the physics of the problem of interest can be addressed by truncating the number of
 bosons per site to $N_b$ ({\em i.e.} $N_b$ is taken large enough), the full Hilbert space can be replaced by the finite size $\tilde{ {\cal{H}} }$ space
 and the operators $\Phi$ and $\Pi$ can be replaced by $\tilde{\Phi}$ and, respectively $\tilde{\Pi}$.
 The operators $\tilde{\Phi}$ and $\tilde{\Pi}$ act on the subspace spanned by $\{\ket{\tphi_n}\}_{n<N_b}$
 as the field operators $\Phi$ and $\Pi$ act on the subspace spanned by $\{\ket{n}\}_{n<N_b}$.
 The situation is illustrated in \cref{fig:lediag}.

 Nevertheless, the high-energy eigenvectors of the finite space $\tilde{ {\cal{H}} }$
 have very different properties then the corresponding eigenvectors of
 the full Hilbert space. For example, one can see
 in  \cref{fig:hgdisc}-(b) that
 the $\tilde{H}_{h}$ eigenvector coefficients $\braket{\tvphi_i}{\tphi_{57}}$ (circle symbols)  do not sample
 the HG function $\phi_{57}(\vphi)$ (solid line), since $\phi_{57}(\vphi)$
 does not belong to the low-energy subspace  when $N_\vphi=64$.
 When doing numerical simulations
 one has to make sure that $N_b$ and $N_\vphi$ are sufficiently large that the high-energy
 subspace contribution to the physical problem can be safely neglected.
 This will be discussed more in Section~\ref{sec:val}.

 An interesting property of the discrete harmonic oscillator Hamiltonian $\dH_h$,  \cref{eq:dhosc},
 is that it commutes with the FFT. By writing
 \begin{align}
 \label{eq:hosc_fft}
 \tilde{H}_{h}&=\frac{1}{2} m_0^2 \left(\tF \tilde{\Phi}^2 \tF^{-1}+\frac{1}{2}  \tilde{\Phi}^2\right)=
 \frac{1}{2} m_0^2 \left( \tF^{-1} \tilde{\Phi}^2 \tF+\frac{1}{2}  \tilde{\Phi}^2\right),
  \end{align}
  \noindent it is easy to see that $[\dH,\tF]=0$.
  The last equality in  \cref{eq:hosc_fft} is a consequence of the parity inversion symmetry of $
  \dH$. All  eigenvectors $\{\ket{\tphi_n}\}_n$
  of $\dH$ (the ones belonging to the high-energy subspace too) are eigenvectors
  of the finite Fourier transform. This is just the discrete version of the
  HG functions' property of being eigenvectors of both the harmonic oscillator
  Hamiltonian and the continuous Fourier transform.\\

 \subparagraph{Error analysis.}

\begin{figure}[tb]
    \begin{center}
        \includegraphics*[width=5in]{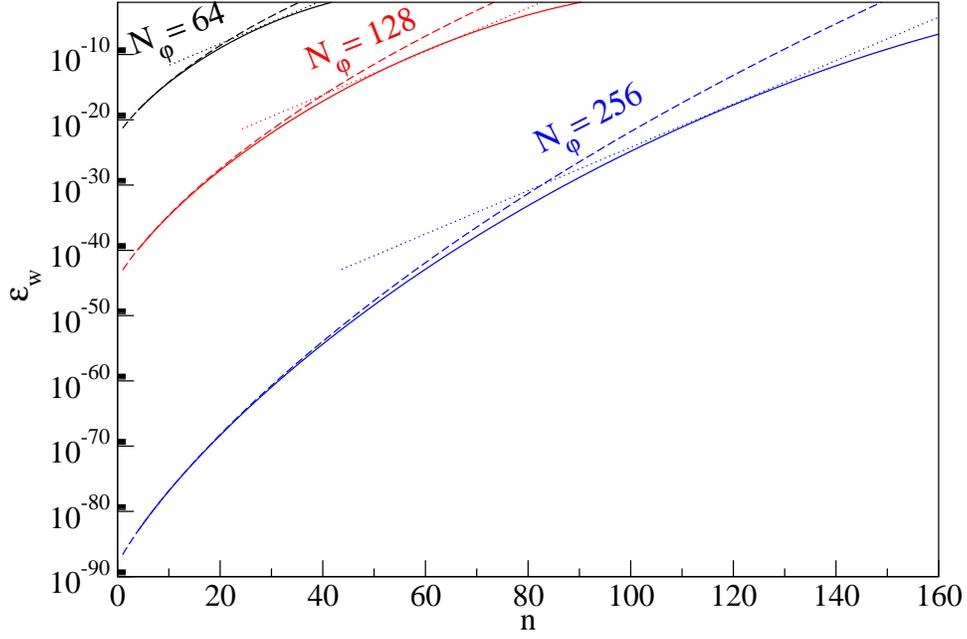}
        \caption{Tail weight $\epsilon_w(n)$,  \cref{eq:tail_hgn}, (solid lines) of  HG functions
        versus $n$ for
        $N_\vphi=64$ (upper, black), $N_\vphi=128$ (middle, red) and $N_\vphi=256$ (lower, blue) discretization points.
         \cref{eq:envsNxNb} (dashed lines) is a good approximation for small $n \lesssim 0.3N_\vphi$.
        For larger $n$, $\epsilon_w(n)=\epsilon_c(n-2)/\left(1.5\sqrt{n\left(n-1\right)}\right)$
        with $\epsilon_c(n)$ given by  \cref{eq:epsNphiNb}, (dotted lines) provides a better  bound for the error.}
        \label{fig:wnerror}
    \end{center}
\end{figure}

We argued previously that the errors of the finite representation
are of the same order of magnitude as the weight $\epsilon_w(n, L)$
of the HG functions with $n \le N_b+2$ outside the interval $\left[-F,F\right]$.
In this section we investigate numerically
the errors involved in the construction of the finite representation.

\Cref{fig:wnerror} shows the tail weight of the HG functions,
$\epsilon_w\left[n,L(N_\vphi)\right]$ (see \cref{eq:Nvphi_L,eq:tail_hgn}),
as a function of $n$ for  $N_\vphi=64$, $N_\vphi=128$ and $N_\vphi=256$.
The tail weight is obtained by numerical integration
of  \cref{eq:tail_f}. For comparison, the tail weight approximation
obtained from \cref{eq:Nvphi_L,eq:wncst},
\begin{align}
\label{eq:envsNxNb}
\epsilon_w(n, N_\vphi) \approx \frac{1}{\pi \sqrt{\pi}}e^{-\frac{\pi}{4}N_\vphi}e^{\frac{2n-1}{4}\ln N_\vphi} e^{-\frac{n}{2} \ln n}
e^{\frac{n}{2}\left(\ln \pi+1 \right)}e^{-\frac{1}4 \ln n},
\end{align}
\noindent is shown with dashed lines.  \Cref{eq:envsNxNb}
is a good approximation for the tail weight for $n \lesssim 0.3N_\vphi$ and overestimates
$\epsilon_w(n)$ at larger values of $n$.

 \begin{figure}[tb]
     \begin{center}
         \includegraphics*[width=5in]{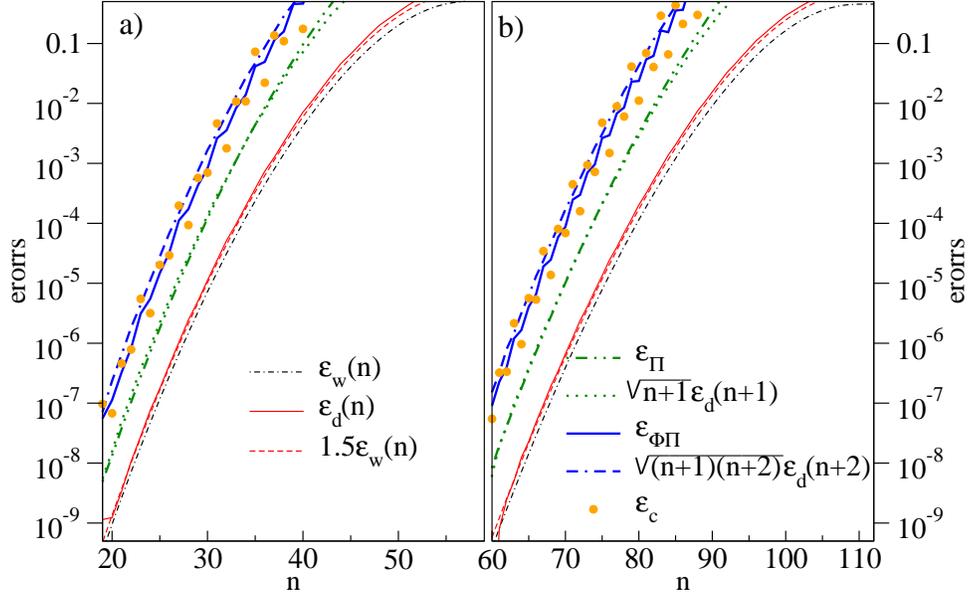}
         \caption{The tail weight, $\epsilon_w(n)$,   \cref{eq:tail_hgn}, of HG functions (dash-dot black),
         the difference between the discretized HG and
         the discrete harmonic oscillator, $\epsilon_d(n)$,  \cref{eq:err_vhg} (thin solid red),
          the error associated with the $\tPi$ operator $\epsilon_\Pi(n)$,  \cref{eq:err_Pi}, (dash-dot-dot green),
          the error associated with the $\tPhi\tPi$ operator $\epsilon_{\Phi\Pi}(n)$,  \cref{eq:err_phiPi}, (thick solid blue),
         and the commutation relation error, $\epsilon_c(n)$,  \cref{eq:err_c}, (orange circle symbols).  In good approximation $\epsilon_{d}(n) \approx 1.5 \epsilon_w(n)$ (dashed red line),  $\epsilon_{\Pi}(n) \approx \sqrt{n+1}\epsilon_d(n+1)$ (dotted green line), $\epsilon_{\Phi \Pi}(n) \approx \epsilon_c \approx  \sqrt{(n+1) (n+2)}\epsilon_d(n+2)$ (dot-dash-dash blue line). The size of the finite Hilbert space is (a) $N_\vphi=64$ and (b) $N_\vphi=128$.}
         \label{fig:errors}
     \end{center}
 \end{figure}

Nonzero $\epsilon_w(n)$ causes a finite difference between the discretized HG functions
$\ket{\tilde{n}}$ defined by  \cref{eq:tn}, and the eigenvectors of the discrete harmonic
oscillator, $\ket{\tphi_n}$. Employing exact diagonalization
to calculate $\ket{\tphi_n}$ we find that
\begin{align}
\label{eq:err_vhg}
\epsilon_{d}(n)=\left|\left| \ket{\tilde{n}} -\ket{\tphi_n}\right|\right|,
\end{align}
\noindent is proportional to $\epsilon_w(n)$, \textit{i.e.}
$\epsilon_{d}(n) \approx 1.5 \epsilon_w(n)$, as illustrated with thin continuous red and dashed red lines in  \cref{fig:errors}.

 Each time the field operators $\tPhi$ and $\tPi$ act on the eigenvector $\ket{\tphi_n}$ of $\tilde{H}_{h}$,
 the errors are amplified approximately by a factor of $\sqrt{n+1}$. This can be understood from \cref{eq:rec_phi1,eq:rec_kappa1} when one replaces $\ket{\widetilde{n+1}}$ with
 $\ket{\tphi_{n+1}}+\left(\ket{\widetilde{n+1}}-\ket{\tphi_{n+1}}\right)$. The leading error
 associated with the finite magnitude of $\ket{\widetilde{n+1}}-\ket{\tphi_{n+1}}$ is magnified by a factor
 $\sqrt{n+1}$. Numerical calculations agree with this assertion.
For example, the state $\tPi\ket{\tphi_n}$ behaves as $\Pi\ket{n}$ up to an error,
 \begin{align}
 \label{eq:err_Pi}
 \epsilon_{\Pi}(n)=\left|\left| \tPi \ket{\tphi_n} -\frac{-i}{\sqrt{2}} \left(\sqrt{n}\ket{\tphi_{n-1}} -\sqrt{n+1}\ket{\tphi_{n+1}} \right)   \right|\right|.
 \end{align}
\noindent As shown in  \cref{fig:errors} with dash-dot-dot green and dotted green lines,  $\epsilon_{\Pi}(n) \approx \sqrt{n+1}\epsilon_d(n+1)$.
The same conclusion is valid for the error associated with the behavior of the state
 $\tPhi\ket{\tphi_n}$ (not shown).

 The error associated to the commutation relation, $\left[\tPhi, \tPi\right]\ket{\tphi_n}$,
 is comparable with the errors associated to the states $\tPhi \tPi \ket{\tphi_n}$ and $\tPi \tPhi \ket{\tphi_n}$.
 Figure~\ref{fig:errors}
 shows
 \begin{align}
 \label{eq:err_phiPi}
 \epsilon_{\Phi \Pi}(n)=\left|\left|\tPhi \tPi \ket{\tphi_n} -\frac{i}{2} \left(-\sqrt{(n-1)n}\ket{\tphi_{n-2}} +\sqrt{(n+1) (n+2)}\ket{\tphi_{n+2}} + \ket{\tphi_{n}}\right)   \right|\right|,
 \end{align}
 \noindent with a thick solid blue line, and
 \begin{align}
 \label{eq:err_c}
 \epsilon_c=\left|\left|\left(\left[\tPhi, \tPi \right] -i\right) \ket{\tphi_n}\right|\right|,
 \end{align}
 \noindent with orange dots. We find (see also the dot-dash-dash blue line) that
 \begin{align}
 \label{eq:err_cw}
\epsilon_{\Phi \Pi}(n) \approx \epsilon_c \approx  \sqrt{(n+1) (n+2)}\epsilon_d(n+2).
\end{align}

Since $\epsilon_w(n, N_\vphi)$
increases with increasing $n$,
for a finite representation of size $N_\vphi$ and  cutoff $N_b$, the leading error is of the order of
$\epsilon_w(N_b+2, N_\vphi)$.
For a given cutoff $N_b$,  the error can be reduced exponentially by increasing the number of discretization
 points $N_{\vphi}$,  $\epsilon \propto \epsilon_w(N_b+2, N_\vphi) \le e^{-\frac{\pi}{4}N_\vphi+\frac{\left(2N_b+3\right)}{4}\ln N_\vphi}$,
 as  \cref{eq:envsNxNb} and the numerical results shown in \cref{fig:wnerror} imply.

 For fixed accuracy, an increase of the low-energy subspace requires an increase  of $N_{\vphi}$.
 For small $N_b/N_\vphi$
 the dependence  between $N_b$ and $N_\vphi$ at fixed error is given by   \cref{eq:envsNxNb}.
 The region where the accuracy is of order $\epsilon=10^{-3} \sim 10^{-5}$ is of practical interest
 for simulations. In this region $N_b/N_\vphi \in  \approx \left[0.3,0.7\right]$ and
  \cref{eq:envsNxNb} overestimates the errors. Numerical investigations and arguments based on the
 WKB approximation~\cite{macridin_pra_2018,macridin_prl_2018} indicate
 that, in this region
 \begin{align}
 \label{eq:LvsNb}
 N_{\vphi}\approx c_1+c_2 N_b,
 \end{align}
 \noindent where $c_1$ and $c_2$ are accuracy dependent parameters. At fixed accuracy, there is a linear dependence between the size
 $N_{\vphi}$ of the finite space $\tilde{ {\cal{H}} }$ and the boson cutoff number $N_b$.
 For example, we find that  the number of discretization points
 $N_{\vphi} \approx 32+1.5 N_{b}$ for an accuracy $\epsilon=10^{-3}$~\cite{macridin_pra_2018}.
 In practice, for many problems of interest, such as scalar $\Phi^4$ theory and electron-phonon systems,
   the representation in the field amplitude basis requires only one more qubit per harmonic oscillator than the
 representation in the boson number basis.

 Numerical investigations in the region with the error range
 $\epsilon \in \left[ 10^{-5}, 10^{-3}\right]$~\cite{macridin_pra_2018},
 yield the following upper bound for the error associated with the
 commutation relation  (\cref{eq:err_c}),
 \begin{align}
 \label{eq:epsNphiNb}
 \epsilon_c < 10e^{ -\left(0.51 N_{\vphi}-0.765 N_b\right) }.
 \end{align}
 \noindent In  \cref{fig:wnerror}, we show with
 dotted lines
 $\epsilon_w(n)=\epsilon_c(n-2)/\left(1.5\sqrt{n\left(n-1\right)}\right)$
 (see the numerical dependence between $\epsilon_d$ and $\epsilon_c$ in  \cref{eq:err_cw}) where $\epsilon_c$ is given by   \cref{eq:epsNphiNb}.

\subsubsection{Representation of the lattice Hilbert space}
\label{ssec:latH}

The construction of the lattice representation is a straightforward extension of the local representation construction. The lattice Hilbert space is a direct product of $N$ local infinite Hilbert spaces,
\begin{align}
{\cal{H}}=\prod_{j=1}^{N}  \otimes {\cal{H}}_j\equiv  L^2(\mathbb{R}^N),
\end{align}
\noindent where $N$ represents the number of the lattice sites. The finite
size Hilbert space of dimension $\left(N_\vphi\right)^N$,
\begin{align}
\dH=\prod_{j=1}^N \otimes \dH_j,
\end{align}
\noindent with $\dH_j$ being the local Hilbert spaces of dimension $N_\vphi$ constructed in \cref{ssec:local}, is a representation of the lattice  low-energy subspace with maximum $N_b$ bosons per site. The Hilbert space $\dH$ is spanned by the vectors
\begin{align}
\label{eq:latt_phiij}
\left\{\ket{\tvphi_{i_1}}_1\ket{\tvphi_{i_2}}_2...\ket{\tvphi_{i_N}}_N \right\}, ~~~~~~\text{with    }~~~~ i_j=\overline{-\frac{N_\vphi-1}{2},\frac{N_\vphi-1}{2}}.
\end{align}
\noindent The discrete field operators are defined as
\begin{align}
\label{eq:tPhi_j}
\tPhi_j \ket{\tvphi_{i_1}}_1..\ket{\tvphi_{i_j}}_j...\ket{\tvphi_{i_N}}_N&= i_j  \Delta_\vphi \ket{\tvphi_{i_1}}_1..\ket{\tvphi_{i_j}}_j...\ket{\tvphi_{i_N}}_N \\
\label{eq:tPi_j}
 \tPi_j\ket{\tvphi_{i_1}}_1..\ket{\tkappa_{m_j}}_j...\ket{\tvphi_{i_N}}_N&= m_j  \Delta_\kappa \ket{\tvphi_{i_1}}_1..\ket{\tkappa_{m_j}}_j...\ket{\tvphi_{i_N}}_N
\end{align}
\noindent where
\begin{align}
\ket{\tvphi_{i_1}}_1..\ket{\tkappa_{m_j}}_j...\ket{\tvphi_{i_N}}_N= \frac{1}{\sqrt{ N_{\vphi} } }
\sum_{i_j=-\frac{N_{\vphi}-1}{2}}^{\frac{N_{\vphi}-1}{2}}
\sum \ket{\tvphi_{i_1}}_1..\ket{\tvphi_{i_j}}_j...\ket{\tvphi_{i_N}}_N e^{i \frac{2 \pi i_j m_j}{N_{\vphi}} }
\end{align}
\noindent is obtained via a local Fourier transform at site $j$. The conjugate-field operator
can be written as
\begin{align}
\tPi_j&= m_0 \tF_j \tPhi_j \tF^{-1}_j,
\end{align}
\noindent where
\begin{align}
\tF_j=I_1 \otimes I_2 \otimes ...I_{j-1}\otimes (\tF)_j \otimes I_{j+1} ... \otimes I_N
\end{align}
is the finite Fourier transform acting on the local Hilbert space $\dH_j$.

On the subspace spanned by
\begin{align}
\{ \ket{\tphi_{n_1}}_1 \ket{\tphi_{n_2}}_2...\ket{\tphi_{n_N}}_N \}_{{n_1},{n_2},...,{n_N}<N_b},
\end{align}
\noindent where $\ket{\tphi_{n}}_j \in \dH_j$ is the $n$'s' eigenvector of a discrete harmonic oscillator
Hamiltonian~\eqref{eq:dhosc}, the field operators satisfy
\begin{align}
\left[\tPhi_i, \tPi_j\right]=\delta_{ij}\left(iI_i+\O(\epsilon)\right),
\end{align}
\noindent where $\O(\epsilon)$ represents the error
of constructing local representations and was discussed in \cref{ssec:local}.
With $\O(\epsilon)$ accuracy, the algebra generated by the field operators is isomorphic with
the algebra generated by the continuous field operators
when restricted to the low-energy subspace defined by
$n_j<N_b$ at every $j$ site.

\subsection{Accurately sampled states not contained
in the low-energy subspace}
\label{ssec:leval}

We have described how to map a low-energy subspace of the infinite Hilbert space onto a low-energy subspace of a finite Hilbert space. The dimension $N_\vphi$
of the local finite Hilbert space depends on
the dimension $N_b$ of the low-energy subspace and the accuracy
$\epsilon$.

While an accurate representation of the low-energy subspace implies accurate sampling of the low-energy wavefunctions, the converse is not necessarily true. Good sampling of a wavefunction does not imply that the wavefunction belongs to the low-energy subspace.
There are functions
 that can be  sampled with $\epsilon$-accuracy in $N_\vphi$ points and
do not  belong  to the low-energy subspace of dimension $N_b(N_\vphi, \epsilon)$.
Since the high-energy subspace projection of these wavefunctions
is significant, the actions of the  discrete field operators on them yield uncontrollable errors. Therefore, it is important to verify that the system wavefunction has a boson distribution that is below the cutoff.  We describe how this can be accomplished with quantum simulations in \cref{sec:val}.

To emphasize this point,
we present examples of wavefunctions with small tail weights outside sampling intervals
that can be sampled accurately on $N_\vphi$ discretization points, but have a significant high-energy weight
and therefore cannot be represented accurately on a finite Hilbert space of dimension  $N_\vphi$.

\begin{figure}[tb]
    \begin{center}
        \includegraphics*[width=5in]{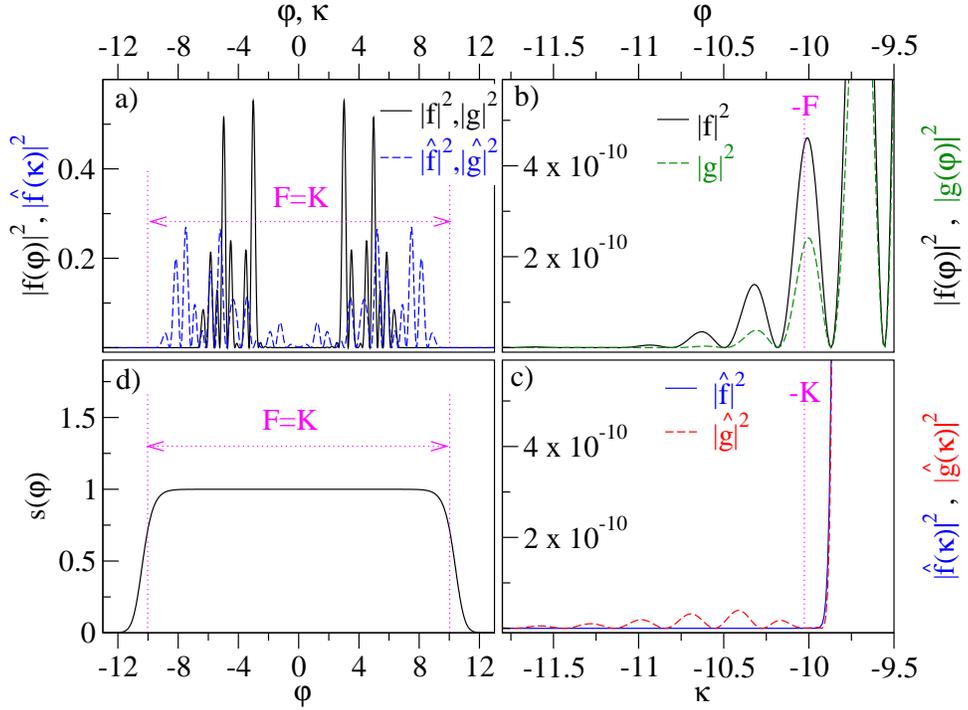}
        \caption{(a) The square amplitude of the functions $f(\vphi)$
        (\cref{eq:fbad_asym})
        and $g(\vphi)$ (\cref{eq:gbad}) (solid black)
        and of their Fourier transforms $\hat{f}(\kappa)$
        and, respectively, $\hat{g}(\kappa)$ (dashed blue). At this scale, $f$ and $g$ are indistinguishable.
        (b)  $|f(\vphi)|^2$ (solid black) and  $|g(\vphi)|^2$ (dashed green) for $\vphi<-F$.  (c) $|\hat{f}(\kappa)|^2$ (solid blue) and
        $|\hat{g}(\kappa)|^2$ (dashed red) for $\kappa<-K$. The weights of both $f$ and $g$ outside the
        sampling interval is small, $\approx \O(10^{-5})$.
        (d) The function $s(\vphi)$ ( \cref{eq:sbad}) used to define $g(\vphi)$.
         $s(\vphi)$, and implicitly $g(\vphi)$, decay exponentially fast at large $|\vphi|$.}
        \label{fig:fbad_tails}
    \end{center}
\end{figure}

For the first example, we consider a band-limited function $f(\vphi)$ (see  \cref{eq:NS_K,eq:phiiuk})
\begin{align}
\label{eq:f}
f(\vphi)=\sum_{i=-\frac{N_\vphi-1}{2}}^{\frac{N_\vphi-1}{2}}  a_i u_K(\vphi-\vphi_i)\,
\end{align}
\noindent where we take $F=K=\sqrt{\pi N_\vphi/2}$, (see  \cref{eq:Ncphi}),
with $N_\vphi=64$.
As described in \cref{ap:fbad}, the coefficients $a_i$ are chosen such that the behavior as $\varphi\to\infty$ is
\begin{align}
\label{eq:fbad_asym}
f(\vphi) \sim c_f \sin \left(\frac{\pi \vphi} {\Delta_\vphi} - \frac{\pi}{2} \right)\frac{\Delta_\vphi^8} {\pi  \vphi^8 }+\O \left( \frac{\Delta_\vphi^{10}} {\pi  \vphi^{10} } \right),
\end{align}
\noindent where $c_f$ is a normalization constant, \emph{i.e.}
the function decays as $|\vphi|^{-8}$ with increasing $|\vphi|$.
The square amplitudes  $|f(\vphi)|^2$ and $|\hat{f}(\kappa)|^2$ are plotted in  \cref{fig:fbad_tails}.
For $\vphi<-F$, we have $|f(\vphi)|^2 \approx \O(10^{-10})$ as can be seen in  \cref{fig:fbad_tails}-(b).
The weight outside the interval $\left[-F, F \right]$ is $||w_F^f|| \approx 1.1 \times 10^{-5}$.
Since the function is band-limited, $\hat{f}(\kappa)=0$ for $|\kappa|>K$. By construction, the Finite Fourier
transform connects
the sets $\{f(\vphi_i)\}_{i=\overline{-\left(N_\vphi-1\right)/2,\left(N_\vphi-1\right)/2 }}$ and
$\{\hf(\kappa_p)\}_{p=\overline{-\left(N_\vphi-1\right)/2,\left(N_\vphi-1\right)/2 }}$ without error, since,
in the sampling points, the
function coincides with the aliased function (see  \cref{eq:faliased,eq:hfaliased}).

\begin{figure}[tb]
    \begin{center}
        \includegraphics*[width=5in]{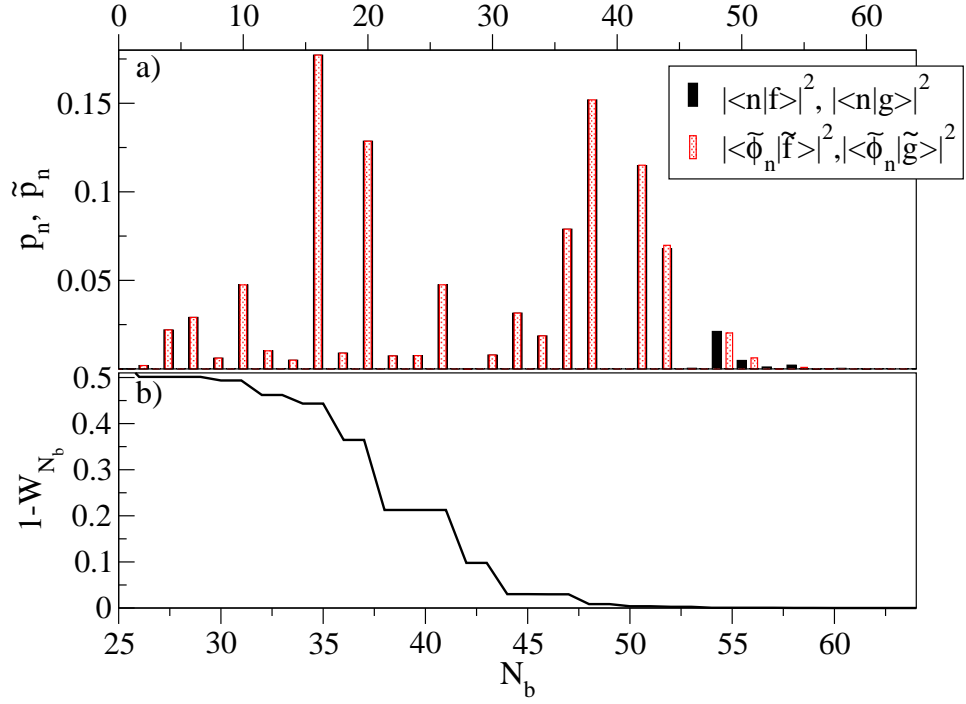}
        \caption{(a)  Boson distribution of the wavefunctions $f(\vphi)$  and
        $g(\vphi)$ (solid black).  Discrete harmonic oscillator eigenstate distribution of the discretized states $\ket{\tilde{f}}$  and $\ket{\tilde{g}}$ (hashed red). At large $n$ the boson
        and the discrete harmonic oscillator eigenstate distributions differ significantly.
        (b) The high-energy weight of the wavefunctions $f(\vphi)$  and $g(\vphi)$
        versus the cutoff $N_b$. $50\%$ ($20\%$) of the wavefunctions' weight
        belongs to the high-energy subspace spanned by states with $n>30$ ($n>40$).}
        \label{fig:fabd_bosons}
    \end{center}
\end{figure}

Despite the small tail weight and perfect sampling, the wavefunction $f(\vphi)$
cannot be represented accurately on a finite Hilbert space of size $N_\vphi=64$.
To demonstrate this,
we show in  \cref{fig:fabd_bosons}-(a) the boson distribution of the wavefunction $f(\vphi)$,
\begin{align}
\label{eq:bdf4}
p(n)=|\braket{n}{f}|^2=\left|\int \phi_n(\vphi) f(\vphi) d \vphi \right|^2,
\end{align}
\noindent  and in  \cref{fig:fabd_bosons}-(b) the weight $1-W_{N_b}$ of the high-energy subspace versus the cutoff $N_b$, where
\begin{align}
\label{eq:wndef}
W_{N_b}=\sum_{n<N_b}p(n).
\end{align}
\noindent The figure indicates a significant boson distribution for $n>N_b=30$. In fact we observe that $50\%$ of the wavefunction belongs  to the subspace spanned by boson states with $n>30$
and $20\%$ of the wavefunction belongs to the subspace spanned by boson states with $n>40$.
However, according to the data presented in  \cref{fig:errors}-(a),
the boson number states with $n>30$ cannot be represented with $\O(10^{-5})$ accuracy on $N_\vphi=64$ discretization points.

Due to the significant high-energy weight of $f(\vphi)$, the representation of the function on a finite space with $N_\vphi=64$,
\begin{align}
\label{eq:fbad_disc}
\ket{\tilde{f}}=\sum_{i=-\frac{N_\vphi-1}{2}}^{i=\frac{N_\vphi-1}{2}} f(\vphi_i) \ket{\tvphi_i},
\end{align}
\noindent yields uncontrollable errors
when measurements are taken.
For example, the boson distribution calculated on the finite Hilbert space using the discrete representation  $\tilde{f}$ and the harmonic oscillator eigenstates $\tilde{\phi_n}$,
\begin{align}
\label{eq:dbdf4}
\tilde{p}(n)=|\braket{\tphi_n}{\tilde{f}}|^2,
\end{align}
\noindent is different from the real boson distribution given by  \cref{eq:bdf4},
as illustrated in  \cref{fig:fabd_bosons}-(a).

Since the asymptotic behavior of the wavefunction might impact
significantly its boson distribution, we consider a second example obtained by multiplying $f(\vphi)$ with the exponentially decaying function $s(\vphi)$
 \begin{align}
 \label{eq:gbad}
 g(\vphi) &=c_g f(\vphi) s(\vphi), \\
 \label{eq:sbad}
 s(\vphi)&= \frac{1} { \left(e^{-\frac{\vphi+L}{\sigma}}+1\right)^2\left(e^{\frac{\vphi-L}{\sigma}}+1\right)^2}.
 \end{align}
 In  \cref{eq:gbad}, $c_g$ is a normalization constant and, in  \cref{eq:sbad}, we take $\sigma=0.4$.
 The function $s(\vphi)$, plotted in  \cref{fig:fbad_tails}-(d), takes the value $1$ almost everywhere
 inside the interval $\left[-F, F\right]$ and decays  exponentially outside this interval
 ($s(\vphi) \propto \mathrm{exp}(-\vphi^2/\sigma^2)$ at large $|\vphi|$). Unlike $f(\vphi)$, which might
 be considered a specially chosen case, $g(\vphi)$ is a more common example. It is not band-limited or field-limited and has exponentially decaying tails.
However, at the scale shown in  \cref{fig:fbad_tails}-(a),
the functions $f(\vphi)$ and $g(\vphi)$ are indistinguishable.
The difference between $f(\vphi)$ and $g(\vphi)$ can be seen in  \cref{fig:fbad_tails}-(b).
The difference between their Fourier transforms can be seen in  \cref{fig:fbad_tails}-(c).
The tail weight of $g(\vphi)$ outside  $\left[-F, F\right]$ is $||w_F^g||\approx 6 \times 10^{-6}$.
 Unlike $\hat{f}(\kappa)$, the conjugate variable function  $\hat{g}(\kappa)$ is not zero for $\kappa>|K|$.
 However, its tail weight is small,  $||w_K^g|| \approx 6.2 \times 10^{-6}$.
 Within accuracy $\O(10^{-5})$, the discrete representation of $g(\vphi)$
 is the same as the one for $f(\vphi)$, $\ket{\tilde{g}} \approx \ket{\tilde{f}}$.

 Despite the different asymptotic behavior of the functions $f(\vphi)$ and $g(\vphi)$ at large argument,
 the difference between the boson distribution of these two functions functions is very small, indistinguishable
 on the scale shown in  \cref{fig:fabd_bosons}. The differences are noticeable for $n>80$ where the boson
 weight is small, of the order $\O(10^{-10})$ (not shown).
 All the conclusions we drew about $f(\vphi)$ are valid for $g(\vphi)$ too.
 The wavefunction  $g(\vphi)$ is not restricted to the low-energy subspace
 corresponding to $N_\vphi=64$ and accuracy $\O(10^{-5})$ and cannot be represented accurately on
 a finite Hilbert space of size $N_\vphi=64$.  The boson distribution
 $|\braket{n}{g}|^2$ of the wavefunction $g(\vphi)$ differs from the boson distribution
 $|\braket{\tphi_n}{\tilde{g}}|^2$ of the discrete representation.

 These two examples of functions with small tail weight at large argument,
 one band-limited and having algebraic decay and one with exponential decay, that can be
 sampled accurately but cannot be restricted to the low-energy subspace, show that the criteria
 of small  weight at large argument is not enough for determining the size of
 the finite representation.
 It would be useful to have an estimate of the
 Hermite-Gauss functions expansion series for {\em{almost}} band-limited and field-limited functions
 as a function of the tail weights and the cutoff $N_b$,
 \begin{align}
 \label{eq:estimnb}
 \left|\left|\ket{f}-\sum_{n=0}^{N_b} c_n \ket{n}\right|\right|\approx {\cal{E}}(N_b, ||w_F^f||, ||w_K^f||,...).
 \end{align}
\noindent  Such an expression could be used to
estimate the cutoff $N_b$ and the number of the discretization points necessary for an accurate representation by measuring the field and conjugate-field distributions.
We are not aware if an estimation like   \cref{eq:estimnb} exists in the literature. It is possible that combining the estimation
 of the prolate spheroidal wavefunctions expansion of {\em{almost}} band limited functions~\cite{Landau_Pollak_3_1962}
 with the estimation of the Hermite-Gauss function expansion of prolate spheroidal wavefunctions~\cite{osipov2013prolate}
 would yield an useful expression, but the problem requires further investigation.

\section{Sampling parameters and the boson mass choice}
\label{sec:bmass}

As discussed previously,
the  low-energy subspace of a bosonic field can be mapped
accurately onto a low-energy subspace of a finite Hilbert space.
The dimension $N_\vphi$ of the local finite Hilbert space
is monotonically increasing with the low-energy subspace dimension $N_b$.
The boson number states and implicitly
the cutoff $N_b$ are dependent on the mass parameter $m_0$ (see
 \cref{eq:phi_aa}).
The definition of the finite Hilbert space $\dH$
and of the discrete field operators  depends on $m_0$ too, as implied by \cref{eq:tphi,eq:pi_field}.
There are many possible finite representations of the bosonic field that correspond to different
choices of the boson mass. The optimal representation
is the one that requires the smallest cutoff $N_b$ for the ground state
and for the low-energy excitations  of the system.

\subsection{Squeezed boson states}
\label{ssec:squeezedb}

\begin{figure}[tb]
    \begin{center}
        \includegraphics*[width=5in]{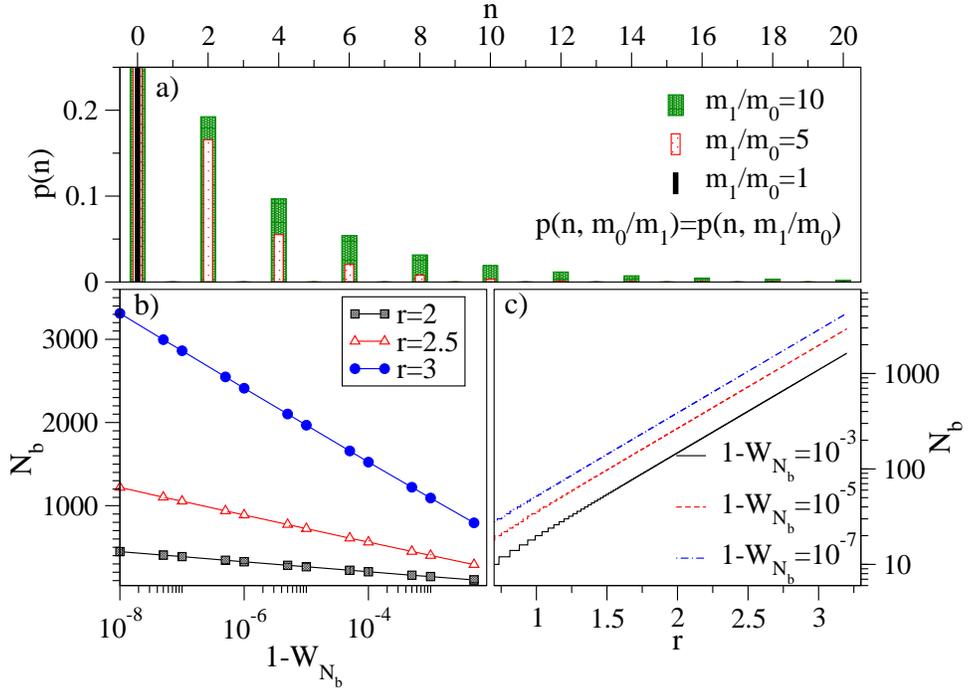}
        \caption{(a) $m_1$-boson distribution of the harmonic oscillator ground state for different values of the ratio $m_1/m_0$.
        (b) Boson cutoff number $N_b$ versus the high-energy subspace weight $1-W_{N_b}$ for different values of the
        squeezing parameter $r=\frac{1}{2}\ln\left(\frac{m_1}{m_0}\right)$.
        (c) Boson cutoff number $N_b$ versus the squeezing parameter $r$ for different values of $1-W_{N_b}$.
        Numerical fitting yields $N_b \approx \left[-0.595-0.477 \ln \left(1-W_{N_b}\right)\right]\displaystyle\frac{m_1}{m_0}$.}
        \label{fig:squeeze}
    \end{center}
\end{figure}

To represent the ground state of a harmonic oscillator with mass
$m_0$, the optimal choice for the boson mass is simply $m_0$,
because for this choice the ground state has zero bosons (the ground state is the vacuum).
However, other choices for the mass parameter can be taken, but they require more discretization
points for a specified accuracy, as we  discuss below.   We work
through this case as a prelude to more complicated Hamiltonians where
the optimal choice of mass is not obvious.

The Hamiltonian~\eqref{eq:hosc} can be re-written as
\begin{align}
\label{eq:hosc_m1}
 H_h&=\frac{m_1}{4}\left(\frac{m_0^2}{m_1^2}-1\right)\left(a^{\dagger}_{1}a^{\dagger}_{1}+a_{1}a_{1}\right)+\frac{m_1}{4}\left(\frac{m_0^2}{m_1^2}+1\right)\left(2 a^{\dagger}_{1} a_{1}+1 \right).
\end{align}
\noindent where the mass $m_1$-bosons are defined by
\begin{align}
\label{eq:aa1}
a^{\dagger}_1= \frac{1}{\sqrt{2}}\left( \sqrt{m_1} \Phi -i \frac{1}{\sqrt{m_1}}\Pi \right)~~~\text{and}~~~
a_1= \frac{1}{\sqrt{2}} \left( \sqrt{m_1} \Phi +i \frac{1}{\sqrt{m_1}}\Pi \right).
\end{align}
The relation between the mass $m_1$-bosons and the mass $m_0$ ones is given by the squeezing operation
\begin{align}
a^{\dagger}_0= S(r)^{\dagger}a^{\dagger}_1 S(r)
\end{align}
\noindent where
\begin{align}
\label{eq:rsqueez}
S(r)=e^{\frac{1}{2}r\left({a_1^{\dagger}}^2-a^2_1\right)}~~~\text{and}~~~
r=\frac{1}{2}\ln\left(\frac{m_1}{m_0}\right).
\end{align}

In the basis $\{\ket{n}_1\}_n$, where $\ket{n}_1=\frac{1}{\sqrt{n!}}a_1^{\dagger n}\ket{0}_1$ is the state with $n$
$m_1$-bosons,  the harmonic oscillator ground state $\ket{0}$ is
a squeezed vacuum state~\cite{gerry_knight_2004},
\begin{align}
\label{eq:m1vac}
\ket{0}\equiv\ket{0}_0&=S^{\dagger}(r)\ket{0}_1 =\sum_{n=0}^{\infty} C_{n}\ket{n}_1,\\
\label{eq:c2m}
C_{2n} &= \left(-1\right)^n\frac{\sqrt{\left(2n\right)!}}{2^n n!}\frac{\left(\tanh r \right)^n}{\sqrt{\cosh r}},~~~\text{ }~~~C_{2n+1}=0.
\end{align}
\noindent The magnitude of the coefficients $C_{2n}$ in  \cref{eq:c2m} decrease rapidly with
increasing $n$.  For any small $\epsilon$ a cutoff $N_b$ can be introduced such that the
the harmonic oscillator ground state has $\epsilon$ probability to have more
than $N_b$ $m_1$-bosons.

The cutoff $N_b$  increases with increasing or decreasing $\frac{m_1}{m_0}$.
In  \cref{fig:squeeze}-(a) we plot
the $m_1$-boson distribution, $p\left(n;\frac{m_1}{m_0}\right)=p\left(n;\frac{m_0}{m_1}\right)=\left|C_{n}\right|^2$
as a function of $n$ for different values of
$\frac{m_1}{m_0}$.
When $m_1=m_0$ the distribution has  $p(0)=1$
and $p(n>1)=0$, since the ground state is the $m_0$-bosons vacuum.
The distribution weight at large $n$ increases with increasing $\frac{m_1}{m_0}$ or $\frac{m_0}{m_1}$.
The cutoff $N_b$ is defined by requiring  that $1-W_{N_b} =\epsilon$, where
$W_{N_b}$ is the weight of the low-energy subspace spanned by the boson number states below the cutoff $N_b$ (see  \cref{eq:wndef}) and $\epsilon$ is the desired truncation error. In  \cref{fig:squeeze}-(b) we show
$N_b$ versus $\epsilon=1-W_{N_b}$ for different values of the squeezing parameter $r$.
The cutoff $N_b$ increases logarithmically with decreasing
$1-W_{N_b}$.
In  \cref{fig:squeeze}-(c) we show the cutoff $N_b$ versus  $r$ for different values of $1-W_{N_b}$.
The cutoff $N_b$ increases exponentially with increasing  $r$, which implies linear dependence of $N_b$ on
the boson mass $m_1$. Numerical fitting yields
$N_b \approx \left[-0.595-0.477 \ln \epsilon\right]\frac{m_1}{m_0}$.
Since the number of the discretization points $N_\vphi$ needed
to represent the low-energy subspace increases monotonically with $N_b$,
a boson mass choice $m_1 \ne m_0$ is not optimal.

This same conclusion
%The non-optimal choice $m_1 \ne m_0$ of the boson mass for the finite representation of the harmonic oscillator ground state
can be inferred just by analyzing the Nyquist-Shannon sampling parameters
of the harmonic oscillator  wavefunctions $\phi_0(\vphi)$ and $\hat{\phi}_0(\kappa)$.
For a given  number $N_\vphi$ of discretization  points,
the $m_0$-sampling  implies the sampling intervals (see \cref{eq:Ldef,eq:Nvphi_L})
\begin{align}
\label{eq:FKvsm}
F_0=\sqrt{\frac{\pi N_\vphi}{2 m_0}}, ~~~~K_0=\sqrt{\frac{\pi N_\vphi  m_0}{2}},
\end{align}
\noindent which yield equal tail weights $||w_{F_0}^{\phi_0}||=||w_{K_0}^{\phi_0}||$.
For $m_1$-sampling one has
\begin{align}
F_1= F_0\sqrt{\frac{m_0}{m_1}},~~~~K_1=K_0\sqrt{\frac{m_1}{m_0}}.
\end{align}
\noindent For $m_1>m_0$, the field sampling interval decreases
while the conjugate-field sampling interval increases by a factor $\sqrt{m_1/m_0}$.
Consequently the tail weight $||w_{F_1}^{\phi_0}|| \gg  ||w_{F_0}^{\phi_0}||$ increases exponentially, while
$||w_{K_1}^{\phi_0}|| \ll ||w_{K_0}^{\phi_0}||$ decreases
exponentially (since the tail weights have an exponential dependence on the sampling intervals' length).
Similarly, when $m_0>m_1$
the tail weight $||w_{F_1}^{\phi_0}|| \ll ||w_{F_0}^{\phi_0}||$ and
$||w_{K_1}^{\phi_0}|| \gg ||w_{K_0}^{\phi_0}||$.
In both cases, because of the large increase of one of the tail weights, the Finite Fourier transform that connects
the field and the conjugate-field sampling sets  yields a much larger error (see \cref{eq:fftdif1,eq:fftdif2}) than in the case of $m_0$-sampling. Since the error in constructing
the finite Hilbert space representation is proportional to the error
introduced by the Finite Fourier transform (see  \cref{eq:tn}), sampling corresponding
to $m_1 \ne m_0$ implies larger errors than $m_0$-sampling.

\subsection{Sampling intervals}
\label{ssec:samplingint}

The sampling and discretization intervals depend on the boson mass and the number of the discretization points, in accordance with \cref{eq:delta_x,eq:deltakappa,eq:FKvsm}.
The ratio of the sampling intervals and, as well, the ratio of the discretization intervals, equal the boson mass
\begin{align}
\label{eq:mKF}
m=\frac{K}{F}=\frac{\Delta_\kappa}{\Delta_\vphi}.
\end{align}

By definition, the optimal boson mass requires the minimal number of the discretization points for an accurate representation.
In principle, for a  specified accuracy, the optimal boson mass
can be obtained by minimizing the cutoff $N_b$
of the wavefunction's expansion in the boson number basis.
However, this is not easy to accomplish,
since the extraction of $N_b$ from quantum simulations is laborious, as  will be discussed in Section~\ref{sec:val}.

Nevertheless, instead of finding the boson mass for optimal representation, one can ask about the boson mass that yields optimal sampling.  Adjusting parameters for optimal sampling in quantum simulations is much easier than optimizing for the smallest cutoff $N_b$, as will be discussed in Section~\ref{sec:val}.
The sampling accuracy of a wavefunction is determined by the wavefunction behavior outside the field and the conjugate-field sampling intervals.
For a specified accuracy $\epsilon$, the sampling intervals parameters $F$ and $K$  should be chosen
such that (see  \cref{eq:tail_f} and  \cref{eq:tail_hf})
\begin{align}
\label{eq:FK_opt}
||w_F^{\phi}||=||w_K^{\phi}||=\epsilon.
\end{align}
\noindent  This choice will
provide, via  \cref{eq:Ncphi},
the minimum number of discretization points required for
a sampling approximation with $\O(\epsilon)$ accuracy.

Do the sampling intervals $F$ and $K$ determined by imposing   \cref{eq:FK_opt} yield  the optimal boson mass through  \cref{eq:mKF}?
While we do not know  the answer in general,
numerical checks show that the answer is \textit{yes} in many cases.
That is the case of the harmonic oscillator, as was already discussed
in Section~\ref{ssec:squeezedb}. We also found the answer to be \textit{yes} for small size $\Phi^4$ scalar field models that we can solve numerically using exact diagonalization methods.
In the following, we  present two relevant $\Phi^4$ scalar field examples .

\subsubsection{Local \texorpdfstring{$\Phi^4$} ~~scalar field}
\label{ssec:1phi4}

\begin{figure}[tb]
    \begin{center}
        \includegraphics*[width=5in]{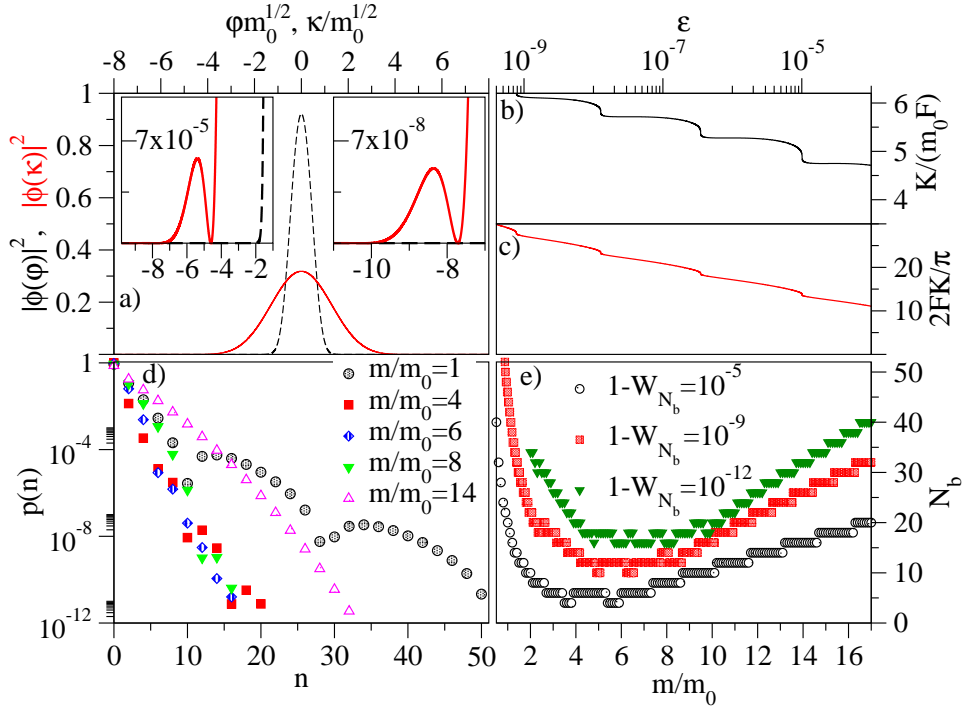}
        \caption{Local $\phi^4$ field theory,  \cref{eq:aho}, with
          $\displaystyle\frac{g}{m_0^3}=100$.
          (a) Field and conjugate
        field distributions, $|\Phi(\vphi)|^2$ (dashed black) and
        $|\hat\Phi(\kappa)|^2$ (solid red), respectively, in the ground state.
        Insets: The conjugate-field distribution has an oscillatory behavior at large $|\kappa|$. (b) The ratio
        of the sampling intervals widths $\displaystyle\frac{K}{F}$ versus the tail weight
        $\epsilon$, calculated by employing \cref{eq:FK_opt}.
        $\displaystyle\frac{K}{F}$ is larger than the bare mass $m_0$,
        and increases logarithmically with increasing the accuracy.
        (c) The number of the required discretization points $N_\vphi=\displaystyle\lceil\frac{2}{\pi}FK\rceil$ increases logarithmically
        with decreasing the tail weight.
        (d) $m$-boson distribution for different choices of $\displaystyle\frac{m}{m_0}$. (e) The low-energy subspace
        cutoff $N_b$ versus $\displaystyle\frac{m}{m_0}$ for different truncation errors $1-W_{N_b}$. The optimal
        boson mass is found when $N_b$ is minimum.}
        \label{fig:ahogg100}
    \end{center}
\end{figure}

The first example is a strong interacting local $\Phi^4$ field model,
equivalent to an anharmonic oscillator, with
the Hamiltonian
\begin{align}
\label{eq:aho}
H=\frac{\Pi^2}{2}+\frac{1}{2}m_0^2 \Phi^2+\frac{g}{4!} \Phi^4.
\end{align}

Figure~\ref{fig:ahogg100} (a) shows the field and the conjugate-field distribution
of the ground state of the Hamiltonian~\eqref{eq:aho} for interaction strength $\frac{g}{m_0^3}=100$.
One effect of the interaction is to narrow
the field distribution $|\Phi(\vphi)|^2$  and to widen the conjugate-field distribution
$|\hat{\Phi}(\kappa)|^2$ compared to the non-interacting case.  The interaction
also affects the field distributions behavior at large argument, as can  be seen in the insets.
The wavefunction $\hat{\Phi}(\kappa)$ has an  oscillatory behavior at large $|\kappa|$.

Optimal sampling implies
a ratio ($\propto K/F$) larger than the bare mass $m_0$,
because the $|\hat{\Phi}(\kappa)|^2$ distribution is wider than the $|\Phi(\vphi)|^2$ one.
Figure~\ref{fig:ahogg100}-(b) shows the ratio of the sampling intervals $\frac{1}{m_0}\frac{K}{F}$ versus the tail weight $\epsilon$, where
$F$ and $K$ are determined by  \cref{eq:FK_opt}.
The ratio $\frac{1}{m_0}\frac{K}{F}$ is dependent on $\epsilon$, and increases logarithmically
(and non-uniformly due to the oscillatory behavior of $|\hat{\Phi}(\kappa)|^2$)
with increasing the accuracy, from  $\frac{1}{m_0}\frac{K}{F} \approx 4$ for an accuracy
$\epsilon=10^{-3}$ to $\frac{1}{m_0}\frac{K}{F} \approx 6$ for $\epsilon=10^{-9}$.
The number of discretization
points $N_\vphi=\lceil 2KF/\pi \rceil$,  necessary to sample the local $\Phi^4$ field  ground state
increases logarithmically with increasing the accuracy, as can be seen in  \cref{fig:ahogg100}-(c).

To calculate the boson
distribution, we diagonalize numerically
the Hamiltonian~\eqref{eq:aho} in the boson number basis.
Figure~\ref{fig:ahogg100}-(d) shows the boson distribution, $p(n)$, as function of $n$
for different choices of the boson mass.
In all cases, the boson distribution decreases rapidly with increasing number of bosons.
We find that the largest decreasing slope  occurs when the boson mass $m/m_0  \in \approx \left[4,8\right]$.
Figure~\ref{fig:ahogg100}-(e) shows the cutoff $N_b$ versus the
boson mass for different truncation errors $1-W_{N_b}$. Remember that $1-W_{N_b}$,
with $W_{N_b}$ defined by  \cref{eq:wndef}, is the weight of the subspace spanned by the boson number states above the cutoff.
The optimal boson mass occurs at the minimum of $N_b(m/m_0)$.
For a truncation error $1-W_{N_b} \approx 10^{-5}$
we find $m/m_0 \approx 4$.  The optimal boson mass increases to $m/m_0 \approx 8$
with decreasing the truncation error to $1-W_{N_b}\approx 10^{-12}$.

The optimal boson mass determined by minimizing
$N_b$ is in agreement with the boson  mass calculated by
minimizing the sampling errors of $\Phi(\vphi)$ and $\hat{\Phi}(\kappa)$.
Since the truncation error given by the weight of the subspace
spanned by the boson number states above the cutoff
is not the same as the sampling error determined by the wavefunction's weight outside the sampling intervals,
a quantitative comparison between $\frac{K}{F}$ plotted in  \cref{fig:ahogg100}-(b) and an
optimal boson mass extracted from  \cref{fig:ahogg100}-(e) is not meaningful.
However, we found in both cases that the optimal boson mass is in the same range, $m/m_0 \in \left[4,8\right]$,
and that it increases when increasing the accuracy of the approximation.

\subsubsection{Two-site \texorpdfstring{$\Phi^4$} ~~scalar field}

\label{ssec:2phi4}

\begin{figure}[tb]
    \begin{center}
        \includegraphics*[width=5in]{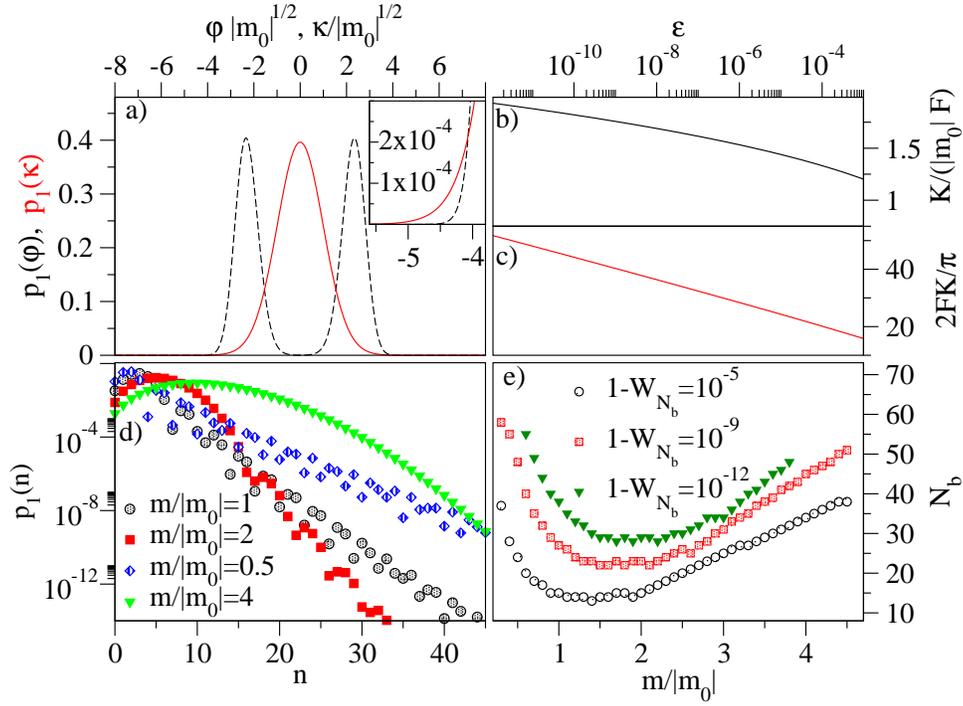}
        \caption{Two-site $\Phi^4$ field theory,  \cref{eq:2phi4},
          with $m_0^2=-1$ and $\displaystyle\frac{g}{|m_0|^3}=2$ and $h=1$
         (a) Local field,  \cref{eq:p1phi}, (dashed black) and conjugate-field,  \cref{eq:p1kappa}, (continous red) distributions in the
         ground state.
         Inset: At large argument, the field distribution decays faster than the conjugate-field distribution.
         (b) The ratio of the sampling intervals widths $\displaystyle\frac{K}{F}$ versus the tail weight.
         $\displaystyle\frac{K}{F}$ is larger than  $|m_0|$, and increases logarithmically with increasing accuracy.
         (c) The number of required discretization points $N_\vphi=\displaystyle\lceil\frac{2}{\pi}FK\rceil$ increases logarithmically
         with decreasing the tail weight.
         (d) $m$-boson local distribution for different choices of $\displaystyle\frac{m}{|m_0|}$. (e) The low-energy subspace
         cutoff $N_b$ versus $\displaystyle\frac{m}{|m_0|}$ for different truncation errors $1-W_{N_b}$. The optimal
         boson mass is found when $N_b$ is minimum.}
        \label{fig:gg2m2m2}
    \end{center}
\end{figure}

The next example is a two site $\Phi^4$ field theory,
\begin{align}
\label{eq:2phi4}
H=\sum_{i=1,2}\left(\frac{\Pi_i^2}{2}+\frac{1}{2}m_0^2 \Phi_i^2+\frac{g}{4!} \Phi_i^4\right)-h \Phi_1\Phi_2,
\end{align}
\noindent with $m_0^2=-1$, $\displaystyle\frac{g}{|m_0|^3}=2$
and $\displaystyle\frac{h}{|m_0|^2}=1$. The coupling between the fields operators at neighboring sites is a consequence
of the gradient term, $(\nabla \Phi)^2$, present in the Lagrangian of a continuous $\Phi^4$ field theory.
Although no real broken symmetry occurs for a two-site system,
the negative value of $m_0^2$ yields interesting behavior relevant for
exploring models with a broken symmetry phase.
The field in the ground state has a two-peak structure and the excitation gap is small.

The local field distribution,
\begin{align}
\label{eq:p1phi}
p_1(\vphi)=\opmatrix{\vphi}{\rho_1}{\vphi},
\end{align}
\noindent and the local conjugate-field distribution
\begin{align}
\label{eq:p1kappa}
p_1(\kappa)=\opmatrix{\kappa}{\rho_1}{\kappa},
\end{align}
\noindent are plotted in  \cref{fig:gg2m2m2}-(a). In  \cref{eq:p1phi,eq:p1kappa}
$\rho_1$ is  the local density matrix
\begin{align}
\label{eq:rho1}
\rho_1&= \text{Tr}_2(\ket{\phi}\bra{\phi}),
\end{align}
\noindent obtained by tracing out the degrees of freedom at site $2$, while $\ket{\phi}$ in  \cref{eq:rho1} is  the
ground state of the Hamiltonian~\eqref{eq:2phi4}.

Since the sampling errors of lattice wavefunctions depend on the tail weights of the local distributions (see \cref{app:tailslattice}), the sampling intervals lengths are determined by imposing
$w_{1F}^{\phi}=w_{1K}^{\phi}=\epsilon$,
where
\begin{align}
   {w_{1F}^\phi}^2&=\int_{-\infty}^{-F} d \vphi \opmatrix{\vphi}{\rho_1}{\vphi}+\int_{F}^{\infty} d \vphi \opmatrix{\vphi}{\rho_1}{\vphi}, \\
   {w^\phi_{1K}}^2&=\int_{-\infty}^{-K}  \opmatrix{\kappa}{\rho_1}{\kappa} d \kappa +
    \int_{K}^{\infty}  \opmatrix{\kappa}{\rho_1}{\kappa}d \kappa,
\end{align}
\noindent (see also
 \cref{eq:ntail_vphi,eq:ntail_kappa}).
As can be seen in the inset of  \cref{fig:gg2m2m2}-(a), the local field distribution decays more rapidly with increasing argument than the conjugate-field one.
 The ratio of the sampling intervals widths, $\displaystyle\frac{1}{|m_0|}\frac{K}{F}$, versus
 the tail weight  is plotted in  \cref{fig:gg2m2m2}-(b).
It  increases logarithmically with decreasing tail weight, from
$\displaystyle \frac{1}{|m_0|}\frac{K}{F} \approx 1.2$ when the tail weight is $\epsilon \approx 10^{-4}$ to
$\displaystyle \frac{1}{|m_0|}\frac{K}{F} \approx 2$ for  a tail weight  $\epsilon \approx 10^{-12}$.
The number of discretization points, $N_\vphi=\lceil\frac{2}{\pi}FK\rceil$,
increases logarithmically with the accuracy, as shown in  \cref{fig:gg2m2m2}-(c).

The local boson distribution,
\begin{align}
\label{eq:p1n}
p_1(n)=\opmatrix{n}{\rho_1}{n},
\end{align}
\noindent for different choices of the boson mass is shown in  \cref{fig:gg2m2m2}-(d).
The boson distribution decreases rapidly with increasing number of bosons.
The largest decreasing slope  is observed for $ {m}/{|m_0|}  \approx 2$.
The cutoff $N_b$ versus the boson mass
is shown in  \cref{fig:gg2m2m2}-(e) for different values of the truncation error $1-W_{N_b}$.
For a truncation error $1-W_{N_b} \approx 10^{-5}$
we find the optimal boson mass to be $m/|m_0| \approx 1.2$.  The optimal boson mass increases to $m/|m_0| \approx 2$
with decreasing the truncation error to $1-W_{N_b}\approx 10^{-12}$.

As in the case of the local $\Phi^4$ field example,
the boson optimal mass calculated by minimizing
$N_b$ is in agreement with the boson  mass
that minimizes  the sampling errors of the local field distributions
$p_1(\vphi)$ and $p_1(\kappa)$. In both cases, the boson mass is in the same range, $m/|m_0| \in \left[1.2,2\right]$,
and it increases when increasing the accuracy of the approximation.

Note that the optimal mass from our analysis is not determined by the standard deviation of the field distributions but by the field and conjugate-field distributions' behavior at large argument.
The ratio of the standard deviations in some mean-field theory approaches is related to the value of the boson mass.
Our results suggest that the mean-field solutions obtained in this way
are not very good approximations to the optimal mass.

\section{Post-simulation discretization validation and parameters
adjustment}
\label{sec:val}

For an accurate simulation, the low-energy subspace
should be large enough to  contain the relevant physics.
The number $N_\vphi$  of discretization points per lattice site
and the boson mass determine the low-energy subspace, but the optimal
values for these parameters are not known \textit{a priori}.
Therefore, it is important to
determine \textit{a posteriori} whether the chosen simulation's parameters are good and to
have procedures to adjust them for optimal performance.

Note that when sufficient quantum computational resources are available, in order to estimate the accuracy of the simulation's results, one can run simulations for subsequently increasing values of $N_\vphi$ and  analyze the  results' convergence properties.
However, this approach does not provide direct information about optimal discretization intervals
and likely will result in sub-optimal use of the available resources.

\subsection{Local measurements}
\label{ssec:meas}

The results of a quantum simulation are obtained by measuring
the state of the qubits in the computational basis. Not all information
about the system is easily accessible from quantum simulations.
To validate the choice of discretization parameters in our
simulation,
we  only need measurements of the local field distribution, the local conjugate-field distribution and the local boson distribution. Fortunately, these observables
can be measured relatively easily. We discuss their measurements below.

%%\paragraph{Qubit representation.}
The implementation of quantum algorithms for bosonic fields is described at length in Refs.~\cite{macridin_prl_2018, macridin_pra_2018, Li_2021}. Here we present only the minimum information necessary
to understand the measurements methods.
For every lattice site, $n_q=\log(N_\vphi)$ qubits are assigned
and the discrete field eigenvector $\ket{\tvphi_{i}}_j$
is mapped to
\begin{align}
\ket{\tvphi_{i}}_j\equiv\ket{x^{i}_0,...x^{i}_{n_q-1}}_j,
\end{align}
\noindent where $x^{i}_r\in\{0,1\}$ and $j=\overline{1,N}$ is the site label.
The field operators (see   \cref{eq:tphi} and  \cref{eq:tPhi_j}) are defined by
\begin{align}
 \tPhi_j\ket{\tvphi_{i}}_j=\vphi_{i}\ket{\tvphi_{i}}_j~~~\text{ with }~~~
\vphi_{i} =\Delta_\vphi\left(\sum_{r=0}^{n_q-1} 2^r x^{i}_r-\frac{N_\vphi-1}{2}\right).
\end{align}
\noindent

\begin{figure}[tb]
    \begin{center}
        \includegraphics*[width=5in]{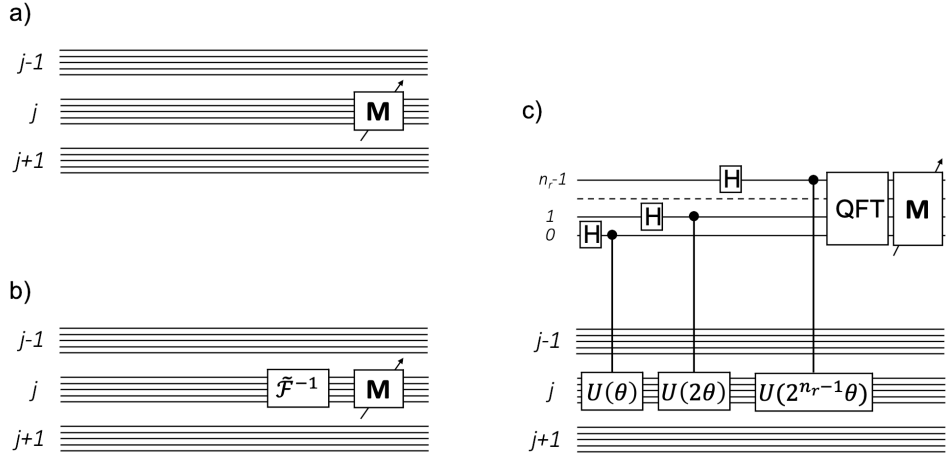}
        \caption{At every lattice site $n_q=\log\left(N_\vphi\right)$ qubits are assigned to represent
        the field. a) The field amplitude distribution at site $j$ can is obtained by direct measurement
        of the qubits assigned for the site $j$. b) The conjugate-field amplitude distribution
        requires an inverse Fourier transform, $\tF^{-1}$, see  \cref{eq:fft}, at $j$ before measurement.
        c) Quantum phase estimation algorithm for measuring the boson distribution at site $j$. An
        ancillary register of size $n_r=n_q+1$ is used to store the phase factors associated with the evolution of the system
        under the action of a local discrete harmonic oscillator Hamiltonian ( \cref{eq:dhosc_m}).
        }
        \label{fig:meas}
    \end{center}
\end{figure}

The field distribution at site $j$ is given by
\begin{align}
\label{eq:pphi2}
p_j(\vphi_{i})=\opmatrix{\tvphi_{i}}{\rho_j}{\tvphi_{i}},
\end{align}
\noindent and is obtained by the direct measurement of the qubits assigned to represent the
field at site $j$, as shown in  \cref{fig:meas}-(a).

The conjugate-field distribution at site $j$ is given by
\begin{align}
\label{eq:pkappa2}
p_j(\kappa_{p})=\opmatrix{\tkappa_{p}}{\rho_j}{\tkappa_{p}},
\end{align}
\noindent where $\{\ket{\tkappa_{p}}_j\}_{p}$ are obtained by applying a local Fourier
transform ({\em i.e.} a $n_q$-qubit Fourier transform at site $j$) to $\{\ket{\tvphi_{i}}_j\}_{i}$,
as described by  \cref{eq:fft_im}.
The measurement of this distribution requires an inverse Fourier transform, $\tF^{-1}$
(see  \cref{eq:fft}), at site $j$ before measuring the
qubits, as shown in  \cref{fig:meas}-(b).

The finite representation of the boson occupation number distribution (\textit{i.e.} the probability of the discrete harmonic oscillator eigenstates) at site $j$ is given by
\begin{align}
\label{eq:pjn}
p_j(n)=\opmatrix{\tphi_n}{\rho_j}{\tphi_n}.
\end{align}
If we write the system's wavefunction as
\begin{align}
\ket{\phi}=\sum_{e} \sum_{n=0}^{N_\vphi-1}c_{en}\ket{e}\ket{\tphi_n}_j,
\end{align}
\noindent where $\{\ket{e}\}$ is an arbitrary basis for the whole system with the site $j$ excluded,
the boson distribution is
\begin{align}
\label{eq:pjn1}
p_j(n)=\sum_{e} |c_{en}|^2.
\end{align}
The probability to have bosons above the cutoff $N_b$ is given by
\begin{align}
\label{eq:whe}
\epsilon_H=\sum_{n = N_b}^{N_\vphi-1} p_j(n)=\sum_{e} \sum_{n = N_b}^{N_\vphi-1} |c_{en}|^2.
\end{align}
The bosonic field representation is accurate when
$\epsilon_H$ is negligible.

We present two methods for the measurement of the local boson distribution. The first method
employs quantum state tomography (QST) for the local density matrix $\rho_j$. As described
in~\cite{Altepeter2004,nielsen_chuang_2010}, $\rho_j$ can be written as
\begin{align}
\label{eq:qstrho}
\rho_j=\frac{1}{2^{n_q}}\sum_{v_0,...v_{n_q-1}=0}^3 s_{v_0,...v_{n_q-1}} P_{v_0,...v_{n_q-1}}.
\end{align}
\noindent The {\em Pauli strings} $P_{v_0,...v_{n_q-1}} \equiv \sigma^j_{v_0} \otimes \sigma^j_{v_1}\otimes...\otimes\sigma^j_{v_{n_q-1}}$ are products of Pauli matrices. The single-qubit operator $\sigma^j_{v_q}$, acting  on the qubit $q \in \left\{0,1,...,n_q-1\right\}$ belonging to the local register at site $j$, takes four possible values, $\sigma^j_{v_q} \in \{I, \sigma_{x}, \sigma_y, \sigma_z\}$. The coefficients $s_{v_0,...v_{n_q-1}} =\text{Tr}\left(P_{v_0,...v_{n_q-1}}\rho_j\right)$
are determined by measuring the corresponding Pauli strings. Similar measurements of the Pauli strings are also
employed in Variational Quantum Eigensolver algorithms~\cite{McClean_NJP_2016}. Since the number of the independent coefficients defining $\rho_j$ is $4^{n_q}-1$, the number of measurements required for QST scales exponentially with $n_q$. This put a severe limitation on QST with large $n_q$~\cite{Cramer_nature_2010,Lanyon_naturephys_2017,titchener_npj_2018}. However, the current experimental development~\cite{haffner_nature_2002,Song_prl_2017,titchener_npj_2018} indicates that QST for $n_q \le 8$ (which we believe is large enough for addressing most interesting boson problems) will be feasible in the near future.

Once the local density matrix is determined, its elements in the computational basis
can be easily calculated, since this implies evaluating  the matrix elements of the Pauli strings in the computational basis. Finally, the boson distribution is given by
\begin{align}
\label{eq:pjntomot}
p_j(n)=\sum_{i,l=0}^{N_\vphi-1}\braket{\tphi_n}{\tvphi_{i}}\opmatrix{\tvphi_{i}}{\rho_j}{\tvphi_{l}}\braket{\tvphi_{l}}{\tphi_n},
\end{align}
\noindent where the coefficients $\braket{\tphi_n}{\tvphi_{i}}$ are obtained from the exact diagonalization of the discrete harmonic oscillator Hamiltonian~\eqref{eq:dhosc}.

 The second  method for the measurement of  the boson distribution at the lattice site $j$
employs quantum phase estimation (QPE)~\cite{nielsen_chuang_2010,cleve_1998_procRsocA}
 measurements for the discrete harmonic oscillator
 \begin{align}
 \label{eq:dhosc_m}
 \tilde{H}_{h1}&=\frac{1}{2}\tilde{\Pi}^2+\frac{1}{2} m_0^2 \tilde{\Phi}^2-\frac{1}{2}m_0=H_h-\frac{1}{2}m_0,
  \end{align}
  \noindent where we subtract the constant term $\frac{1}{2}m_0$ for convenience.
  The eigenvalues of
  the Hamiltonian~\eqref{eq:dhosc_m} have the following property (within the desired accuracy of the finite representation approximation)
  \begin{align}
  \label{eq:enm}
  \tilde{E}_n=m_0 n,~~~~\text{for}~~~n < N_b,\\
  \label{eq:enhe}
  \tilde{E}_n \ne m_0 n,~~~~\text{for}~~~n \ge N_b.
  \end{align}
  \noindent For example, see the eigenvalues of the discrete harmonic oscillator for $N_\vphi=64$ and $N_\vphi=128$
  plotted in Fig.1-(a) of Ref~\cite{macridin_pra_2018}.

The time evolution  operator corresponding to Hamiltonian~\eqref{eq:dhosc_m}
\begin{align}
\label{eq:evolHh}
U(\theta)\equiv e^{-i 2 \pi \theta H_{h1}}
\end{align}
\noindent  can be implemented using Trotterization methods, as described
in Ref~\cite{macridin_prl_2018, macridin_pra_2018, Li_2021}. The operator~\eqref{eq:evolHh}
acts only on the $n_q$ qubits assigned to the field at the site $j$.

The implementation of the phase estimation algorithm
is illustrated in  \cref{fig:meas}-(c). An ancillary register of $n_r$ qubits is used.
On every ancillary qubit, a Hadamard gate is applied. Next, for every  qubit
 $m$ from the ancillary register (with $m=\overline{0,n_r-1}$), a control-$U(2^{m}\theta)$ gate,
acting on the ancilla qubit $m$ and the local field register at site $j$, is applied.

The state of the system together with the ancillas changes from
\begin{align}
\ket{\phi}\ket{0}_a=\sum_{e} \sum_{n=0}^{N_\vphi-1}c_{en}\ket{e}\ket{\tphi_n}_j\ket{0}_a,
\end{align}
\noindent where  $\ket{0}_a$ is the ancillary register state, to
\begin{align}
\label{eq:qpei}
\sum_{e} \sum_{n=0}^{N_\vphi-1}c_{en}\ket{e}\ket{\tphi_n}_j  \frac{1}{2^{\frac{n_r}{2}}} \sum_{x=0}^{2^{n_r}-1} \ket{x}_a e^{-i 2 \pi\theta \tilde{E}_n x},
\end{align}
\noindent after applying the Hadamard and the $CU$ operators. In  \cref{eq:qpei}, $\ket{x}_a$ is the binary
representation on qubits of the integer $x=\overline{0,2^{n_r}-1}$.
To distinguish between the phase factors corresponding to all
eigenvalues  of the Hamiltonian~\eqref{eq:dhosc_m}, the parameter $\theta$ should be chosen such that
\begin{align}
\label{eq:thetabound}
\theta  < \frac{1}{\Delta E},~~~\text{where}~~ \Delta E =\max_n{\tilde{E}_{n}}-\min_n{\tilde{E}_{n}}
\end{align}
\noindent is the range of the Hamiltonian~\eqref{eq:dhosc_m} spectrum.

After the Quantum Fourier transform is applied on the ancilla register,
the state described previously by  \cref{eq:qpei} becomes
\begin{align}
\label{eq:chi_ancilla1}
\ket{\chi} \equiv
\sum_{e}\sum_{n=0}^{N_\vphi-1} \sum_{k=0}^{2^{n_r}-1} c_{en} a_{nk} \ket{e}\ket{\tphi_n}_j \ket{k}_a
\end{align}
\noindent where $\ket{k}_a$ is the binary representation on qubits of the integer $k=\overline{0,2^{n_r}-1}$
and
\begin{align}
\label{eq:ank}
a_{nk} =\frac{1}{2^{n_r}}\sum_{x=0}^{2^{n_r}-1} e^{-i \frac{2 \pi}{2^{n_r}} \left(2^{n_r}\theta \tilde{E}_n-k\right) x}.
\end{align}
The probability to measure the integer $k$ on the ancilla register is given by
\begin{align}
\label{eq:pk}
p(k)=\sum_{e}\sum_{n=0}^{N_\vphi-1} |c_{en} a_{nk}|^2.
\end{align}

If we choose
\begin{align}
\label{eq:thetachoice}
\theta&=\frac{1}{m_0 2^{n_r}},
\end{align}
\noindent then
\begin{align}
\label{eq:ank1}
a_{nk} =\frac{1}{2^{n_r}}\sum_{x=0}^{2^{n_r}-1} e^{-i \frac{2 \pi}{2^{n_r}} \left(\frac{\tilde{E}_n}{m_0}-k\right) x}.
\end{align}

The choice of $\theta$ given by  \cref{eq:thetachoice} is convenient since
$\tilde{E}_n/m_0=n$ for $n<N_b$. Thus, for
$n<N_b$  \cref{eq:ank} is a Kronecker delta function, $a_{nk}=\delta_{nk}$.
The probability to measure an integer $k \ge N_b$ in the ancilla register reduces to
\begin{align}
\label{eq:pkh}
p(k)=\sum_{e}\sum_{n = N_b}^{N_\vphi-1}|c_{en}|^2|a_{nk}|^2,~~~~\text{for}~~~k \ge N_b,
\end{align}
\noindent  since the terms in  \cref{eq:pk} with $n<N_b$ are zero.
Since $|a_{nk}| \le 1$ (see Appendix~\ref{ap:bdm}), we have the following inequality
\begin{align}
\label{eq:probk}
p(k) \le \sum_{e}\sum_{n=N_b}^{N_\vphi-1}|c_{en}|^2 =\epsilon_H,~~~~\text{for}~~~k \ge N_b.
\end{align}
\noindent For any $k \ge N_b$, the probability to measure $k$ is smaller than
the  probability to have more than $N_b$ bosons. Thus
\begin{align}
\label{eq:ehb1}
\epsilon_H \ge \max_{k\ge N_b} p(k) \equiv p_{1\text{max}}.
\end{align}

The probability to measure any integer $k \ge N_b$ in the ancilla register is given by
\begin{align}
\label{eq:prob_he}
p_{\text{all}}=\sum_{k = N_b}^{2^{n_r}-1} p(k)
=\sum_e \sum_{n = N_b}^{N_\vphi-1} |c_{en}|^2 \sum_{k = N_b}^{2^{n_r}-1} |a_{nk}|^2
\ge \frac{4}{\pi^2} \sum_e \sum_{n = N_b}^{N_\vphi-1}|c_{en}|^2  = \frac{4}{\pi^2} \epsilon_H.
\end{align}
\noindent In  \cref{eq:prob_he}, we used
\begin{align}
\label{eq:ankmax}
\sum_{k = N_b}^{2^{n_r}-1} |a_{nk}|^2 \ge \max_{k \ge N_b} |a_{nk}|^2 \ge \frac{4}{\pi^2}~~~~\text{for}~~~n>N_b,
\end{align}
\noindent which is proven in Appendix~\ref{ap:bdm} (see  \cref{eq:ankmaxapp}).

Combining  \cref{eq:ehb1} and  \cref{eq:prob_he}, the probability to have more than $N_b$ bosons
 is bounded as
\begin{align}
\label{eq:ehbounds}
p_{1\text{max}} \le \epsilon_H \le \frac{\pi^2}{4}p_{\text{all}}.% \approx 2.4674 P_H.
\end{align}

According to  \cref{eq:ehbounds}, the discretization parameters $N_\vphi$ and $m_0$ used for bosonic field representation
are valid if there is a negligible probability to measure
integers larger than the cutoff $N_b(N_\vphi)$ on the ancillary registry.

\begin{table}
\begin{tabular}{ |c |c | c | c| c | c| c |}
\hline
 $N_\vphi$ & 32 & 64 & 128 & 256 & 512 & 1024\\
 \hline
 $\Delta E /m_0$ &  42.319 & 89.396 & 185.376 & 379.976 & 772.944 & 1564.233 \\
 \hline
 $N_b$ ($\epsilon_c<10^{-4}$) & 10 & 30 & 74 & 164 & 353 & 741\\
 \hline
\end{tabular}
\caption{Middle row: Energy range of the discrete harmonic oscillator for different values
of $N_\vphi$, calculated using exact diagonalization. Bottom row: Boson cutoff
number  corresponding to the commutation relation error ( \cref{eq:err_c}) $\epsilon_c \approx 10^{-4}$.
}
\label{table:de}
\end{table}

The size of the ancillary register is determined by  \cref{eq:thetabound} and  \cref{eq:thetachoice},
 \begin{align}
\label{eq:nrchoice}
 n_r& \ge\log\left(\left[\frac{\Delta E}{m_0}\right]\right).
\end{align}
\noindent The number of ancillary qubits scales logarithmically
with the energy range of the discrete
harmonic oscillator Hamiltonian. The values of the energy range $\Delta E$ corresponding to
different $N_\vphi$ are given in Table~\ref{table:de}. We find that
$\Delta E/m_0 < 2 N_\vphi$ for $N_\vphi \le 1024$ (and probably true for larger values of $N_\vphi$ as well but numerical checks are necessary for confirmation). In practice the number of ancillary qubits
required for the QPE register is
\begin{align}
n_r=n_q+1.
\end{align}

Measuring energies in QPE algorithms with $2^{-n}$ accuracy and with $1-\epsilon$
probability requires registers
of size $n_r=n+\log\left[2+1/(2\epsilon)\right]$~\cite{nielsen_chuang_2010,cleve_1998_procRsocA},
thus larger than in our case when $\epsilon \lesssim 10^{-2}$. In our case, the goal of the QPE measurement is
not to estimate the energies of $H_h$ (which we know from exact diagonalization of the
finite Hamiltonian matrix) but to measure the
boson distribution and especially the
probability to have states with the number of bosons
larger than $N_b$. When the probability to have bosons
above the cutoff is negligible, {\em i.e.} $\epsilon_H \approx 0$,
the boson distribution can be measured with  high precision. This is
true because
the energies of the states with $n<N_b$ are proportional to $n$ (see  \cref{eq:enm}),   \cref{eq:ank1}
becomes a Kronecker delta function and the probability to measure $k<N_b$ on the ancillary register becomes equal
to the probability to have $n$ bosons (see  \cref{eq:pjn1}),
\begin{align}
\label{eq:pklessnb}
p(k=n)=\sum_{e}|c_{en}|^2=p_j(n)~~~\text{for}~~~k<N_b~~~\text{when}~~\epsilon_H=0.
\end{align}

\subsection{Simulation guideline for parameters' validation and adjustment}
\label{ssec:guide}

In this section, we present a guideline for quantum simulations of bosonic fields. The main goal
 is to provide  a practical  procedure for adjusting $N_\vphi$ and boson mass $m$ for optimal performance.
Let's  assume for now that the system has translational symmetry and the local measurements yield identical results at all sites.

\begin{itemize}
\item If $10$ or less bosons per site is expected to be adequate to capture the low-energy physics, start with $N_\vphi=32$ discretization points per lattice site.  Otherwise start with a larger $N_\vphi$. Equation~\eqref{eq:epsNphiNb}
can be used to determine the dependence $N_b(N_\vphi, \epsilon)$.
In Table~\ref{table:de} we provide the value of $N_b$ for different $N_\vphi$ when the accuracy is of order $\O(10^{-4})$.

\item Start with a  boson mass $m=m_0+\delta m$, where $m_0$ is the bare mass and
$\delta m$ is the mean-field contribution.

\item After the system state is prepared on qubits, measure the local field distribution, $p_j(\vphi_{i})$,
and the conjugate-field distribution, $p_j(\kappa_{p})$ at the arbitrary site $j$, as described in Section~\ref{ssec:meas}.

\item Determine the coefficients $\beta_\vphi$ and $\beta_\kappa$ such that
the probability to measure the field outside the range $\left[-\beta_\vphi F, \beta_\vphi F\right]$
and, respectively, the probability to measure the conjugate-field outside the range
$\left[-\beta_\kappa K, \beta_\kappa K\right]$ are smaller than $\epsilon$,
 \begin{align}
 \sum_{i \text{ for }|\vphi_{i}| > \beta_{\vphi} F} p_j(\vphi_{i})<\epsilon\\
 \sum_{p \text{ for }|\kappa_{p}| > \beta_{\kappa} K} p_j(\kappa_{p})<\epsilon.
 \end{align}
\noindent If both $\beta_\vphi \le f_c$
and $\beta_\kappa \le f_c$
the wavefunction sampling is
accurate. The parameter $0<f_c<1$
should be chosen to ensure confidence that the
distribution weights at large argument are
$\epsilon$ small. When $f_c$ is very large the confidence is low and when $f_c$ is very small
resources are wasted. We believe that an acceptable  range value for $f_c$ is $\left[0.6, 0.8\right]$.

The factors $\beta_\vphi$ and $\beta_\kappa$ can be modified
 by changing the mass factor since they depend on the intervals' widths $F \propto m^{-1/2}$ and $K \propto m^{1/2}$ (see  \cref{eq:FKvsm}). A change of the boson mass by a factor $\mu$, $m \rightarrow \mu m$,
 implies $\beta_\vphi \rightarrow \sqrt{\mu} \beta_\vphi $ and
 $\beta_\kappa \rightarrow \left(1/\sqrt{\mu}\right) \beta_\kappa$.

\item
If $\beta_\vphi \beta_\kappa \le f_c^2$ and $\beta_\vphi \approx \beta_\kappa$ the guess of the initial mass was close to optimal.
If $\beta_\vphi \beta_\kappa \le f_c^2$ and $\beta_\vphi \not\approx \beta_\kappa$
adjust the boson mass by multiplying it with a factor of $\mu=\beta_\kappa/\beta_\vphi$. The new boson mass
determines the optimal sampling discretization intervals.

\item The case $\beta_\vphi \beta_\kappa > f_c^2$  means that both the field and the
conjugate-field distributions close to the sampling intervals' edges are significant and cannot be adjusted properly by increasing one sampling interval and decreasing the other via boson mass scaling.
The number $N_\vphi$ of the discretization
points should be increased by at
least a factor of $\beta_\vphi \beta_\kappa/f_c^2$.

At this point the parameters $N_\vphi$ and $m$ are good for optimal field sampling.
However, as shown in Section~\ref{ssec:leval}, accurate field
sampling does not necessary implies wavefunction containment to the low-energy subspace.

\item Measure the local boson distribution as described in Section~\ref{ssec:meas}.

\item If the probability to measure integers $k \ge N_b$ are larger than $\epsilon$,
increase $N_\vphi$ (and implicitly $N_b(N_\vphi,\epsilon)$) until the probability to measure integers
$k \ge N_b$ are smaller than $\epsilon$.

At this point the finite representation of the bosonic field
defined by the parameters $N_\vphi$ and $m$ should be
close to optimal for an accuracy $\O(\epsilon)$.

\end{itemize}

In case the wavefunction has no translational symmetry, measurements at all sites are necessary for the validation and adjustment of the discretization parameters.
The parameters $N_\vphi$ and $m$ should be chosen to provide accurate sampling and
a boson distribution contained to the low-energy subspace at all sites. In this case, the global optimal $m$  might not be optimal at every site.

In many simulations, the system's wavefunction changes in time under the action of the evolution operators.
This might be the case for adiabatic continuation or for studying non-equilibrium
physics, for example.
In principle, measurements for the validation of the discretization parameters
should be taken at every time step to make sure that
the number of bosons above the cutoff is always smaller than $\epsilon$. However, in practice, it
is not necessary to take discretization validation measurements at every Trotter step. The effect
of one Trotter step is of the order of the step size and, therefore, is small.
Likely, it will be sufficient to take discretization validation measurements at a rather small number of time points,
as long as the boson distribution is well below $N_b$
for these measurements.

Use of the optimal parameters will yield
the highest precision results for the computational resources available, but this can be challenging in practice. However,
accurate, error-controlled quantum simulations can  still be performed without
adjusting the parameters to their optimal value as long as the problem we address can be restricted to the low-energy subspace.
Adjusting the boson mass to the one optimizing the sampling of the wavefunction
might increase the
precision of the simulations even when the mass is not optimal.

\section{Discussion of Future Applications}
\label{sec:disc}

 In this paper, we used the boson number basis to construct a local
 finite Hilbert space.  A low-energy subspace was defined by introducing a cutoff in this basis.
A different denumerable basis, for example $\{\ket{\alpha_n}\}$, might
be considered for constructing a finite representation, following a
similar procedure.
However, this change is not trivial, and would require the investigation
of the  Nyquist-Shannon sampling properties of $\braket{\vphi}{\alpha_n}$ and $\braket{\kappa}{\alpha_n}$, knowledge of the
recurrence relations for $\vphi\braket{\vphi}{\alpha_n}$ and
$\kappa\braket{\kappa}{\alpha_n}$,
(similar to the ones given by  \cref{eq:hg_rec_phi} and, respectively,   \cref{eq:hg_rec_kappa}) and measurement methods for the local distribution $\opmatrix{\alpha_n}{\rho_j}{\alpha_n}$. We mention this as a topic for future investigation.

Quantum mechanical problems written in
the {\it first quantization} formalism can be simulated on a quantum computer by employing
the discretization methods developed for the bosonic fields. The position $X_j$ and the momentum $P_j$ operators (here $j$ is an arbitrary label)  entering the first quantization Hamiltonian $H(X_1,X_2,..,P_1,P_2...)$
play the same role as the field operators $\Phi$
and $\Pi$, since they obey the canonical commutation relation $\left[X_j, P_l\right]=iI\delta_{jl}$.
The field variable $\vphi$ becomes the position variable $x$ while the conjugate-field variable $\kappa$ becomes the momentum variable $p$.
The system's wavefunction is discretized in the
position and momentum bases. For a general
interaction potential $V(X_1, X_2,..)$, a qubit implementation of the corresponding Trotter step operator requires the calculation of the phase factor proportional to $V(x_1, x_2,..)$ for each qubit configuration $\ket{x_1,x_2,...}$. This can be challenging when the computation resources are finite, being of similar difficulty as designing a quantum circuit to calculate the function $V(x_1, x_2,..)$ ~\cite{haner_2018,bhaskar_2015}.
However, when the potential can be approximated by
a truncated  Taylor expansion, the implementation reduces to a number of Trotter steps for the
monomial terms appearing in the expansion.
The Trotter step corresponding to a monomial term with  degree $r$
(for example $X_iX_{i+1}...X_{i+r}$) requires $\O(n_q^{r})$ two-qubit
gates~\cite{macridin_pra_2018}. Special care should also be taken to
ensure that the number of the discretization points is large enough
such that the action of $V(X_1, X_2,..)$
does not violate the low-energy subspace constraints.

As for many quantum algorithms, the main limitation for the implementation of bosonic quantum algorithms on present-day quantum hardware is  the two-qubit gate fidelity.
Finite coherence time and control error restrict the maximum number of two-qubit gates to be less than 100 for quantum simulation algorithms implemented on state-of-art quantum processors \cite{google2020hartree,jurcevic2021demonstration}. This is not adequate for large bosonic field simulations, considering that a Trotter step requires $N \times 50 \sim N \times 10000$
two-qubit gates, where $N$ is the lattice size (depending on the interaction type and strength).
Problems which require time evolution
simulations with thousands or millions of Trotter steps can probably
be addressed only after error-corrected quantum technology is
developed. However we are optimistic that interesting problems, such
as the one-dimensional $\phi^4$ model and polaron and bipolaron
models, can be addressed on near-future quantum computers that can run circuits with thousands of two-qubit gates. For example,
for problems where the cutoff $N_b \le 10$ and the accuracy is $\epsilon \approx 10^{-2}$, we estimate
Trotter steps requiring $N \times 50 \sim N \times 100$ two-qubits  gates. These problems can be simulated on near-future
hardware by employing noise mitigation techniques~\cite{kim_arxiv_2021,google_arxiv_2020} and variational algorithms which only
require the implementation of a few Trotter steps~\cite{cirstoiu_npj_2020, benedetti_prr_2021}.

\section{Conclusions}
\label{sec:conclusions}

In this work, we address the representation of lattice bosonic fields on the finite Hilbert space of quantum computers. An accurate representation \textit{i)} implies
accurate storage of the wavefunction on qubits  and \textit{ii)}  requires
definition of qubit field operators whose action on the qubit wavefunction reproduces the action of the real field operators.
We construct a finite representation for
the low-energy subspace spanned by states
with the number of bosons per site below a cutoff $N_b$.
Since the lattice Hilbert space is a direct product of local Hilbert spaces, the representation of the lattice Hilbert space is a direct product of  local Hilbert space representations.

A local Hilbert space is infinite dimensional and equivalent to the space of the square integrable functions.
The construction of the finite representation for a local Hilbert space is based on Nyquist-Shannon sampling properties
of square integrable functions.
Because the weight of these functions vanishes at large argument, they can be sampled with controlled accuracy both in a finite set of field variable points and in a finite set of conjugate-field variable points. Within the same level of accuracy as the  sampling approximation,
the two sampling sets, of field and of conjugate-field points, are connected by a Finite Fourier transform. The accuracy of the sampling approximation is determined by the weight of the functions outside
the sampling intervals. The errors decrease with increasing the width of the sampling intervals and vanish in the infinite width limit.

By exploiting the sampling properties of the Hermite-Gauss functions,
we construct a finite Hilbert space of dimension $N_\vphi$
and define discrete field operators $\tPhi$ and $\tPi$
such that, within $\O(\epsilon)$ accuracy, the operators $\tPhi$ and $\tPi$ act on the
subspace spanned by the first $N_b<N_\vphi$ eigenstates of
the discrete harmonic oscillator in the same way  the field operators, $\Phi$ and, respectively $\Pi$,
act on the  subspace spanned by the first $N_b$ harmonic oscillator eigenvectors.
As long as the relevant physics of the system is restricted to the low-energy subspace defined
by the cutoff $N_b$, the low-energy subspace can be mapped to the low-energy subspace
of the finite Hilbert space.

We investigate analytically and numerically the different errors associated with the sampling of the HG functions
and with the action of the discrete field operators
on the eigenstates of the discrete harmonic oscillator.
These errors are proportional to the tail weight of the HG functions. The accuracy of the finite representation
is of the same order as the weight of the
HG function of the order $N_b+2$ outside the
sampling interval.
The errors are reduced exponentially by increasing the number of the discretization points. For fixed accuracy, the number of discretization points increases
linearly with the size $N_b$ of the low-energy subspace.

The definition of the finite Hilbert space and of the discrete field operators depends on the boson mass.
The optimal boson mass is the one that requires the smallest number of discretization points for a given accuracy.
While a calculation of the optimal boson mass
by minimizing the low-energy cutoff $N_b$ is difficult
in quantum simulations, finding a boson mass that minimizes
the sampling errors of the system wavefunction is much easier.
The boson mass optimizing the wavefunction's sampling equals
the ratio of the sampling intervals that yield $\epsilon$ small tail weights. For scalar
$\Phi^4$ models on small lattices, we find that the boson mass optimizing
the sampling is a good approximation for the optimal boson mass.

The states belonging to the low-energy subspace are sampled accurately. However, the converse is not true: accurate sampling does not  necessarily imply  that the states belong to the low-energy subspace. We present two examples of functions
that are sampled with high accuracy but have a significant high-energy component.
As a consequence,  local boson distribution measurements are necessary
to  validate the
discretization parameters of a quantum simulation.

We present a guideline to validate and adjust
the discretization parameters $N_\vphi$ and $m$
that determine the accuracy of the simulation for optimal performance.
The guideline requires
measurements of the local field, local conjugate-field and local boson distributions.
The field and conjugate field measurements are done by measuring
the qubits assigned to represent the field. For the
measurement of the local boson distribution we present two methods.
The first employs quantum state tomography of the $n_q$-qubit register assigned to describe the boson field at a particular lattice site. The second method employs the QPE technique for a discrete harmonic oscillator evolution operator acting on the local $n_q$ qubit register. The QPE measurements require
an ancillary register of size $n_q+1$.
The probability to measure bosons states above the cutoff is bounded by the probability to measure integers above the cutoff in the ancillary register.
When the bosons number states above the cutoff have negligible weight, the local boson distribution can be measured with high precision.
The guideline's first part explains how, based on the field
and conjugate-field distribution measurements,
the discretization parameters can be optimized for
optimal sampling.
The validation of the discretization parameters is finally done by measuring  the local boson distribution. The parameters are valid if the probability to measure bosons above the cutoff is negligible. Otherwise the number of the discretization points should be increased.

The methodology presented here can be applied to quantum problems written in the first quantization formalism, since the position and momentum operators obey the same commutation relation as the field and conjugate field operators.

The idea of using an interaction-dependent boson mass to represent the system's relevant degrees of freedom in the most efficient way is not new. When the optimal mass is used, the state of the system has the smallest number of excitations per site above the boson vacuum.
This might be related to the renormalization theory method of
using an interaction-dependent physical mass in the diagrammatic calculations.
Instead of the bare mass, which has no real physical significance,
the physical mass absorbs many divergent diagrams from the diagrammatic expansion. In our case,
a large number of bosonic excitations are absorbed by redefining the boson mass.

\section{Acknowledgments}

A. M. is partially supported by the DOE/HEP QuantISED program grant of the theory  consortium "Intersections of QIS and Theoretical Particle Physics" at Fermilab."
A.C.Y.L.\ and S.M.\ are partially supported by the DOE/HEP QuantISED program grant "HEP Machine Learning and Optimization Go Quantum", identification number 0000240323.
This manuscript has been authored by Fermi Research Alliance, LLC under Contract No. DE-AC02-07CH11359 with the U.S. Department of Energy, Office of Science, Office of High Energy Physics.

\appendix
\section{Nyquist-Shannon sampling with half-integer summation indices}
\label{app:NShlf}

Let $f(\vphi)$ be a band-limited function, {\em i.e.} $\hf(\kappa)=0$  for $|\kappa|>K$,
where $\hf(\kappa)$ is the Fourier transform
of $f(\vphi)$ defined by   \cref{eq:ftdef}
and $K$ is a positive real number.

The anti-periodicity of the function defined as
\begin{align}
\label{eq:fap}
\hf_{ap}(\kappa)&=\hf(\kappa) ~~\text{for}~~~~ \kappa \in \left[-K,K\right],\\
\hf_{ap}(\kappa+2K)&=-\hf_{ap}(\kappa),
\end{align}
\noindent implies
\begin{align}
\hf_{ap}(\kappa)&= \frac{\Delta_\vphi}{\sqrt{2 \pi}}\sum_{i=-\infty}^{\infty} f(\vphi_i)e^{-i \kappa \vphi_i},
~~~\text{with}~~~\vphi_i=\left(i+\frac{1}{2}\right)\frac{\pi}{K}=\left(i+\frac{1}{2}\right)\Delta_\vphi,\\
f(\vphi_i)&=\frac{1}{\sqrt{2 \pi}}\int_{-K}^{K} \hf_{ap}(\kappa) e^{i \kappa \vphi_i} d \kappa.
\end{align}

Since $\hf(\kappa)$ has support only on the interval $\left[-K, K \right]$, it can be written as
\begin{align}
\label{eq:hf_R}
\hf(\kappa)=\hf_{ap}(\kappa) R_K(\kappa),
\end{align}
\noindent where $R_K(\kappa)$ is the rectangular function defined as
\begin{align}
 R_K(\kappa)=\left\{
  \begin{array}{ll}
  1 &~~\text{for}~~~~ \kappa \in \left[-K, K\right]\\
  0 &~~\text{for}~~~~ |\kappa|>K
  \end{array}
 \right. .
 \end{align}
 \noindent The Fourier transforms of $R_K(\kappa)$ is proportional to the {\em sinc} function
 $u_K(\vphi)$ (see  \cref{eq:phiiuk}),
\begin{align}
\label{eq:ftuk}
\frac{1}{\sqrt{2 \pi}} \int_{-K}^{K}  e^{i \kappa \vphi} d \kappa =\sqrt{2 \pi} \frac{\sin L \vphi}{\pi \vphi}=\frac{\sqrt{2 \pi}}{\Delta_\vphi} \sinc\bigg(\frac{\vphi}{\Delta_\vphi}\bigg) \equiv \frac{\sqrt{2 \pi}}{\Delta_\vphi} u_K(\vphi).
\end{align}

The function $f(\vphi)$ is obtained by performing an inverse Fourier transform of  \cref{eq:hf_R},
\begin{align}
f(\vphi)=\frac{1}{\sqrt{2 \pi}}  \int_{-\infty}^{\infty} d \kappa  e^{i \kappa \vphi}
\frac{\Delta_\vphi}{\sqrt{2 \pi}}\sum_{i=-\infty}^{\infty} f(\vphi_i)e^{-i \kappa \vphi_i} R_K(k)=
\sum_{i=-\infty}^{\infty} f(\vphi_i) u_K(\vphi-\vphi_i).
\end{align}
Any band-limited function can be reconstructed from its values on an infinite and discrete set of sampling points,
$\{\vphi_i=\left(i+\frac{1}{2}\right) \Delta_\vphi\}_{i=\overline{-\infty,\infty}}$.

\section{Sampling error}

\subsection{Local wavefunctions}
\label{app:tails}

Consider the function $f(\vphi) \in S(\mathbb{R})$.
The difference between $f(\vphi)$ and $\tilde{f_\vphi}(\vphi)$ defined by  \cref{eq:NS_app1} is
\begin{align}
\label{eq:diff_fh}
f(\vphi)-\tilde{f_\vphi}(\vphi)&= f(\vphi)-\sum_{i=-\infty}^{\infty} f(\vphi_i)u_K(\vphi-\vphi_i)+
\sum_{|i|>\frac{N_\vphi-1}{2}} f(\vphi_i)u_K(\vphi-\vphi_i)\\ \nonumber
&= f(\vphi)- \opmatrix{\vphi}{P_K}{f}+ \sum_{i=-\infty}^{\infty} \left(\opmatrix{\vphi_i}{P_K}{f}-f(\vphi_i)\right)u_K(\vphi-\vphi_i)+
\sum_{i=-\infty}^{\infty} w_F^f(\vphi_i)u_K(\vphi-\vphi_i)\\  \nonumber
&=w_K^f(\vphi)-\sum_{i=-\infty}^{\infty} w_K^f(\vphi_i)u_K(\vphi-\vphi_i)+\sum_{i=-\infty}^{\infty} w_F^f(\vphi_i)u_K(\vphi-\vphi_i)
\end{align}
\noindent where $w_F^f$ and $w_K^f$ were defined by  \cref{eq:tailF} and  \cref{eq:cutk_function},
respectively. In the second line of  \cref{eq:diff_fh},
we added and subtracted the band-limited term $\opmatrix{\vphi}{P_K}{f}$.

 \cref{eq:diff_fh} can be written as
\begin{align}
\label{eq:diff_fh1}
\ket{f}-\ket{\tilde{f_\vphi}}=\ket{w_K^f}-\ket{v}+\ket{t}
\end{align}
\noindent with
\begin{align}
\braket{\vphi}{v}=\sum_{i=-\infty}^{\infty} w_K^f(\vphi_i)u_K(\vphi-\vphi_i)\\
\braket{\vphi}{t}=\sum_{i=-\infty}^{\infty} w_F^f(\vphi_i)u_K(\vphi-\vphi_i).
\end{align}
\noindent The sampling error is bounded as:
\begin{align}
\label{eq:diff_f}
|| f- \tilde{f_{\vphi}}|| \le || w_K^f|| +||v|| + ||t||.
\end{align}

To estimate $||v||$, we
write $\ket{v}$ in the conjugate-field basis.
Using  \cref{eq:ftuk}, we have
\begin{align}
\label{eq:vk}
v(\kappa)&=\sum_{i=-\infty}^{\infty} w_K^f(\vphi_i) \frac{\Delta_\vphi}{\sqrt{2 \pi}} R_K(\kappa) e^{-i \kappa \vphi_i}
= \frac{1}{2K}
\int dq w_K^f(q) R_K(\kappa) \sum_{i=-\infty}^{\infty}e^{i (q -\kappa)\vphi_i}\\ \nonumber
&=\sum_{n=-\infty}^{\infty}\int dq w_K^f(q) R_K(\kappa) \left(-1\right)^n\delta(\kappa-q+ 2nK)=R_K(\kappa)\sum_{n=-\infty}^{\infty} \left(-1\right)^n w_K^f(\kappa+2nK)
\end{align}
\noindent The $\left(-1\right)^n$ factor is a consequence of the half-integer values of the summation index $i$ in  \cref{eq:vk}.

The vector $v(\kappa)$ can be written as
\begin{align}
\label{eq:vsum}
v(\kappa)&=\sum_{n=-\infty}^{\infty} v_n(\kappa)
\end{align}
\noindent where
\begin{align}
v_n(\kappa)=R(\kappa)\left(-1\right)^n w_K^f(\kappa+2nK).
\end{align}
\noindent  Note that $v_0(\kappa)=R(\kappa)w_K^f(\kappa)=0$, since  $w_K^f(\kappa)=\opmatrix{\kappa}{Q_K}{f}=0$ for $\kappa \in \left[-K,K\right]$.

For $n \ne 0$
\begin{align}
||v_n||^2 &= \int_{-K}^{K} |w_K^f(\kappa+2nK)|^2 d \kappa=\int_{2nK-K}^{2nK+K} |w_K^f(\kappa)|^2 d \kappa.
\end{align}

Now consider the function $\kappa\hf(\kappa)$.
Since $\hf(\kappa) \in S(\mathbb{R}) \Rightarrow  \kappa\hf(\kappa)\in S(\mathbb{R})$.
The  tail weight of $\kappa\hf(\kappa)$ outside the interval $\left[-K,K\right]$, denoted  by $r_K^f$ is
\begin{align}
\label{eq:rfk}
(r_K^f)^2
&=\int_{-\infty}^{\infty} \kappa^2 |w_K^f(\kappa)|^2 d \kappa= \sum_{n=-\infty}^{\infty}
\int_{2nK-K}^{2nK+K} \kappa^2 |w_K^f(\kappa)|^2 d \kappa.
\end{align}
Since $k^2 \ge K^2 c^2(n)$
for  $k \in \left[2nK-K, 2nK+K\right]$ and
\begin{align}
\label{eq:cn}
 c(n)=\left\{
  \begin{array}{ll}
  2n-1 &~~\text{for}~~~~ n>0 \\
  2n+1 &~~\text{for}~~~~ n<0
  \end{array}
 \right. ,
 \end{align}
the following inequality is true
\begin{align}
\label{eq:rfineq}
(r_K^f)^2& \ge K^2 \sum_{n=-\infty}^{\infty}
 c(n)^2\int_{2nK-K}^{2nK+K} |w_K^f(\kappa)|^2
= K^2 \sum_{n=-\infty}^{\infty}
  c(n)^2 ||v_n||^2.
\end{align}

Employing  \cref{eq:vsum}, the Cauchy-Schwartz inequality and \cref{eq:rfineq},  one gets
\begin{align}
\label{eq:vbound}
 \left|\left|v\right|\right| \le
\sum_{n\ne 0}\left|\left|v_n\right|\right| \left|c(n)\right|\frac{1}{\left|c(n)\right|}
&\le  \sqrt{ \sum_{n \ne 0} c(n)^2 \left|\left|v_n\right|\right|^2} \sqrt{ \sum_{n \ne 0} \frac{1}{c(n)^2}}=\frac{\pi}{2}\sqrt{ \sum_{n \ne 0} c(n)^2  \left|\left|v_n\right|\right|^2 }
\le \frac{\pi r_K^f}{2K}.
\end{align}
\noindent In  \cref{eq:vbound} we
used
\begin{align}
    \sum_{n \ne 0} \frac{1}{c(n)^2}=2\sum_{n >1 }\frac{1}{\left(2n-1\right)^2}=\frac{\pi^2}{4}.
\end{align}

 The square norm $||t||^2$ is given by
 \begin{align}
 \label{eq:t2norm}
 ||t||^2&=\int_{-\infty}^{\infty} d \vphi \sum_{i,j=-\infty}^{\infty} w_F^f(\vphi_i)^* w_F^f(\vphi_j)\int_{-\infty}^{\infty}  u_K(\vphi-\vphi_i)u_K(\vphi-\vphi_j)d \vphi\\ \nonumber
 &=\Delta_\vphi \sum_{i=-\infty}^{\infty} |w_F^f(\vphi_i)|^2.
 \end{align}
 \noindent Note that the sum over $i$ in  \cref{eq:t2norm} is just the Riemann approximation
 of the integral $\int |w_F^f(\vphi)|^2 d \vphi$. Using the Euler-Maclaurin integration
 rule~\cite{devries2011first}, one gets the following approximation
  \begin{align}
 \label{eq:t2approx}
 ||t||^2\approx ||w_F^f||^2 +\frac{\Delta_\vphi}{2} \left(|f(-F)|^2+|f(F)|^2\right)+\O(\Delta^2_\vphi).
 %+ \frac{\Delta^2_\vphi}{24}\left(f^\prime(-F)-f^\prime(F)\right)+\O(\Delta^4_\vphi)
 \end{align}

 Equations ~\eqref{eq:diff_f}, ~\eqref{eq:vbound} and~\eqref{eq:t2approx} imply
 \begin{align}
 \label{eq:diff_ftf}
 ||f-\tilde{f_\vphi}|| \lesssim  ||w_K^f||+||w_F^f|| + \frac{\pi r_K^f }{2K} +
 \sqrt{\frac{\pi}{2K}\left(|f(-F)|^2+|f(F)|^2\right)}.
 \end{align}

  \subsection{Lattice wavefunctions}
  \label{app:tailslattice}

   For a wavefunction $f(\bm{\vphi})\equiv f(\vphi_1,\vphi_2,...,\vphi_N) \in S(\mathbb{R}^N)$,
let's consider  a sampling interval $
\left[-\bm{F},\bm{F} \right] \equiv \left[-F,F \right]^N \subset  \mathbb{R}^N
$, the projector
\begin{align}
\label{eq:lattice_Pphi}
P_{\bF}=\int_{-F}^{F}...\int_{-F}^{F}  \ket{\bm{\vphi}}\bra{\bm{\vphi}} d\bm{\vphi},
\end{align}
\noindent and the tail vector
\begin{align}
\label{eq:lattice cutphi_function}
\ket{w_{\bF}^f}=\left(1-P_{\bF} \right) \ket{f} \equiv Q_{\bF}\ket{f}.
\end{align}
Analogously, for the conjugate-field function
$\hf(\bm{\kappa}) \equiv \hf(\kappa_1,\kappa_2,...,\kappa_N) \in S(\mathbb{R}^N)$
let's consider the sampling interval
$
\left[-\bm{K},\bm{K} \right] \equiv  \left[-K, K\right]^N \subset  \mathbb{R}^N
$, the projector
\begin{align}
\label{eq:lattice_Pkappa}
P_{\bK}=\int_{-K}^{K}...\int_{-K}^{K}  \ket{\bm{\kappa}}\bra{\bm{\kappa}} d\bm{\kappa}
\end{align}
\noindent and the tail vector
\begin{align}
\label{eq:lcutk_function}
\ket{w_{\bK}^f}=\left(1-P_{\bK} \right) \ket{f} \equiv Q_{\bK}\ket{f}.
\end{align}

As for the $1$-dimensional functions, when $K$ is large
$\ket{f}\approx P_{\bK} \ket{f}$, the Nyquist-Shannon theorem can be employed
and $f(\bphi)$ can be approximated by a infinite series expansion of {\em sinc}
functions products. When $F$ is large, the series can be truncated
to $\left(N_\vphi\right)^N$ terms,
\begin{align}
\label{eq:NS_nf}
f(\bphi) \approx \tilde{f}_\vphi(\bm{\vphi}) = \sum_{i_1=-\frac{N_{\vphi}-1}{2}}^{\frac{N_{\vphi}-1}{2}}
...\sum_{i_N=-\frac{N_{\vphi}-1}{2}}^{\frac{N_{\vphi}-1}{2}} f(\vphi_{i_1},...,
\vphi_{i_N})u_K(\vphi_{1}-\vphi_{i_1})...u_K(\vphi_{N}-\vphi_{i_N})
\end{align}
\noindent where
  $\vphi_{i_j}=i_j \Delta_{\vphi}$, $\Delta_{\vphi}=\frac{\pi}{K}$  and $N_{\vphi}=\lceil \frac{2}{\pi}KF \rceil$.

  Analogously to  \cref{eq:diff_fh}, the difference between a
  $N$-dimensional function $f(\bphi) \in S(\mathbb{R}^N)$
and its truncated Nyquist-Shannon sampled approximation $\tilde{f}_\vphi(\bphi)$ defined by  \cref{eq:NS_nf} is given by

 \begin{align}
 \label{eq:diff_fhlattice}
  f(\bphi)-\tilde{f}_\vphi(\bphi)
%   f(\bphi)-\opmatrix{\bphi}{P_K}{f}+
%   \sum_{\bi=-\infty}^{\infty} \left(\opmatrix{\bphi_{\bi}}{P_K}{f}- f(\bphi_{\bi})\right) u_{\bK}(\bphi-\bphi_{\bi})
%   +\sum_{\bi=-\infty}^{\infty} w_{\bF}^f(\bphi_{\bi})u_{\bK}(\bphi-\bphi_{\bi})\\ \nonumber
%   &=w_{\bK}^f(\bphi)-\sum_{\bi=-\infty}^{\infty}w_{\bK}^f(\bphi_{\bi})u_{\bK}(\bphi-\bphi_{\bi})+\sum_{\bi=-\infty}^{\infty} w_{\bF}^f(\bphi_{\bi})u_{\bK}(\bphi-\bphi_{\bi})\\ \nonumber
  =w_{\bK}^f(\bphi)-v(\bphi)+t(\bphi)
  \end{align}
  \noindent where
  \begin{align}
  \label{eq:vdeflattice}
  v(\bphi) &= \sum_{\bi=-\infty}^{\infty}w_{\bK}(\bphi_{\bi})u_{\bK}(\bphi-\bphi_{\bi}),\\
  \label{eq:tdeflattice}
  t(\bphi) &= \sum_{\bi=-\infty}^{\infty} w_{\bF}(\bphi_{\bi})u_{\bK}(\bphi-\bphi_{\bi}).
  \end{align}
\noindent  The following notation was used in \cref{eq:vdeflattice,eq:tdeflattice}
  \begin{align}
  \nonumber
  u_{\bK}(\bphi)&=u_{K}(\vphi_1)u_{K}(\vphi_2)...u_{K}(\vphi_N)\\ \nonumber
  \bi&=\{i_1, i_2,...,i_N\}.
  \end{align}

The norm of the tail vector $\ket{w_{\bK}^f}$ is bounded as
\begin{align}
  \label{eq:wknormlattice}
  ||w_{\bK}^f||^2 &\le \sum_{j=1}^N \int d \kappa_1 \left(\int_{-\infty}^{-K}d \kappa_j +\int_{K}^{\infty} d \kappa_j \right)
  ... \int d \kappa_N |\hf(\kappa_1,...,\kappa_j,...,\kappa_N)|^2\\ \nonumber
  &= \sum_{j=1}^N\left(\int_{-\infty}^{-K}  \opmatrix{\kappa_j}{\rho_j^f}{\kappa_j}d \kappa_j +\int_{K}^{\infty}   \opmatrix{\kappa_j}{\rho_j^f}{\kappa_j} d \kappa_j\right) = \sum_{j=1}^N {w_{jK}^f}^2
  \end{align}
  \noindent where
  \begin{align}
\label{eq:rhoj}
\rho_j^f=\text{Tr}_{1,2,...j-1,j+1...,N} \left(\ket{f}\bra{f}\right),
\end{align}
\noindent is the local density operator at site $j$ obtained by
tracing over all other sites and
\begin{align}
\label{eq:ntail_kappa}
 {w^f_{jK}}^2&=\int_{-\infty}^{-K}  \opmatrix{\kappa}{\rho^f_j}{\kappa} d \kappa +
    \int_{K}^{\infty}  \opmatrix{\kappa}{\rho^f_j}{\kappa}d \kappa.
\end{align}
 \noindent is the tail weight of $\hf(\bk)$
 at site $j$.

To estimate $||v||$, we
write $\ket{v}$ in the conjugate-field basis.
The Fourier transform of  \cref{eq:vdeflattice} yields
  \begin{align}
  \label{eq:vbk}
  v(\bk)
=\sum_{\bm{n}} v_{\bm{n}}(\bk)
  \end{align}
  \noindent where
  \begin{align}
  v_{\bm{n}}(\bk) =R_{\bK}(\bk)\left(-1\right)^{n_1+n_2+...n_N} w_{\bK}(\bk+2\bm{n}\bK),
  \end{align}
  \noindent with $R_{\bK}(\bk)=1$ for $\bk \in \left[-\bK,\bK\right]$ and zero otherwise.
The norm of $v_{ \bm{n} }$ is
\begin{align}
    \norm{ v_{\bm{n} } }^2=\int_{-K}^{K} d\kappa_1...\int_{-K}^{K} d\kappa_N \absv{w_{\bK}(\bk+2\bm{n}\bK)}^2
    =\int_{2\bm{n} \bK-\bK}^{2 \bm{n} \bK+\bK}  \absv{w_{\bK}(\bk)}^2 d \bk
\end{align}

Now consider the function $\bk\hf(\bk)\equiv \kappa_1...\kappa_n\hf(\kappa_1,...,\kappa_n) \in S(\mathbb{R}^N) $.

The  tail weight of $\bk\hf(\bk)$ outside the interval $\left[-\bK,\bK\right]$, denoted  by $r_{\bK}^f$ is
\begin{align}
   {r_{\bK}^f}^2
   =\sum_{\bm{n}} \int_{2n_1 K-K}^{2 n_1 K+K} d \kappa_1 \kappa_1^2...\int_{2 n_N K-K}^{2 n_N K+K} d \kappa_N \kappa_N^2 \absv{w_{\bK}(\bk)}^2.
\end{align}
\noindent The following inequality is true
\begin{align}
    {r_{\bK}^f}^2 \ge K^{2N}\sum_{\bm{n}} c(\bm{n})^2 \int_{2\bm{n} \bK-\bK}^{2 \bm{n} \bK+\bK}  \absv{w_{\bK}(\bk)}^2 d \bk= K^{2N}\sum_{\bm{n}} c(\bm{n})^2 \norm{v_{\bm{n}}}^2
\end{align}
\noindent where
\begin{align}
 \label{eq:cnmulti}
    c(\bm{n})=c(n_1) c(n_2)...c(n_N),
\end{align}
\noindent with $c(n)$  defined by  \cref{eq:cn}.
Employing the Cauchy-Schwartz inequality
\begin{align}
    \sum_{\bm{n}}  \left|v_{\bm{n}}(\bk)\right|=\sum_{\bm{n}}  \left|v_{\bm{n}}(\bk)\right| \left|c(\bm{n})\right| \frac{1}{\left|c(\bm{n})\right|}
    \le \sqrt{\sum_{\bm{n}} \left|v_{\bm{n}}(\bk)\right|^2 \left|c(\bm{n})\right|^2}\sqrt{\sum_{\bm{n}}\frac{1}{\left|c(\bm{n})\right|^2}}
\end{align}
\noindent and
\begin{align}
    \sum_{\bm{n} \ne \bm{0}}\frac{1}{\absv{c(\bm{n})}^2}=
    \left(2\sum_{n>0}\frac{1}{\left(2n-1\right)^2}+1\right)^N-1=\left(\frac{\pi^2}{4}+1\right)^N-1,
\end{align}
\noindent one gets
\begin{align}
\label{eq:vnormlattice}
  \norm{v} \le  \left(\frac{\pi^2}{4}+1\right)^{N/2} \frac{r_{\bK}^f}{K^{N}}.
\end{align}

Using  \cref{eq:tdeflattice} and the orthogonality properties of {\em sinc} functions,
one gets
  \begin{align}
  \label{eq:tnormlattice}
  ||t||^2= \int  |t(\bphi)|^2 d \bphi= \Delta_{\vphi}^N\sum_{\bi=-\infty}^{\infty}|w_{\bF}(\bphi_{\bi})|^2
  \end{align}

  \begin{align}
  \label{eq:tnormlattice2}
  ||t||^2 &\le \sum_{j=1}^{N}\Delta_{\vphi}^N\sum_{i_1=\-\infty}^{\infty}...\sum_{|i_j|>\frac{N_\vphi-1}{2}}
  ...\sum_{i_N=\-\infty}^{\infty}
  |f(\vphi_{i_1},...,\vphi_{i_j},...,\vphi_{i_N}) |^2\\ \nonumber
  &\approx \sum_{j=1}^{N} \int_{-\infty}^{\infty} d \vphi_1 ... \sum_{|i_j|>\frac{N_\vphi-1}{2}}\Delta_{\vphi}
  ...\int_{-\infty}^{\infty} d \vphi_N f(\vphi_{1},...,\vphi_{i_j},...,\vphi_{N}) |^2\\ \nonumber
  &\approx \sum_{j=1}^{N} \sum_{|i_j|>\frac{N_\vphi-1}{2}}\Delta_{\vphi}
  \opmatrix{\vphi_{i_j}}{\rho^f_j}{\vphi_{i_j}}\\ \nonumber
  &\approx \sum_{j=1}^{N} \left[{w_{jF}^f}^2+\frac{\Delta_{\vphi}}{2} \left(\opmatrix{-F}{\rho^f_j}{-F}+\opmatrix{F}{\rho^f_j}{F}\right)\right].
  \end{align}
  \noindent where
  \begin{align}
  \label{eq:ntail_vphi}
  {w_{jF}^f}^2=\int_{-\infty}^{-F} d \vphi \opmatrix{\vphi}{\rho^f_j}{\vphi}+\int_{F}^{\infty} d \vphi \opmatrix{\vphi}{\rho^f_j}{\vphi},
  \end{align}
  \noindent is the tail weight of $f(\bphi)$ at site $j$.
  Analogously to  \cref{eq:t2approx}, in  \cref{eq:tnormlattice2} we used the Euler-Maclaurin integration rule to approximate the Riemann sum with the integral.

  Employing \cref{eq:diff_fhlattice,eq:wknormlattice,eq:vnormlattice,eq:tnormlattice2},
  one has
  \begin{align}
  ||f-\tilde{f}_\vphi|| \le \sum_{j=1}^N  \left[w_{jK}^f + w_{jF}^f + \sqrt{\frac{\pi}{2K}\left(\opmatrix{-F}{\rho^f_j}{-F}+\opmatrix{F}{\rho^f_j}{F}\right)}\right]+\left(\frac{\pi^2}{4}+1\right)^{N/2} \frac{r_{\bK}^f}{K^{N}}.
  \end{align}

   Similarly, $\ket{f}$  can be approximated by the field-limited function $\ket{\tilde{f}_\kappa}$ defined as
\begin{align}
\label{eq:NS_fkappa}
\hf(\bk) \approx \tilde{f}_\kappa(\bm{\kappa}) =
\sum_{p_1=-\frac{N_{\vphi}-1}{2}}^{\frac{N_{\vphi}-1}{2}}
...\sum_{p_N=-\frac{N_{\vphi}-1}{2}}^{\frac{N_{\vphi}-1}{2}} \hf(\kappa_{p_1},...,
\kappa_{p_N})u_F(\kappa_{1}-\kappa_{p_1})...u_F(\kappa_{N}-\kappa_{p_N})
\end{align}
\noindent where
  $\kappa_{p_j}=p_j \Delta_{\kappa}$, $\Delta_{\kappa}=\frac{\pi}{F}$.
  The error of the approximation is bounded as
  \begin{align}
  \label{eq:ffkappadifflat}
  ||f-\tilde{f}_\kappa|| \le \sum_{j=1}^N  \left[w_{jF}^f + w_{jK}^f + \sqrt{\frac{\pi}{2F}\left(\opmatrix{-K}{\rho^f_j}{-K}+\opmatrix{K}{\rho^f_j}{K}\right)}\right]+\left(\frac{\pi^2}{4}+1\right)^{N/2} \frac{r_{\bF}^f}{F^{N}}.
  \end{align}
  \noindent where $r_{\bF}^f$ is the  weight of $\vphi_1\vphi_2...\vphi_Nf(\vphi_1,...,\vphi_N)$ outside the $N$-dimensional sampling interval $\left[-\bF, \bF\right]$,
  \begin{align}
  {r_{\bF}^f}^2=\int d \vphi_1... \int d \vphi_N \vphi_1^2 ... \vphi_N^2 \absv{w_{\bF}(\vphi_1, ..., \vphi_N)}^2.
  \end{align}

 \section{Aliasing and Finite Fourier transform}
 \subsection{Aliased functions}
 \label{app:FFTaliasing}

Consider a function $f(\vphi) \in L^2(\mathbb{R})$ and its Fourier transform
$\hf(\kappa)$ given by  \cref{eq:ftdef}.
Consider also a set of $N_\vphi$ field  sampling points $\{\vphi_i=i \Delta_\vphi\}_i$
 with $i=\overline{-\frac{N_\vphi-1}{2}, \frac{N_\vphi-1}{2} }$ and a set of $N_\vphi$ conjugate-field sampling points $\{\kappa_p=p \Delta_\kappa\}_p$ with $p=\overline{-\frac{N_\vphi-1}{2}, \frac{N_\vphi-1}{2} }$, where the discretization intervals are chosen such that $\Delta_\vphi \Delta_\kappa=\frac{2 \pi}{N_\vphi}$.

 Here we will show that  the aliased functions
 at the sampling points,
 \begin{align}
 \label{eq:faliased}
 f_a(\vphi_i)&= \sqrt{\Delta_\vphi} \sum_{n=-\infty}^{\infty} \left(-1\right)^n f(\vphi_i+nN_\vphi\Delta_\vphi)\\
 \label{eq:hfaliased}
 \hf_a(\kappa_p)&=\sqrt{\Delta_\kappa}\sum_{n=-\infty}^{\infty}  \left(-1\right)^n \hf(\kappa_p+nN_\vphi\Delta_\kappa)
 \end{align}
 \noindent are related via a Finite Fourier transform, {\em i.e.}
 \begin{align}
 \label{eq:fft1}
 \hf_a(\kappa_p)= (\tF f_a)(\kappa_p) \equiv \frac{1}{\sqrt{N_\vphi}}\sum_{j=-\frac{N_\vphi-1}{2}}^{\frac{N_\vphi-1}{2}} f_a(\vphi_j)  e^{-i \kappa_p \vphi_j},
 \end{align}
 \noindent and
 \begin{align}
 \label{eq:fft2}
 f_a(\vphi_j)= (\tF^{-1} \hf_a)(\vphi_j) \equiv \frac{1}{\sqrt{N_\vphi}}\sum_{p=-\frac{N_\vphi-1}{2}}^{\frac{N_\vphi-1}{2}} \hf_a(\kappa_p)  e^{i \kappa_p \vphi_j}.
 \end{align}
\noindent The proof of  \cref{eq:fft1} and  \cref{eq:fft2} is similar to the one presented in Ref.~\cite{cooley_ieee_1967} and is sketched below.

 The value of the function $f(\vphi)$ at   $\{\vphi_i=\left(i+1/2 \right)\Delta_\vphi\}_{i=\overline{-\infty,\infty}}$  is given by
 \begin{align}
 \label{eq:fsampling1}
 f(\vphi_i)=\frac{1}{\sqrt{2 \pi}} \sum_{n=-\infty}^{\infty} \int_{-K+2nK}^{K+2nK} \hf(\kappa)e^{i \kappa \vphi_i} d\kappa
 =\frac{1}{\sqrt{2 \pi}} \sum_{n=-\infty}^{\infty}  \int_{-K}^{K} \left(-1\right)^n \hf(\kappa+nN_{\vphi} \Delta_{\kappa})e^{i \kappa \vphi_i} d \kappa.
 \end{align}
 \noindent where $K=N_{\vphi} \Delta_{\kappa}/2$.  \cref{eq:fsampling1} reads
 \begin{align}
 f(\vphi_i)&=\frac{1}{\sqrt{2 \pi \Delta_\kappa}} \int_{-K}^{K} \hf_a(\kappa) e^{i \kappa \vphi_i},
 \end{align}
 \noindent where
 \begin{align}
 \label{eq:fa}
 \hf_a(\kappa)&\equiv\sqrt{\Delta_\kappa}\sum_{n=-\infty}^{\infty}  \left(-1\right)^n \hf(\kappa+nN_{\vphi} \Delta_{\kappa}).
 \end{align}

 Since $\hf_a(\kappa)$ defined by  \cref{eq:fa} is anti-periodic, {\em i.e.} $\hf_a(\kappa)=-\hf_a(\kappa+N_{\vphi} \Delta_{\kappa})$, it can be written as
 \begin{align}
 \hf_a(\kappa)= \sqrt{\Delta_\kappa}\frac{\Delta_\vphi}{\sqrt{2 \pi }}\sum_{i=-\infty}^{\infty} f(\vphi_i)  e^{-i \kappa \vphi_i}.
 \end{align}

 \noindent The value of $\hf_a(\kappa)$ at the sampling points
 $\{\kappa_p\}_{p=\overline{-\frac{N_\vphi-1}{2}, \frac{N_\vphi-1}{2} }}$ is given by
 \begin{align}
 \hf_a(\kappa_p)= \sqrt{\Delta_\vphi} \sqrt{\frac{\Delta_\vphi \Delta_\kappa}{2 \pi}} \sum_{n=-\infty}^{\infty} \sum_{i=-\frac{N_\vphi-1}{2}}^{\frac{N_\vphi-1}{2}} \left(-1\right)^n f(\vphi_i+nN_\vphi\Delta_\vphi)  e^{-i \kappa_p \vphi_i}
 \end{align}
 \noindent which implies  \cref{eq:fft1}.
Analogously,  \cref{eq:fft2} can be derived.

 \subsection{Finite Fourier transform approximation for the continuous Fourier transform}
 \label{app:FFTerror}

The difference between the vector
defined by the sampling points of
a function and the vector defined by the
aliased function is given by the function's values outside the sampling interval. For example,
  \begin{align}
 \label{eq:ffadif1}
 \Delta_\vphi \sum_{i=-\frac{N_\vphi-1}{2}}^{\frac{N_\vphi-1}{2}} |f_a(\vphi_i)-f(\vphi_i)|^2 &=\Delta_\vphi \sum_{n=-\infty}^{\infty} \sum_{i=-\frac{N_\vphi-1}{2}}^{\frac{N_\vphi-1}{2}}
  |w_F^f(\vphi_i+n N_\vphi \Delta_\vphi)|^2
 = \Delta_\vphi \sum_{i=-\infty}^{\infty} |w_F^f(\vphi_i)|^2  \\ \nonumber
 &\approx ||w_F^f||^2+\frac{\Delta_\vphi}{2} \left(|f(-F)|^2+|f(F)|^2\right),
 \end{align}
 \noindent where the same approximation as in  \cref{eq:t2approx} was made. Similarly,
 \begin{align}
 \label{eq:ffadif2}
 \Delta_\kappa \sum_{p=-\frac{N_\vphi-1}{2}}^{\frac{N_\vphi-1}{2}} |\hf_a(\kappa_p)-\hf(\kappa_p)|^2 &=\Delta_\kappa \sum_{n=-\infty}^{\infty} \sum_{p=-\frac{N_\vphi-1}{2}}^{\frac{N_\vphi-1}{2}}
  |w_K^f(\kappa_p+n N_\vphi \Delta_\kappa)|^2
 = \Delta_\kappa \sum_{p=-\infty}^{\infty} |w_K^f(\kappa_p)|^2 \\ \nonumber
 &\approx ||w_K^f||^2+\frac{\Delta_\kappa}{2} \left(|\hf(-K)|^2+|\hf(K)|^2\right).
 \end{align}

  The difference between the Finite Fourier transform of the set $\{f(\vphi_i)\}_i$ and the vector defined by the function $\hf(\kappa)$  at the conjugate-field sampling points is given by
 \begin{align}
 \label{eq:FFTdif}
 \Delta_\kappa  \sum_{p=-\frac{N_\vphi-1}{2}}^{\frac{N_\vphi-1}{2}} |\tF f(\kappa_p)-\hf(\kappa_p) |^2
 &= \Delta_\kappa  \sum_{p=-\frac{N_\vphi-1}{2}}^{\frac{N_\vphi-1}{2}} |(\tF f)(\kappa_p)- (\tF f_a)(\kappa_p)
 +\hf_a(\kappa_p) - \hf(\kappa_p) |^2 \\ \nonumber
 &\le 2 \Delta_\kappa  \sum_{p=-\frac{N_\vphi-1}{2}}^{\frac{N_\vphi-1}{2}} |(\tF f)(\kappa_p)- (\tF f_a)(\kappa_p)|^2
 + 2 \Delta_\kappa  \sum_{p=-\frac{N_\vphi-1}{2}}^{\frac{N_\vphi-1}{2}} |\hf_a(\kappa_p) -\hf(\kappa_p) |^2\\ \nonumber
 &= 2 \Delta_\vphi \sum_{i=-\frac{N_\vphi-1}{2}}^{\frac{N_\vphi-1}{2}} |f(\vphi_i)-f_a(\vphi_i)|^2
 + 2 \Delta_\kappa  \sum_{p=-\frac{N_\vphi-1}{2}}^{\frac{N_\vphi-1}{2}} |\hf_a(\kappa_p) -\hf(\kappa_p) |^2\\ \nonumber
 &\approx 2 \left(||w_F^f||^2+ ||w_K^f||^2\right) \\ \nonumber
 &+\frac{\pi}{K} \left(|f(-F)|^2+|f(F)|^2\right)+ \frac{\pi}{F} \left(|\hf(-K)|^2+|\hf(K)|^2\right)
 \end{align}
 \noindent  In the first line of  \cref{eq:FFTdif},  we added and subtracted the aliased function $\hf_a(\kappa_p)=(\tF f_a)(\kappa_p)$. In the last line of
  \cref{eq:FFTdif}, we used
  \cref{eq:ffadif1} and  \cref{eq:ffadif2}.

\section{Band-limited wavefunction with large number of bosons}
\label{ap:fbad}

For our example in Section~\ref{ssec:leval}, we construct a band-limited function
\begin{align}
\label{eq:f_app}
f(\vphi)=\sum_i a_i u_K(\vphi-\vphi_i)\,
\end{align}
\noindent where we take $F=K=\sqrt{\pi N_\vphi/2}$, (see  \cref{eq:Ncphi}),
with $N_\vphi=64$.

When the summation over $i$ is restricted to a finite set, $|f(\vphi)|$  decays as least as $|\vphi|^{-1}$ with increasing $|\vphi|$ (since $u_K(\vphi) \propto \vphi^{-1}$, see  \cref{eq:ftuk}).  As described below, we choose the coefficients $a_i$ such that $|f(\vphi)|$ decays
 as $|\vphi|^{-8}$ at large $|\vphi|$. Let's first take all the coefficients $a_i=0$ except for the one corresponding to
the indices $i=\pm q_1$, (where $q_1$ is an arbitrary half-integer). If $a_{\pm q_1}=1$,
one gets
\begin{align}
\label{eq:fq1}
f_{q_1}(\vphi)&= \frac{1}{\sqrt{2\Delta_\vphi}}\left( \frac{\sin \left(\frac{\pi \vphi} {\Delta_\vphi} -q_1 \pi \right)}{\frac{\pi \vphi}
 {\Delta_\vphi} -q_1 \pi} + \frac{\sin \left(\frac{\pi \vphi} {\Delta_\vphi} +q_1 \pi \right)}
 {\frac{\pi \vphi} {\Delta_\vphi} +q_1 \pi} \right) \\ \nonumber
 &=- 2 \frac{1}{\sqrt{2\Delta_\vphi}}  \sin \left(\frac{\pi \vphi} {\Delta_\vphi} - \frac{\pi}{2} \right)
 \left(  q_1 \frac{\Delta_\vphi^2} {\pi \vphi^2}
  + 6 q_1^3 \frac{\Delta_\vphi^4} {\pi \vphi^4 }  +120 q_1^5 \frac{\Delta_\vphi^6} {\pi \vphi^6 }
 +5040 q_1^7 \frac{\Delta_\vphi^8} {\pi  \vphi^8 } + \O\left(\frac{\Delta_\vphi^{10}} { \vphi^{10} }\right)\right).
 \end{align}
 \noindent The function $f_{q_1}(\vphi)$ decays as $|\vphi|^{-2}$ with increasing $|\vphi|$.
 We define our function as
 \begin{align}
 \label{eq:fbad}
 f(\vphi)=c_1f_{q_1}(\vphi)+c_2f_{q_2}(\vphi)+c_3f_{q_3}(\vphi)+c_4f_{q_4}(\vphi)
 +c_5f_{q_5}(\vphi)+c_6f_{q_6}(\vphi)+c_7f_{q_7}(\vphi)+c_8f_{q_8}(\vphi),
 \end{align}
 \noindent where $q_1,...,q_8$ are half-integer smaller than $N_\vphi/3$ and $c_1,...,c_8$
 are chosen such that the terms proportional to $|\vphi|^{-2}$, $|\vphi|^{-4}$
 and $|\vphi|^{-6}$ cancel out. The function can be written as
 \begin{align}
 \label{eq:fbad_asym_app}
 f(\vphi) = c_f \sin \left(\frac{\pi \vphi} {\Delta_\vphi} - \frac{\pi}{2} \right)\frac{\Delta_\vphi^8} {\pi  \vphi^8 }+\O \left( \frac{\Delta_\vphi^{10}} {\pi  \vphi^{10} } \right).
 \end{align}
\noindent where $c_f$ is a normalization constant term depending on $q_1,...,q_8$ and $\Delta_\vphi$.

\section{Inequalities for local boson distribution measurement}
\label{ap:bdm}

The probability to measure a certain integer $k$ on the ancillary register depends on the
quantity $a_{nk}$ defined by  \cref{eq:ank1}.

We have
\begin{align}
\label{eq:ankless1}
|a_{nk}| =\left|\frac{1}{2^{n_r}}\sum_{x=0}^{2^{n_r}-1}
e^{-i \frac{2 \pi}{2^{n_r}} \mu_{nk} x}\right|
=\frac{1}{2^{n_r}} \frac{|\sin\left(\pi \mu_{nk}\right)|}{|\sin\left(\frac{\pi \mu_{nk}}{2^{n_r}}\right)|}
\end{align}
\noindent with
\begin{align}
\mu_{nk}=\frac{\tilde{E}_n}{m_0}-k.
\end{align}

The following properties of $a_{nk}$ are true:
\begin{itemize}
\item $|a_{nk}|=1$ when $\mu_{nk}=0$. It can be checked  by direct substitution in the first part of  \cref{eq:ankless1}.

\item $|a_{nk}|\le 1$.
It follows from the inequality $|\sin(Mx)| \le M|\sin(x)|$, which holds for
any integer $M>1$ and any $x$ (it can be easily proven by induction).  In  \cref{eq:ankless1} one needs to take $M=2^{n_r}$ and $x=\pi \mu_{nk}/2^{n_r}$.

\item $|a_{nk}|\ge 2/\pi$ when $|\mu_{nk}| \le 1/2$. The proof is similar to the one in Refs~\cite{cleve_1998_procRsocA,nielsen_chuang_2010} for estimating the probability to measure the nearest integer
to the phase factor in  a QPE algorithm.
The inequality $|x| \ge |\sin(x)|$  implies
\begin{align}
|a_{nk}|
\ge \frac{1}{2^{n_r}} \frac{|\sin\left(\pi \mu_{nk}\right)|}{|\frac{\pi \mu_{nk}}{2^{n_r}}|}
=\frac{|\sin\left(\pi \mu_{nk}\right)|}{|\pi \mu_{nk}|} .
\end{align}
Furthermore, the inequality $|\sin(x)| \ge | 2x/\pi|$,  which holds for $|x| \le \pi/2$
(on the interval $\left[0,\pi/2\right]$ $\sin(x)$, is above the line connecting $\left(0,0\right)$
and $\left(\pi/2,1\right)$) implies
\begin{align}
\label{eq:akge2pi}
|a_{nk}| \ge \frac{2}{\pi}~~~\text{for}~~~|\mu_{nk}| \le \frac{1}{2}.
\end{align}

\end{itemize}

For any $n \ge N_b$ we have $\mu_{nk} \le 1/2$ when $k$ is the nearest integer to $\tilde{E}_n/m_0$.
Thus for any $n \ge N_b$ there is always a $k$ such that  \cref{eq:akge2pi} is true. That implies
\begin{align}
\label{eq:ankmaxapp}
\max_{k \ge N_b} |a_{nk}|^2 \ge \frac{4}{\pi^2}.
\end{align}

%\bibliography{bibmacridin}

\begin{thebibliography}{44}%
\makeatletter
\providecommand \@ifxundefined [1]{%
 \@ifx{#1\undefined}
}%
\providecommand \@ifnum [1]{%
 \ifnum #1\expandafter \@firstoftwo
 \else \expandafter \@secondoftwo
 \fi
}%
\providecommand \@ifx [1]{%
 \ifx #1\expandafter \@firstoftwo
 \else \expandafter \@secondoftwo
 \fi
}%
\providecommand \natexlab [1]{#1}%
\providecommand \enquote  [1]{``#1''}%
\providecommand \bibnamefont  [1]{#1}%
\providecommand \bibfnamefont [1]{#1}%
\providecommand \citenamefont [1]{#1}%
\providecommand \href@noop [0]{\@secondoftwo}%
\providecommand \href [0]{\begingroup \@sanitize@url \@href}%
\providecommand \@href[1]{\@@startlink{#1}\@@href}%
\providecommand \@@href[1]{\endgroup#1\@@endlink}%
\providecommand \@sanitize@url [0]{\catcode `\\12\catcode `\$12\catcode
  `\&12\catcode `\#12\catcode `\^12\catcode `\_12\catcode `\%12\relax}%
\providecommand \@@startlink[1]{}%
\providecommand \@@endlink[0]{}%
\providecommand \url  [0]{\begingroup\@sanitize@url \@url }%
\providecommand \@url [1]{\endgroup\@href {#1}{\urlprefix }}%
\providecommand \urlprefix  [0]{URL }%
\providecommand \Eprint [0]{\href }%
\providecommand \doibase [0]{http://dx.doi.org/}%
\providecommand \selectlanguage [0]{\@gobble}%
\providecommand \bibinfo  [0]{\@secondoftwo}%
\providecommand \bibfield  [0]{\@secondoftwo}%
\providecommand \translation [1]{[#1]}%
\providecommand \BibitemOpen [0]{}%
\providecommand \bibitemStop [0]{}%
\providecommand \bibitemNoStop [0]{.\EOS\space}%
\providecommand \EOS [0]{\spacefactor3000\relax}%
\providecommand \BibitemShut  [1]{\csname bibitem#1\endcsname}%
\let\auto@bib@innerbib\@empty
%</preamble>
\bibitem [{\citenamefont {Privman}(1990)}]{privman_finite_size}%
  \BibitemOpen
  \bibfield  {author} {\bibinfo {author} {\bibfnamefont {V.}~\bibnamefont
  {Privman}},\ }\href {\doibase 10.1142/1011} {\emph {\bibinfo {title} {Finite
  Size Scaling and Numerical Simulation of Statistical Systems}}}\ (\bibinfo
  {publisher} {WORLD SCIENTIFIC},\ \bibinfo {year} {1990})\ \Eprint
  {http://arxiv.org/abs/https://www.worldscientific.com/doi/pdf/10.1142/1011}
  {https://www.worldscientific.com/doi/pdf/10.1142/1011} \BibitemShut {NoStop}%
\bibitem [{\citenamefont {Cardy}(2012)}]{cardy_finite_size}%
  \BibitemOpen
  \bibfield  {author} {\bibinfo {author} {\bibfnamefont {J.~L.}\ \bibnamefont
  {Cardy}},\ }\href@noop {} {\emph {\bibinfo {title} {Finite Size Scaling}}},\
  Vol.~\bibinfo {volume} {2}\ (\bibinfo  {publisher} {Elsevier, Amsterdam},\
  \bibinfo {year} {2012})\BibitemShut {NoStop}%
\bibitem [{\citenamefont {Jordan}\ \emph {et~al.}(2012)\citenamefont {Jordan},
  \citenamefont {Lee},\ and\ \citenamefont {Preskill}}]{jordan_science_2012}%
  \BibitemOpen
  \bibfield  {author} {\bibinfo {author} {\bibfnamefont {S.~P.}\ \bibnamefont
  {Jordan}}, \bibinfo {author} {\bibfnamefont {K.~S.~M.}\ \bibnamefont {Lee}},
  \ and\ \bibinfo {author} {\bibfnamefont {J.}~\bibnamefont {Preskill}},\
  }\href {\doibase 10.1126/science.1217069} {\bibfield  {journal} {\bibinfo
  {journal} {Science}\ }\textbf {\bibinfo {volume} {336}},\ \bibinfo {pages}
  {1130} (\bibinfo {year} {2012})}\BibitemShut {NoStop}%
\bibitem [{\citenamefont {Jordan}\ \emph {et~al.}(2014)\citenamefont {Jordan},
  \citenamefont {Lee},\ and\ \citenamefont {Preskill}}]{jordan_qic_2014}%
  \BibitemOpen
  \bibfield  {author} {\bibinfo {author} {\bibfnamefont {S.~P.}\ \bibnamefont
  {Jordan}}, \bibinfo {author} {\bibfnamefont {K.~S.~M.}\ \bibnamefont {Lee}},
  \ and\ \bibinfo {author} {\bibfnamefont {J.}~\bibnamefont {Preskill}},\
  }\href@noop {} {\bibfield  {journal} {\bibinfo  {journal} {Quant. Inf.
  Comput.}\ }\textbf {\bibinfo {volume} {14}},\ \bibinfo {pages} {1014}
  (\bibinfo {year} {2014})},\ \Eprint {http://arxiv.org/abs/1112.4833}
  {arXiv:1112.4833 [hep-th]} \BibitemShut {NoStop}%
\bibitem [{\citenamefont {Somma}(2016)}]{somma_qic_2016}%
  \BibitemOpen
  \bibfield  {author} {\bibinfo {author} {\bibfnamefont {R.~D.}\ \bibnamefont
  {Somma}},\ }\href {https://dl.acm.org/doi/10.5555/3179430.3179434} {\bibfield
   {journal} {\bibinfo  {journal} {Quant. Inf. Comput.}\ }\textbf {\bibinfo
  {volume} {16}},\ \bibinfo {pages} {1125} (\bibinfo {year}
  {2016})}\BibitemShut {NoStop}%
\bibitem [{\citenamefont {Macridin}\ \emph
  {et~al.}(2018{\natexlab{a}})\citenamefont {Macridin}, \citenamefont
  {Spentzouris}, \citenamefont {Amundson},\ and\ \citenamefont
  {Harnik}}]{macridin_prl_2018}%
  \BibitemOpen
  \bibfield  {author} {\bibinfo {author} {\bibfnamefont {A.}~\bibnamefont
  {Macridin}}, \bibinfo {author} {\bibfnamefont {P.}~\bibnamefont
  {Spentzouris}}, \bibinfo {author} {\bibfnamefont {J.}~\bibnamefont
  {Amundson}}, \ and\ \bibinfo {author} {\bibfnamefont {R.}~\bibnamefont
  {Harnik}},\ }\href {\doibase 10.1103/PhysRevLett.121.110504} {\bibfield
  {journal} {\bibinfo  {journal} {Phys. Rev. Lett.}\ }\textbf {\bibinfo
  {volume} {121}},\ \bibinfo {pages} {110504} (\bibinfo {year}
  {2018}{\natexlab{a}})}\BibitemShut {NoStop}%
\bibitem [{\citenamefont {Macridin}\ \emph
  {et~al.}(2018{\natexlab{b}})\citenamefont {Macridin}, \citenamefont
  {Spentzouris}, \citenamefont {Amundson},\ and\ \citenamefont
  {Harnik}}]{macridin_pra_2018}%
  \BibitemOpen
  \bibfield  {author} {\bibinfo {author} {\bibfnamefont {A.}~\bibnamefont
  {Macridin}}, \bibinfo {author} {\bibfnamefont {P.}~\bibnamefont
  {Spentzouris}}, \bibinfo {author} {\bibfnamefont {J.}~\bibnamefont
  {Amundson}}, \ and\ \bibinfo {author} {\bibfnamefont {R.}~\bibnamefont
  {Harnik}},\ }\href {\doibase 10.1103/PhysRevA.98.042312} {\bibfield
  {journal} {\bibinfo  {journal} {Phys. Rev. A}\ }\textbf {\bibinfo {volume}
  {98}},\ \bibinfo {pages} {042312} (\bibinfo {year}
  {2018}{\natexlab{b}})}\BibitemShut {NoStop}%
\bibitem [{\citenamefont {Klco}\ and\ \citenamefont
  {Savage}(2019)}]{klco_pra_2019}%
  \BibitemOpen
  \bibfield  {author} {\bibinfo {author} {\bibfnamefont {N.}~\bibnamefont
  {Klco}}\ and\ \bibinfo {author} {\bibfnamefont {M.~J.}\ \bibnamefont
  {Savage}},\ }\href {\doibase 10.1103/PhysRevA.99.052335} {\bibfield
  {journal} {\bibinfo  {journal} {Phys. Rev. A}\ }\textbf {\bibinfo {volume}
  {99}},\ \bibinfo {pages} {052335} (\bibinfo {year} {2019})}\BibitemShut
  {NoStop}%
\bibitem [{\citenamefont {Jaming}\ \emph {et~al.}(2016)\citenamefont {Jaming},
  \citenamefont {Karoui},\ and\ \citenamefont {Spektor}}]{jaming_2016}%
  \BibitemOpen
  \bibfield  {author} {\bibinfo {author} {\bibfnamefont {P.}~\bibnamefont
  {Jaming}}, \bibinfo {author} {\bibfnamefont {A.}~\bibnamefont {Karoui}}, \
  and\ \bibinfo {author} {\bibfnamefont {S.}~\bibnamefont {Spektor}},\ }\href
  {\doibase https://doi.org/10.1016/j.jat.2016.08.002} {\bibfield  {journal}
  {\bibinfo  {journal} {Journal of Approximation Theory}\ }\textbf {\bibinfo
  {volume} {212}},\ \bibinfo {pages} {41} (\bibinfo {year} {2016})}\BibitemShut
  {NoStop}%
\bibitem [{\citenamefont {Slepian}(1976)}]{slepian_ieee_1976}%
  \BibitemOpen
  \bibfield  {author} {\bibinfo {author} {\bibfnamefont {D.}~\bibnamefont
  {Slepian}},\ }\href {\doibase 10.1109/PROC.1976.10110} {\bibfield  {journal}
  {\bibinfo  {journal} {Proceedings of the IEEE}\ }\textbf {\bibinfo {volume}
  {64}},\ \bibinfo {pages} {292} (\bibinfo {year} {1976})}\BibitemShut
  {NoStop}%
\bibitem [{\citenamefont {Landau}\ and\ \citenamefont
  {Pollak}(1962)}]{Landau_Pollak_3_1962}%
  \BibitemOpen
  \bibfield  {author} {\bibinfo {author} {\bibfnamefont {H.~J.}\ \bibnamefont
  {Landau}}\ and\ \bibinfo {author} {\bibfnamefont {H.~O.}\ \bibnamefont
  {Pollak}},\ }\href {\doibase 10.1002/j.1538-7305.1962.tb03279.x} {\bibfield
  {journal} {\bibinfo  {journal} {The Bell System Technical Journal}\ }\textbf
  {\bibinfo {volume} {41}},\ \bibinfo {pages} {1295} (\bibinfo {year}
  {1962})}\BibitemShut {NoStop}%
\bibitem [{\citenamefont {Altepeter}\ \emph {et~al.}(2004)\citenamefont
  {Altepeter}, \citenamefont {James},\ and\ \citenamefont
  {Kwiat}}]{Altepeter2004}%
  \BibitemOpen
  \bibfield  {author} {\bibinfo {author} {\bibfnamefont {J.~B.}\ \bibnamefont
  {Altepeter}}, \bibinfo {author} {\bibfnamefont {D.~F.}\ \bibnamefont
  {James}}, \ and\ \bibinfo {author} {\bibfnamefont {P.~G.}\ \bibnamefont
  {Kwiat}},\ }\enquote {\bibinfo {title} {Qubit quantum state tomography},}\
  in\ \href {\doibase 10.1007/978-3-540-44481-7_4} {\emph {\bibinfo {booktitle}
  {Quantum State Estimation}}},\ \bibinfo {editor} {edited by\ \bibinfo
  {editor} {\bibfnamefont {M.}~\bibnamefont {Paris}}\ and\ \bibinfo {editor}
  {\bibfnamefont {J.}~\bibnamefont {{\v{R}}eh{\'a}{\v{c}}ek}}}\ (\bibinfo
  {publisher} {Springer Berlin Heidelberg},\ \bibinfo {address} {Berlin,
  Heidelberg},\ \bibinfo {year} {2004})\ pp.\ \bibinfo {pages}
  {113--145}\BibitemShut {NoStop}%
\bibitem [{\citenamefont {Nielsen}\ and\ \citenamefont
  {Chuang}(2010)}]{nielsen_chuang_2010}%
  \BibitemOpen
  \bibfield  {author} {\bibinfo {author} {\bibfnamefont {M.~A.}\ \bibnamefont
  {Nielsen}}\ and\ \bibinfo {author} {\bibfnamefont {I.~L.}\ \bibnamefont
  {Chuang}},\ }\href {\doibase 10.1017/CBO9780511976667} {\emph {\bibinfo
  {title} {Quantum Computation and Quantum Information: 10th Anniversary
  Edition}}}\ (\bibinfo  {publisher} {Cambridge University Press},\ \bibinfo
  {year} {2010})\BibitemShut {NoStop}%
\bibitem [{\citenamefont {Cleve}\ \emph {et~al.}(1998)\citenamefont {Cleve},
  \citenamefont {Ekert}, \citenamefont {Macchiavello},\ and\ \citenamefont
  {Mosca}}]{cleve_1998_procRsocA}%
  \BibitemOpen
  \bibfield  {author} {\bibinfo {author} {\bibfnamefont {R.}~\bibnamefont
  {Cleve}}, \bibinfo {author} {\bibfnamefont {A.}~\bibnamefont {Ekert}},
  \bibinfo {author} {\bibfnamefont {C.}~\bibnamefont {Macchiavello}}, \ and\
  \bibinfo {author} {\bibfnamefont {M.}~\bibnamefont {Mosca}},\ }\href
  {http://doi.org/10.1098/rspa.1998.0164} {\bibfield  {journal} {\bibinfo
  {journal} {Proc. R. Soc. Lond. A}\ }\textbf {\bibinfo {volume} {454}},\
  \bibinfo {pages} {339} (\bibinfo {year} {1998})}\BibitemShut {NoStop}%
\bibitem [{\citenamefont {Sawaya}\ \emph {et~al.}(2020)\citenamefont {Sawaya},
  \citenamefont {Menke}, \citenamefont {Kyaw}, \citenamefont {Johri},
  \citenamefont {Aspuru-Guzik},\ and\ \citenamefont
  {Guerreschi}}]{sawaya_npj_2020}%
  \BibitemOpen
  \bibfield  {author} {\bibinfo {author} {\bibfnamefont {N.~P.~D.}\
  \bibnamefont {Sawaya}}, \bibinfo {author} {\bibfnamefont {T.}~\bibnamefont
  {Menke}}, \bibinfo {author} {\bibfnamefont {T.~H.}\ \bibnamefont {Kyaw}},
  \bibinfo {author} {\bibfnamefont {S.}~\bibnamefont {Johri}}, \bibinfo
  {author} {\bibfnamefont {A.}~\bibnamefont {Aspuru-Guzik}}, \ and\ \bibinfo
  {author} {\bibfnamefont {G.~G.}\ \bibnamefont {Guerreschi}},\ }\href
  {\doibase 10.1038/s41534-020-0278-0} {\bibfield  {journal} {\bibinfo
  {journal} {npj Quantum Information}\ }\textbf {\bibinfo {volume} {6}},\
  \bibinfo {pages} {49} (\bibinfo {year} {2020})}\BibitemShut {NoStop}%
\bibitem [{\citenamefont {Barenco}\ \emph {et~al.}(1995)\citenamefont
  {Barenco}, \citenamefont {Bennett}, \citenamefont {Cleve}, \citenamefont
  {DiVincenzo}, \citenamefont {Margolus}, \citenamefont {Shor}, \citenamefont
  {Sleator}, \citenamefont {Smolin},\ and\ \citenamefont
  {Weinfurter}}]{Barenco_pra_1995}%
  \BibitemOpen
  \bibfield  {author} {\bibinfo {author} {\bibfnamefont {A.}~\bibnamefont
  {Barenco}}, \bibinfo {author} {\bibfnamefont {C.~H.}\ \bibnamefont
  {Bennett}}, \bibinfo {author} {\bibfnamefont {R.}~\bibnamefont {Cleve}},
  \bibinfo {author} {\bibfnamefont {D.~P.}\ \bibnamefont {DiVincenzo}},
  \bibinfo {author} {\bibfnamefont {N.}~\bibnamefont {Margolus}}, \bibinfo
  {author} {\bibfnamefont {P.}~\bibnamefont {Shor}}, \bibinfo {author}
  {\bibfnamefont {T.}~\bibnamefont {Sleator}}, \bibinfo {author} {\bibfnamefont
  {J.~A.}\ \bibnamefont {Smolin}}, \ and\ \bibinfo {author} {\bibfnamefont
  {H.}~\bibnamefont {Weinfurter}},\ }\href {\doibase 10.1103/PhysRevA.52.3457}
  {\bibfield  {journal} {\bibinfo  {journal} {Phys. Rev. A}\ }\textbf {\bibinfo
  {volume} {52}},\ \bibinfo {pages} {3457} (\bibinfo {year}
  {1995})}\BibitemShut {NoStop}%
\bibitem [{\citenamefont {Shende}\ \emph {et~al.}(2004)\citenamefont {Shende},
  \citenamefont {Markov},\ and\ \citenamefont {Bullock}}]{shende_pra69_2004}%
  \BibitemOpen
  \bibfield  {author} {\bibinfo {author} {\bibfnamefont {V.~V.}\ \bibnamefont
  {Shende}}, \bibinfo {author} {\bibfnamefont {I.~L.}\ \bibnamefont {Markov}},
  \ and\ \bibinfo {author} {\bibfnamefont {S.~S.}\ \bibnamefont {Bullock}},\
  }\href {\doibase 10.1103/PhysRevA.69.062321} {\bibfield  {journal} {\bibinfo
  {journal} {Phys. Rev. A}\ }\textbf {\bibinfo {volume} {69}},\ \bibinfo
  {pages} {062321} (\bibinfo {year} {2004})}\BibitemShut {NoStop}%
\bibitem [{\citenamefont {Krol}\ \emph {et~al.}(2021)\citenamefont {Krol},
  \citenamefont {Sarkar}, \citenamefont {Ashraf}, \citenamefont {Al-Ars},\ and\
  \citenamefont {Bertels}}]{krol_2021_arxiv}%
  \BibitemOpen
  \bibfield  {author} {\bibinfo {author} {\bibfnamefont {A.~M.}\ \bibnamefont
  {Krol}}, \bibinfo {author} {\bibfnamefont {A.}~\bibnamefont {Sarkar}},
  \bibinfo {author} {\bibfnamefont {I.}~\bibnamefont {Ashraf}}, \bibinfo
  {author} {\bibfnamefont {Z.}~\bibnamefont {Al-Ars}}, \ and\ \bibinfo {author}
  {\bibfnamefont {K.}~\bibnamefont {Bertels}},\ }\href {\doibase
  10.48550/ARXIV.2101.02993} {\  (\bibinfo {year} {2021}),\
  10.48550/ARXIV.2101.02993}\BibitemShut {NoStop}%
\bibitem [{\citenamefont {Li}\ \emph {et~al.}(2021)\citenamefont {Li},
  \citenamefont {Macridin}, \citenamefont {Spentzouris},\ and\ \citenamefont
  {Mrenna}}]{Li_2021}%
  \BibitemOpen
  \bibfield  {author} {\bibinfo {author} {\bibfnamefont {A.}~\bibnamefont
  {Li}}, \bibinfo {author} {\bibfnamefont {A.}~\bibnamefont {Macridin}},
  \bibinfo {author} {\bibfnamefont {P.}~\bibnamefont {Spentzouris}}, \ and\
  \bibinfo {author} {\bibfnamefont {S.}~\bibnamefont {Mrenna}},\ }\href@noop {}
  {\bibfield  {journal} {\bibinfo  {journal} {preprint}\ } (\bibinfo {year}
  {2021})}\BibitemShut {NoStop}%
\bibitem [{\citenamefont {Gieres}(2000)}]{Gieres_2000}%
  \BibitemOpen
  \bibfield  {author} {\bibinfo {author} {\bibfnamefont {F.}~\bibnamefont
  {Gieres}},\ }\href {\doibase 10.1088/0034-4885/63/12/201} {\bibfield
  {journal} {\bibinfo  {journal} {Reports on Progress in Physics}\ }\textbf
  {\bibinfo {volume} {63}},\ \bibinfo {pages} {1893} (\bibinfo {year}
  {2000})}\BibitemShut {NoStop}%
\bibitem [{\citenamefont {Becnel}\ and\ \citenamefont
  {Sengupta}(2015)}]{becnel_2015}%
  \BibitemOpen
  \bibfield  {author} {\bibinfo {author} {\bibfnamefont {J.}~\bibnamefont
  {Becnel}}\ and\ \bibinfo {author} {\bibfnamefont {A.}~\bibnamefont
  {Sengupta}},\ }\href {\doibase 10.3390/math3020527} {\bibfield  {journal}
  {\bibinfo  {journal} {Mathematics}\ }\textbf {\bibinfo {volume} {3}},\
  \bibinfo {pages} {527} (\bibinfo {year} {2015})}\BibitemShut {NoStop}%
\bibitem [{\citenamefont {Melrose}(2017)}]{schwartz_space}%
  \BibitemOpen
  \bibfield  {author} {\bibinfo {author} {\bibfnamefont {R.~B.}\ \bibnamefont
  {Melrose}},\ }\href
  {https://www.freebookcentre.net/maths-books-download/Differential-Analysis-Lecture-notes-by-Richard-B.-Melrose.html}
  {\enquote {\bibinfo {title} {Differential analysis lecture notes},}\ }
  (\bibinfo {year} {2017})\BibitemShut {NoStop}%
\bibitem [{\citenamefont {Suijlekom}(2021)}]{schwartz_space_math}%
  \BibitemOpen
  \bibfield  {author} {\bibinfo {author} {\bibfnamefont {W.~V.}\ \bibnamefont
  {Suijlekom}},\ }\href {https://mathworld.wolfram.com/SchwartzSpace.html}
  {\emph {\bibinfo {title} {Schwartz Space}}}\ (\bibinfo  {publisher} {From
  MathWorld. A Wolfram Web Resource, created by Eric W. Weisstein},\ \bibinfo
  {year} {2021})\BibitemShut {NoStop}%
\bibitem [{\citenamefont {{Shannon}}(1949)}]{Shannon_1949}%
  \BibitemOpen
  \bibfield  {author} {\bibinfo {author} {\bibfnamefont {C.~E.}\ \bibnamefont
  {{Shannon}}},\ }\href {\doibase 10.1109/JRPROC.1949.232969} {\bibfield
  {journal} {\bibinfo  {journal} {Proceedings of the IRE}\ }\textbf {\bibinfo
  {volume} {37}},\ \bibinfo {pages} {10} (\bibinfo {year} {1949})}\BibitemShut
  {NoStop}%
\bibitem [{\citenamefont {Engelberg}(2008)}]{Engelberg_2008}%
  \BibitemOpen
  \bibfield  {author} {\bibinfo {author} {\bibfnamefont {S.}~\bibnamefont
  {Engelberg}},\ }\enquote {\bibinfo {title} {Time-limited functions are not
  band-limited},}\ in\ \href {\doibase 10.1007/978-1-84800-119-0_3} {\emph
  {\bibinfo {booktitle} {Digital Signal Processing: An Experimental
  Approach}}}\ (\bibinfo  {publisher} {Springer London},\ \bibinfo {address}
  {London},\ \bibinfo {year} {2008})\ pp.\ \bibinfo {pages}
  {21--27}\BibitemShut {NoStop}%
\bibitem [{\citenamefont {Gradshteyn}\ and\ \citenamefont
  {Ryzhik}(1980)}]{Gradshteyn_1980}%
  \BibitemOpen
  \bibfield  {author} {\bibinfo {author} {\bibfnamefont {I.}~\bibnamefont
  {Gradshteyn}}\ and\ \bibinfo {author} {\bibfnamefont {I.}~\bibnamefont
  {Ryzhik}},\ }in\ \href {\doibase
  https://doi.org/10.1016/B978-0-12-294760-5.50019-2} {\emph {\bibinfo
  {booktitle} {Table of Integrals, Series, and Products}}},\ \bibinfo {editor}
  {edited by\ \bibinfo {editor} {\bibfnamefont {I.}~\bibnamefont {Gradshteyn}}\
  and\ \bibinfo {editor} {\bibfnamefont {I.}~\bibnamefont {Ryzhik}}}\ (\bibinfo
   {publisher} {Academic Press},\ \bibinfo {year} {1980})\ \bibinfo {note}
  {formula 7.376}\BibitemShut {NoStop}%
\bibitem [{\citenamefont {Osipov}\ \emph {et~al.}(2013)\citenamefont {Osipov},
  \citenamefont {Rokhlin},\ and\ \citenamefont {Xiao}}]{osipov2013prolate}%
  \BibitemOpen
  \bibfield  {author} {\bibinfo {author} {\bibfnamefont {A.}~\bibnamefont
  {Osipov}}, \bibinfo {author} {\bibfnamefont {V.}~\bibnamefont {Rokhlin}}, \
  and\ \bibinfo {author} {\bibfnamefont {H.}~\bibnamefont {Xiao}},\ }\href
  {https://books.google.com/books?id=wkG9BAAAQBAJ} {\emph {\bibinfo {title}
  {Prolate Spheroidal Wave Functions of Order Zero: Mathematical Tools for
  Bandlimited Approximation}}},\ Applied Mathematical Sciences\ (\bibinfo
  {publisher} {Springer US},\ \bibinfo {year} {2013})\BibitemShut {NoStop}%
\bibitem [{\citenamefont {Gerry}\ and\ \citenamefont
  {Knight}(2004)}]{gerry_knight_2004}%
  \BibitemOpen
  \bibfield  {author} {\bibinfo {author} {\bibfnamefont {C.}~\bibnamefont
  {Gerry}}\ and\ \bibinfo {author} {\bibfnamefont {P.}~\bibnamefont {Knight}},\
  }\href {\doibase 10.1017/CBO9780511791239} {\emph {\bibinfo {title}
  {Introductory Quantum Optics}}}\ (\bibinfo  {publisher} {Cambridge University
  Press},\ \bibinfo {year} {2004})\BibitemShut {NoStop}%
\bibitem [{\citenamefont {McClean}\ \emph {et~al.}(2016)\citenamefont
  {McClean}, \citenamefont {Romero}, \citenamefont {Babbush},\ and\
  \citenamefont {Aspuru-Guzik}}]{McClean_NJP_2016}%
  \BibitemOpen
  \bibfield  {author} {\bibinfo {author} {\bibfnamefont {J.~R.}\ \bibnamefont
  {McClean}}, \bibinfo {author} {\bibfnamefont {J.}~\bibnamefont {Romero}},
  \bibinfo {author} {\bibfnamefont {R.}~\bibnamefont {Babbush}}, \ and\
  \bibinfo {author} {\bibfnamefont {A.}~\bibnamefont {Aspuru-Guzik}},\ }\href
  {\doibase 10.1088/1367-2630/18/2/023023} {\bibfield  {journal} {\bibinfo
  {journal} {New Journal of Physics}\ }\textbf {\bibinfo {volume} {18}},\
  \bibinfo {pages} {023023} (\bibinfo {year} {2016})}\BibitemShut {NoStop}%
\bibitem [{\citenamefont {Cramer}\ \emph {et~al.}(2010)\citenamefont {Cramer},
  \citenamefont {Plenio}, \citenamefont {Flammia}, \citenamefont {Somma},
  \citenamefont {Gross}, \citenamefont {Bartlett}, \citenamefont
  {Landon-Cardinal}, \citenamefont {Poulin},\ and\ \citenamefont
  {Liu}}]{Cramer_nature_2010}%
  \BibitemOpen
  \bibfield  {author} {\bibinfo {author} {\bibfnamefont {M.}~\bibnamefont
  {Cramer}}, \bibinfo {author} {\bibfnamefont {M.~B.}\ \bibnamefont {Plenio}},
  \bibinfo {author} {\bibfnamefont {S.~T.}\ \bibnamefont {Flammia}}, \bibinfo
  {author} {\bibfnamefont {R.}~\bibnamefont {Somma}}, \bibinfo {author}
  {\bibfnamefont {D.}~\bibnamefont {Gross}}, \bibinfo {author} {\bibfnamefont
  {S.~D.}\ \bibnamefont {Bartlett}}, \bibinfo {author} {\bibfnamefont
  {O.}~\bibnamefont {Landon-Cardinal}}, \bibinfo {author} {\bibfnamefont
  {D.}~\bibnamefont {Poulin}}, \ and\ \bibinfo {author} {\bibfnamefont {Y.-K.}\
  \bibnamefont {Liu}},\ }\href@noop {} {\bibfield  {journal} {\bibinfo
  {journal} {Nature Communications}\ }\textbf {\bibinfo {volume} {1}},\
  \bibinfo {pages} {149} (\bibinfo {year} {2010})}\BibitemShut {NoStop}%
\bibitem [{\citenamefont {Lanyon}\ \emph {et~al.}(2017)\citenamefont {Lanyon},
  \citenamefont {Maier}, \citenamefont {Holz{\"a}pfel}, \citenamefont
  {Baumgratz}, \citenamefont {Hempel}, \citenamefont {Jurcevic}, \citenamefont
  {Dhand}, \citenamefont {Buyskikh}, \citenamefont {Daley}, \citenamefont
  {Cramer}, \citenamefont {Plenio}, \citenamefont {Blatt},\ and\ \citenamefont
  {Roos}}]{Lanyon_naturephys_2017}%
  \BibitemOpen
  \bibfield  {author} {\bibinfo {author} {\bibfnamefont {B.~P.}\ \bibnamefont
  {Lanyon}}, \bibinfo {author} {\bibfnamefont {C.}~\bibnamefont {Maier}},
  \bibinfo {author} {\bibfnamefont {M.}~\bibnamefont {Holz{\"a}pfel}}, \bibinfo
  {author} {\bibfnamefont {T.}~\bibnamefont {Baumgratz}}, \bibinfo {author}
  {\bibfnamefont {C.}~\bibnamefont {Hempel}}, \bibinfo {author} {\bibfnamefont
  {P.}~\bibnamefont {Jurcevic}}, \bibinfo {author} {\bibfnamefont
  {I.}~\bibnamefont {Dhand}}, \bibinfo {author} {\bibfnamefont {A.~S.}\
  \bibnamefont {Buyskikh}}, \bibinfo {author} {\bibfnamefont {A.~J.}\
  \bibnamefont {Daley}}, \bibinfo {author} {\bibfnamefont {M.}~\bibnamefont
  {Cramer}}, \bibinfo {author} {\bibfnamefont {M.~B.}\ \bibnamefont {Plenio}},
  \bibinfo {author} {\bibfnamefont {R.}~\bibnamefont {Blatt}}, \ and\ \bibinfo
  {author} {\bibfnamefont {C.~F.}\ \bibnamefont {Roos}},\ }\href@noop {}
  {\bibfield  {journal} {\bibinfo  {journal} {Nature Physics}\ }\textbf
  {\bibinfo {volume} {13}},\ \bibinfo {pages} {1158} (\bibinfo {year}
  {2017})}\BibitemShut {NoStop}%
\bibitem [{\citenamefont {Titchener}\ \emph {et~al.}(2018)\citenamefont
  {Titchener}, \citenamefont {Gr{\"a}fe}, \citenamefont {Heilmann},
  \citenamefont {Solntsev}, \citenamefont {Szameit},\ and\ \citenamefont
  {Sukhorukov}}]{titchener_npj_2018}%
  \BibitemOpen
  \bibfield  {author} {\bibinfo {author} {\bibfnamefont {J.~G.}\ \bibnamefont
  {Titchener}}, \bibinfo {author} {\bibfnamefont {M.}~\bibnamefont
  {Gr{\"a}fe}}, \bibinfo {author} {\bibfnamefont {R.}~\bibnamefont {Heilmann}},
  \bibinfo {author} {\bibfnamefont {A.~S.}\ \bibnamefont {Solntsev}}, \bibinfo
  {author} {\bibfnamefont {A.}~\bibnamefont {Szameit}}, \ and\ \bibinfo
  {author} {\bibfnamefont {A.~A.}\ \bibnamefont {Sukhorukov}},\ }\href
  {\doibase 10.1038/s41534-018-0063-5} {\bibfield  {journal} {\bibinfo
  {journal} {npj Quantum Information}\ }\textbf {\bibinfo {volume} {4}},\
  \bibinfo {pages} {19} (\bibinfo {year} {2018})}\BibitemShut {NoStop}%
\bibitem [{\citenamefont {H{\"a}ffner}\ \emph {et~al.}(2005)\citenamefont
  {H{\"a}ffner}, \citenamefont {H{\"a}nsel}, \citenamefont {Roos},
  \citenamefont {Benhelm}, \citenamefont {Chek-al kar}, \citenamefont
  {Chwalla}, \citenamefont {K{\"o}rber}, \citenamefont {Rapol}, \citenamefont
  {Riebe}, \citenamefont {Schmidt}, \citenamefont {Becher}, \citenamefont
  {G{\"u}hne}, \citenamefont {D{\"u}r},\ and\ \citenamefont
  {Blatt}}]{haffner_nature_2002}%
  \BibitemOpen
  \bibfield  {author} {\bibinfo {author} {\bibfnamefont {H.}~\bibnamefont
  {H{\"a}ffner}}, \bibinfo {author} {\bibfnamefont {W.}~\bibnamefont
  {H{\"a}nsel}}, \bibinfo {author} {\bibfnamefont {C.~F.}\ \bibnamefont
  {Roos}}, \bibinfo {author} {\bibfnamefont {J.}~\bibnamefont {Benhelm}},
  \bibinfo {author} {\bibfnamefont {D.}~\bibnamefont {Chek-al kar}}, \bibinfo
  {author} {\bibfnamefont {M.}~\bibnamefont {Chwalla}}, \bibinfo {author}
  {\bibfnamefont {T.}~\bibnamefont {K{\"o}rber}}, \bibinfo {author}
  {\bibfnamefont {U.~D.}\ \bibnamefont {Rapol}}, \bibinfo {author}
  {\bibfnamefont {M.}~\bibnamefont {Riebe}}, \bibinfo {author} {\bibfnamefont
  {P.~O.}\ \bibnamefont {Schmidt}}, \bibinfo {author} {\bibfnamefont
  {C.}~\bibnamefont {Becher}}, \bibinfo {author} {\bibfnamefont
  {O.}~\bibnamefont {G{\"u}hne}}, \bibinfo {author} {\bibfnamefont
  {W.}~\bibnamefont {D{\"u}r}}, \ and\ \bibinfo {author} {\bibfnamefont
  {R.}~\bibnamefont {Blatt}},\ }\href@noop {} {\bibfield  {journal} {\bibinfo
  {journal} {Nature}\ }\textbf {\bibinfo {volume} {438}},\ \bibinfo {pages}
  {643} (\bibinfo {year} {2005})}\BibitemShut {NoStop}%
\bibitem [{\citenamefont {Song}\ \emph {et~al.}(2017)\citenamefont {Song},
  \citenamefont {Xu}, \citenamefont {Liu}, \citenamefont {Yang}, \citenamefont
  {Zheng}, \citenamefont {Deng}, \citenamefont {Xie}, \citenamefont {Huang},
  \citenamefont {Guo}, \citenamefont {Zhang}, \citenamefont {Zhang},
  \citenamefont {Xu}, \citenamefont {Zheng}, \citenamefont {Zhu}, \citenamefont
  {Wang}, \citenamefont {Chen}, \citenamefont {Lu}, \citenamefont {Han},\ and\
  \citenamefont {Pan}}]{Song_prl_2017}%
  \BibitemOpen
  \bibfield  {author} {\bibinfo {author} {\bibfnamefont {C.}~\bibnamefont
  {Song}}, \bibinfo {author} {\bibfnamefont {K.}~\bibnamefont {Xu}}, \bibinfo
  {author} {\bibfnamefont {W.}~\bibnamefont {Liu}}, \bibinfo {author}
  {\bibfnamefont {C.-p.}\ \bibnamefont {Yang}}, \bibinfo {author}
  {\bibfnamefont {S.-B.}\ \bibnamefont {Zheng}}, \bibinfo {author}
  {\bibfnamefont {H.}~\bibnamefont {Deng}}, \bibinfo {author} {\bibfnamefont
  {Q.}~\bibnamefont {Xie}}, \bibinfo {author} {\bibfnamefont {K.}~\bibnamefont
  {Huang}}, \bibinfo {author} {\bibfnamefont {Q.}~\bibnamefont {Guo}}, \bibinfo
  {author} {\bibfnamefont {L.}~\bibnamefont {Zhang}}, \bibinfo {author}
  {\bibfnamefont {P.}~\bibnamefont {Zhang}}, \bibinfo {author} {\bibfnamefont
  {D.}~\bibnamefont {Xu}}, \bibinfo {author} {\bibfnamefont {D.}~\bibnamefont
  {Zheng}}, \bibinfo {author} {\bibfnamefont {X.}~\bibnamefont {Zhu}}, \bibinfo
  {author} {\bibfnamefont {H.}~\bibnamefont {Wang}}, \bibinfo {author}
  {\bibfnamefont {Y.-A.}\ \bibnamefont {Chen}}, \bibinfo {author}
  {\bibfnamefont {C.-Y.}\ \bibnamefont {Lu}}, \bibinfo {author} {\bibfnamefont
  {S.}~\bibnamefont {Han}}, \ and\ \bibinfo {author} {\bibfnamefont {J.-W.}\
  \bibnamefont {Pan}},\ }\href {\doibase 10.1103/PhysRevLett.119.180511}
  {\bibfield  {journal} {\bibinfo  {journal} {Phys. Rev. Lett.}\ }\textbf
  {\bibinfo {volume} {119}},\ \bibinfo {pages} {180511} (\bibinfo {year}
  {2017})}\BibitemShut {NoStop}%
\bibitem [{\citenamefont {Häner}\ \emph {et~al.}(2018)\citenamefont {Häner},
  \citenamefont {Roetteler},\ and\ \citenamefont {Svore}}]{haner_2018}%
  \BibitemOpen
  \bibfield  {author} {\bibinfo {author} {\bibfnamefont {T.}~\bibnamefont
  {Häner}}, \bibinfo {author} {\bibfnamefont {M.}~\bibnamefont {Roetteler}}, \
  and\ \bibinfo {author} {\bibfnamefont {K.~M.}\ \bibnamefont {Svore}},\
  }\href@noop {} {\  (\bibinfo {year} {2018})},\ \Eprint
  {http://arxiv.org/abs/1805.12445} {arXiv:1805.12445 [quant-ph]} \BibitemShut
  {NoStop}%
\bibitem [{\citenamefont {Bhaskar}\ \emph {et~al.}(2015)\citenamefont
  {Bhaskar}, \citenamefont {Hadfield}, \citenamefont {Papageorgiou},\ and\
  \citenamefont {Petras}}]{bhaskar_2015}%
  \BibitemOpen
  \bibfield  {author} {\bibinfo {author} {\bibfnamefont {M.~K.}\ \bibnamefont
  {Bhaskar}}, \bibinfo {author} {\bibfnamefont {S.}~\bibnamefont {Hadfield}},
  \bibinfo {author} {\bibfnamefont {A.}~\bibnamefont {Papageorgiou}}, \ and\
  \bibinfo {author} {\bibfnamefont {I.}~\bibnamefont {Petras}},\ }\href@noop {}
  {\  (\bibinfo {year} {2015})},\ \Eprint {http://arxiv.org/abs/1511.08253}
  {arXiv:1511.08253 [quant-ph]} \BibitemShut {NoStop}%
\bibitem [{\citenamefont {Quantum}\ \emph {et~al.}(2020)\citenamefont
  {Quantum}, \citenamefont {Collaborators} \emph {et~al.}}]{google2020hartree}%
  \BibitemOpen
  \bibfield  {author} {\bibinfo {author} {\bibfnamefont {G.~A.}\ \bibnamefont
  {Quantum}}, \bibinfo {author} {\bibnamefont {Collaborators}},  \emph
  {et~al.},\ }\href {\doibase 10.1126/science.abb9811} {\bibfield  {journal}
  {\bibinfo  {journal} {Science}\ }\textbf {\bibinfo {volume} {369}},\ \bibinfo
  {pages} {1084} (\bibinfo {year} {2020})}\BibitemShut {NoStop}%
\bibitem [{\citenamefont {Jurcevic}\ \emph {et~al.}(2021)\citenamefont
  {Jurcevic}, \citenamefont {Javadi-Abhari}, \citenamefont {Bishop},
  \citenamefont {Lauer}, \citenamefont {Bogorin}, \citenamefont {Brink},
  \citenamefont {Capelluto}, \citenamefont {G{\"u}nl{\"u}k}, \citenamefont
  {Itoko}, \citenamefont {Kanazawa} \emph
  {et~al.}}]{jurcevic2021demonstration}%
  \BibitemOpen
  \bibfield  {author} {\bibinfo {author} {\bibfnamefont {P.}~\bibnamefont
  {Jurcevic}}, \bibinfo {author} {\bibfnamefont {A.}~\bibnamefont
  {Javadi-Abhari}}, \bibinfo {author} {\bibfnamefont {L.~S.}\ \bibnamefont
  {Bishop}}, \bibinfo {author} {\bibfnamefont {I.}~\bibnamefont {Lauer}},
  \bibinfo {author} {\bibfnamefont {D.~F.}\ \bibnamefont {Bogorin}}, \bibinfo
  {author} {\bibfnamefont {M.}~\bibnamefont {Brink}}, \bibinfo {author}
  {\bibfnamefont {L.}~\bibnamefont {Capelluto}}, \bibinfo {author}
  {\bibfnamefont {O.}~\bibnamefont {G{\"u}nl{\"u}k}}, \bibinfo {author}
  {\bibfnamefont {T.}~\bibnamefont {Itoko}}, \bibinfo {author} {\bibfnamefont
  {N.}~\bibnamefont {Kanazawa}},  \emph {et~al.},\ }\href {\doibase
  10.1088/2058-9565/abe519} {\bibfield  {journal} {\bibinfo  {journal} {Quantum
  Science and Technology}\ }\textbf {\bibinfo {volume} {6}},\ \bibinfo {pages}
  {025020} (\bibinfo {year} {2021})}\BibitemShut {NoStop}%
\bibitem [{\citenamefont {Kim}\ \emph {et~al.}(2021)\citenamefont {Kim},
  \citenamefont {Wood}, \citenamefont {Yoder}, \citenamefont {Merkel},
  \citenamefont {Gambetta}, \citenamefont {Temme},\ and\ \citenamefont
  {Kandala}}]{kim_arxiv_2021}%
  \BibitemOpen
  \bibfield  {author} {\bibinfo {author} {\bibfnamefont {Y.}~\bibnamefont
  {Kim}}, \bibinfo {author} {\bibfnamefont {C.~J.}\ \bibnamefont {Wood}},
  \bibinfo {author} {\bibfnamefont {T.~J.}\ \bibnamefont {Yoder}}, \bibinfo
  {author} {\bibfnamefont {S.~T.}\ \bibnamefont {Merkel}}, \bibinfo {author}
  {\bibfnamefont {J.~M.}\ \bibnamefont {Gambetta}}, \bibinfo {author}
  {\bibfnamefont {K.}~\bibnamefont {Temme}}, \ and\ \bibinfo {author}
  {\bibfnamefont {A.}~\bibnamefont {Kandala}},\ }\href {\doibase
  10.48550/ARXIV.2108.09197} {\  (\bibinfo {year} {2021}),\
  10.48550/ARXIV.2108.09197}\BibitemShut {NoStop}%
\bibitem [{\citenamefont {Arute}\ \emph {et~al.}(2020)\citenamefont {Arute},
  \citenamefont {Arya}, \citenamefont {Babbush}, \citenamefont {Bacon},
  \citenamefont {Bardin}, \citenamefont {Barends}, \citenamefont {Bengtsson},
  \citenamefont {Boixo}, \citenamefont {Broughton}, \citenamefont {Buckley},
  \citenamefont {Buell}, \citenamefont {Burkett}, \citenamefont {Bushnell},
  \citenamefont {Chen}, \citenamefont {Chen}, \citenamefont {Chen},
  \citenamefont {Chiaro}, \citenamefont {Collins}, \citenamefont {Cotton},
  \citenamefont {Courtney}, \citenamefont {Demura}, \citenamefont {Derk},
  \citenamefont {Dunsworth}, \citenamefont {Eppens}, \citenamefont {Eckl},
  \citenamefont {Erickson}, \citenamefont {Farhi}, \citenamefont {Fowler},
  \citenamefont {Foxen}, \citenamefont {Gidney}, \citenamefont {Giustina},
  \citenamefont {Graff}, \citenamefont {Gross}, \citenamefont {Habegger},
  \citenamefont {Harrigan}, \citenamefont {Ho}, \citenamefont {Hong},
  \citenamefont {Huang}, \citenamefont {Huggins}, \citenamefont {Ioffe},
  \citenamefont {Isakov}, \citenamefont {Jeffrey}, \citenamefont {Jiang},
  \citenamefont {Jones}, \citenamefont {Kafri}, \citenamefont {Kechedzhi},
  \citenamefont {Kelly}, \citenamefont {Kim}, \citenamefont {Klimov},
  \citenamefont {Korotkov}, \citenamefont {Kostritsa}, \citenamefont
  {Landhuis}, \citenamefont {Laptev}, \citenamefont {Lindmark}, \citenamefont
  {Lucero}, \citenamefont {Marthaler}, \citenamefont {Martin}, \citenamefont
  {Martinis}, \citenamefont {Marusczyk}, \citenamefont {McArdle}, \citenamefont
  {McClean}, \citenamefont {McCourt}, \citenamefont {McEwen}, \citenamefont
  {Megrant}, \citenamefont {Mejuto-Zaera}, \citenamefont {Mi}, \citenamefont
  {Mohseni}, \citenamefont {Mruczkiewicz}, \citenamefont {Mutus}, \citenamefont
  {Naaman}, \citenamefont {Neeley}, \citenamefont {Neill}, \citenamefont
  {Neven}, \citenamefont {Newman}, \citenamefont {Niu}, \citenamefont
  {O'Brien}, \citenamefont {Ostby}, \citenamefont {Pató}, \citenamefont
  {Petukhov}, \citenamefont {Putterman}, \citenamefont {Quintana},
  \citenamefont {Reiner}, \citenamefont {Roushan}, \citenamefont {Rubin},
  \citenamefont {Sank}, \citenamefont {Satzinger}, \citenamefont {Smelyanskiy},
  \citenamefont {Strain}, \citenamefont {Sung}, \citenamefont {Schmitteckert},
  \citenamefont {Szalay}, \citenamefont {Tubman}, \citenamefont {Vainsencher},
  \citenamefont {White}, \citenamefont {Vogt}, \citenamefont {Yao},
  \citenamefont {Yeh}, \citenamefont {Zalcman},\ and\ \citenamefont
  {Zanker}}]{google_arxiv_2020}%
  \BibitemOpen
  \bibfield  {author} {\bibinfo {author} {\bibfnamefont {F.}~\bibnamefont
  {Arute}}, \bibinfo {author} {\bibfnamefont {K.}~\bibnamefont {Arya}},
  \bibinfo {author} {\bibfnamefont {R.}~\bibnamefont {Babbush}}, \bibinfo
  {author} {\bibfnamefont {D.}~\bibnamefont {Bacon}}, \bibinfo {author}
  {\bibfnamefont {J.~C.}\ \bibnamefont {Bardin}}, \bibinfo {author}
  {\bibfnamefont {R.}~\bibnamefont {Barends}}, \bibinfo {author} {\bibfnamefont
  {A.}~\bibnamefont {Bengtsson}}, \bibinfo {author} {\bibfnamefont
  {S.}~\bibnamefont {Boixo}}, \bibinfo {author} {\bibfnamefont
  {M.}~\bibnamefont {Broughton}}, \bibinfo {author} {\bibfnamefont {B.~B.}\
  \bibnamefont {Buckley}}, \bibinfo {author} {\bibfnamefont {D.~A.}\
  \bibnamefont {Buell}}, \bibinfo {author} {\bibfnamefont {B.}~\bibnamefont
  {Burkett}}, \bibinfo {author} {\bibfnamefont {N.}~\bibnamefont {Bushnell}},
  \bibinfo {author} {\bibfnamefont {Y.}~\bibnamefont {Chen}}, \bibinfo {author}
  {\bibfnamefont {Z.}~\bibnamefont {Chen}}, \bibinfo {author} {\bibfnamefont
  {Y.-A.}\ \bibnamefont {Chen}}, \bibinfo {author} {\bibfnamefont
  {B.}~\bibnamefont {Chiaro}}, \bibinfo {author} {\bibfnamefont
  {R.}~\bibnamefont {Collins}}, \bibinfo {author} {\bibfnamefont {S.~J.}\
  \bibnamefont {Cotton}}, \bibinfo {author} {\bibfnamefont {W.}~\bibnamefont
  {Courtney}}, \bibinfo {author} {\bibfnamefont {S.}~\bibnamefont {Demura}},
  \bibinfo {author} {\bibfnamefont {A.}~\bibnamefont {Derk}}, \bibinfo {author}
  {\bibfnamefont {A.}~\bibnamefont {Dunsworth}}, \bibinfo {author}
  {\bibfnamefont {D.}~\bibnamefont {Eppens}}, \bibinfo {author} {\bibfnamefont
  {T.}~\bibnamefont {Eckl}}, \bibinfo {author} {\bibfnamefont {C.}~\bibnamefont
  {Erickson}}, \bibinfo {author} {\bibfnamefont {E.}~\bibnamefont {Farhi}},
  \bibinfo {author} {\bibfnamefont {A.}~\bibnamefont {Fowler}}, \bibinfo
  {author} {\bibfnamefont {B.}~\bibnamefont {Foxen}}, \bibinfo {author}
  {\bibfnamefont {C.}~\bibnamefont {Gidney}}, \bibinfo {author} {\bibfnamefont
  {M.}~\bibnamefont {Giustina}}, \bibinfo {author} {\bibfnamefont
  {R.}~\bibnamefont {Graff}}, \bibinfo {author} {\bibfnamefont {J.~A.}\
  \bibnamefont {Gross}}, \bibinfo {author} {\bibfnamefont {S.}~\bibnamefont
  {Habegger}}, \bibinfo {author} {\bibfnamefont {M.~P.}\ \bibnamefont
  {Harrigan}}, \bibinfo {author} {\bibfnamefont {A.}~\bibnamefont {Ho}},
  \bibinfo {author} {\bibfnamefont {S.}~\bibnamefont {Hong}}, \bibinfo {author}
  {\bibfnamefont {T.}~\bibnamefont {Huang}}, \bibinfo {author} {\bibfnamefont
  {W.}~\bibnamefont {Huggins}}, \bibinfo {author} {\bibfnamefont {L.~B.}\
  \bibnamefont {Ioffe}}, \bibinfo {author} {\bibfnamefont {S.~V.}\ \bibnamefont
  {Isakov}}, \bibinfo {author} {\bibfnamefont {E.}~\bibnamefont {Jeffrey}},
  \bibinfo {author} {\bibfnamefont {Z.}~\bibnamefont {Jiang}}, \bibinfo
  {author} {\bibfnamefont {C.}~\bibnamefont {Jones}}, \bibinfo {author}
  {\bibfnamefont {D.}~\bibnamefont {Kafri}}, \bibinfo {author} {\bibfnamefont
  {K.}~\bibnamefont {Kechedzhi}}, \bibinfo {author} {\bibfnamefont
  {J.}~\bibnamefont {Kelly}}, \bibinfo {author} {\bibfnamefont
  {S.}~\bibnamefont {Kim}}, \bibinfo {author} {\bibfnamefont {P.~V.}\
  \bibnamefont {Klimov}}, \bibinfo {author} {\bibfnamefont {A.~N.}\
  \bibnamefont {Korotkov}}, \bibinfo {author} {\bibfnamefont {F.}~\bibnamefont
  {Kostritsa}}, \bibinfo {author} {\bibfnamefont {D.}~\bibnamefont {Landhuis}},
  \bibinfo {author} {\bibfnamefont {P.}~\bibnamefont {Laptev}}, \bibinfo
  {author} {\bibfnamefont {M.}~\bibnamefont {Lindmark}}, \bibinfo {author}
  {\bibfnamefont {E.}~\bibnamefont {Lucero}}, \bibinfo {author} {\bibfnamefont
  {M.}~\bibnamefont {Marthaler}}, \bibinfo {author} {\bibfnamefont
  {O.}~\bibnamefont {Martin}}, \bibinfo {author} {\bibfnamefont {J.~M.}\
  \bibnamefont {Martinis}}, \bibinfo {author} {\bibfnamefont {A.}~\bibnamefont
  {Marusczyk}}, \bibinfo {author} {\bibfnamefont {S.}~\bibnamefont {McArdle}},
  \bibinfo {author} {\bibfnamefont {J.~R.}\ \bibnamefont {McClean}}, \bibinfo
  {author} {\bibfnamefont {T.}~\bibnamefont {McCourt}}, \bibinfo {author}
  {\bibfnamefont {M.}~\bibnamefont {McEwen}}, \bibinfo {author} {\bibfnamefont
  {A.}~\bibnamefont {Megrant}}, \bibinfo {author} {\bibfnamefont
  {C.}~\bibnamefont {Mejuto-Zaera}}, \bibinfo {author} {\bibfnamefont
  {X.}~\bibnamefont {Mi}}, \bibinfo {author} {\bibfnamefont {M.}~\bibnamefont
  {Mohseni}}, \bibinfo {author} {\bibfnamefont {W.}~\bibnamefont
  {Mruczkiewicz}}, \bibinfo {author} {\bibfnamefont {J.}~\bibnamefont {Mutus}},
  \bibinfo {author} {\bibfnamefont {O.}~\bibnamefont {Naaman}}, \bibinfo
  {author} {\bibfnamefont {M.}~\bibnamefont {Neeley}}, \bibinfo {author}
  {\bibfnamefont {C.}~\bibnamefont {Neill}}, \bibinfo {author} {\bibfnamefont
  {H.}~\bibnamefont {Neven}}, \bibinfo {author} {\bibfnamefont
  {M.}~\bibnamefont {Newman}}, \bibinfo {author} {\bibfnamefont {M.~Y.}\
  \bibnamefont {Niu}}, \bibinfo {author} {\bibfnamefont {T.~E.}\ \bibnamefont
  {O'Brien}}, \bibinfo {author} {\bibfnamefont {E.}~\bibnamefont {Ostby}},
  \bibinfo {author} {\bibfnamefont {B.}~\bibnamefont {Pató}}, \bibinfo
  {author} {\bibfnamefont {A.}~\bibnamefont {Petukhov}}, \bibinfo {author}
  {\bibfnamefont {H.}~\bibnamefont {Putterman}}, \bibinfo {author}
  {\bibfnamefont {C.}~\bibnamefont {Quintana}}, \bibinfo {author}
  {\bibfnamefont {J.-M.}\ \bibnamefont {Reiner}}, \bibinfo {author}
  {\bibfnamefont {P.}~\bibnamefont {Roushan}}, \bibinfo {author} {\bibfnamefont
  {N.~C.}\ \bibnamefont {Rubin}}, \bibinfo {author} {\bibfnamefont
  {D.}~\bibnamefont {Sank}}, \bibinfo {author} {\bibfnamefont {K.~J.}\
  \bibnamefont {Satzinger}}, \bibinfo {author} {\bibfnamefont {V.}~\bibnamefont
  {Smelyanskiy}}, \bibinfo {author} {\bibfnamefont {D.}~\bibnamefont {Strain}},
  \bibinfo {author} {\bibfnamefont {K.~J.}\ \bibnamefont {Sung}}, \bibinfo
  {author} {\bibfnamefont {P.}~\bibnamefont {Schmitteckert}}, \bibinfo {author}
  {\bibfnamefont {M.}~\bibnamefont {Szalay}}, \bibinfo {author} {\bibfnamefont
  {N.~M.}\ \bibnamefont {Tubman}}, \bibinfo {author} {\bibfnamefont
  {A.}~\bibnamefont {Vainsencher}}, \bibinfo {author} {\bibfnamefont
  {T.}~\bibnamefont {White}}, \bibinfo {author} {\bibfnamefont
  {N.}~\bibnamefont {Vogt}}, \bibinfo {author} {\bibfnamefont {Z.~J.}\
  \bibnamefont {Yao}}, \bibinfo {author} {\bibfnamefont {P.}~\bibnamefont
  {Yeh}}, \bibinfo {author} {\bibfnamefont {A.}~\bibnamefont {Zalcman}}, \ and\
  \bibinfo {author} {\bibfnamefont {S.}~\bibnamefont {Zanker}},\ }\href
  {\doibase 10.48550/ARXIV.2010.07965} {\  (\bibinfo {year} {2020}),\
  10.48550/ARXIV.2010.07965}\BibitemShut {NoStop}%
\bibitem [{\citenamefont {C{\^\i}rstoiu}\ \emph {et~al.}(2020)\citenamefont
  {C{\^\i}rstoiu}, \citenamefont {Holmes}, \citenamefont {Iosue}, \citenamefont
  {Cincio}, \citenamefont {Coles},\ and\ \citenamefont
  {Sornborger}}]{cirstoiu_npj_2020}%
  \BibitemOpen
  \bibfield  {author} {\bibinfo {author} {\bibfnamefont {C.}~\bibnamefont
  {C{\^\i}rstoiu}}, \bibinfo {author} {\bibfnamefont {Z.}~\bibnamefont
  {Holmes}}, \bibinfo {author} {\bibfnamefont {J.}~\bibnamefont {Iosue}},
  \bibinfo {author} {\bibfnamefont {L.}~\bibnamefont {Cincio}}, \bibinfo
  {author} {\bibfnamefont {P.~J.}\ \bibnamefont {Coles}}, \ and\ \bibinfo
  {author} {\bibfnamefont {A.}~\bibnamefont {Sornborger}},\ }\href@noop {}
  {\bibfield  {journal} {\bibinfo  {journal} {npj Quantum Information}\
  }\textbf {\bibinfo {volume} {6}},\ \bibinfo {pages} {82} (\bibinfo {year}
  {2020})}\BibitemShut {NoStop}%
\bibitem [{\citenamefont {Benedetti}\ \emph {et~al.}(2021)\citenamefont
  {Benedetti}, \citenamefont {Fiorentini},\ and\ \citenamefont
  {Lubasch}}]{benedetti_prr_2021}%
  \BibitemOpen
  \bibfield  {author} {\bibinfo {author} {\bibfnamefont {M.}~\bibnamefont
  {Benedetti}}, \bibinfo {author} {\bibfnamefont {M.}~\bibnamefont
  {Fiorentini}}, \ and\ \bibinfo {author} {\bibfnamefont {M.}~\bibnamefont
  {Lubasch}},\ }\href {\doibase 10.1103/PhysRevResearch.3.033083} {\bibfield
  {journal} {\bibinfo  {journal} {Phys. Rev. Research}\ }\textbf {\bibinfo
  {volume} {3}},\ \bibinfo {pages} {033083} (\bibinfo {year}
  {2021})}\BibitemShut {NoStop}%
\bibitem [{\citenamefont {DeVries}\ and\ \citenamefont
  {Hasbun}(2011)}]{devries2011first}%
  \BibitemOpen
  \bibfield  {author} {\bibinfo {author} {\bibfnamefont {P.}~\bibnamefont
  {DeVries}}\ and\ \bibinfo {author} {\bibfnamefont {J.}~\bibnamefont
  {Hasbun}},\ }\href {https://books.google.com/books?id=2ec2r6YCJeAC} {\emph
  {\bibinfo {title} {A First Course in Computational Physics}}}\ (\bibinfo
  {publisher} {Jones \& Bartlett Learning},\ \bibinfo {year}
  {2011})\BibitemShut {NoStop}%
\bibitem [{\citenamefont {Cooley}\ \emph {et~al.}(1967)\citenamefont {Cooley},
  \citenamefont {Lewis},\ and\ \citenamefont {Welch}}]{cooley_ieee_1967}%
  \BibitemOpen
  \bibfield  {author} {\bibinfo {author} {\bibfnamefont {J.}~\bibnamefont
  {Cooley}}, \bibinfo {author} {\bibfnamefont {P.}~\bibnamefont {Lewis}}, \
  and\ \bibinfo {author} {\bibfnamefont {P.}~\bibnamefont {Welch}},\ }\href
  {\doibase 10.1109/TAU.1967.1161904} {\bibfield  {journal} {\bibinfo
  {journal} {IEEE Transactions on Audio and Electroacoustics}\ }\textbf
  {\bibinfo {volume} {15}},\ \bibinfo {pages} {79} (\bibinfo {year}
  {1967})}\BibitemShut {NoStop}%
\end{thebibliography}
%

\end{document}